\documentclass[twocolumn, times]{aastex631}

\usepackage{threeparttable} % for table notes
\usepackage{float}
\usepackage{amsmath}

\usepackage{array, boldline, makecell, booktabs}

\renewcommand{\textbf}[1]{#1}

\begin{document}

%\title{Comparative Cross-correlation Spectroscopy of Hot and Ultra-hot Jupiter Atmospheres using Different Ground-based High-resolution Spectrographs}

\title{A Comparative Simulation Study of Hot and Ultra-hot Jupiter Atmospheres using Different Ground-based High-resolution Spectrographs with Cross-correlation Spectroscopy}

\correspondingauthor{Liton Majumdar}
\email{liton@niser.ac.in, dr.liton.majumdar@gmail.com}

\author[0000-0002-7033-209X]{Dwaipayan Dubey}
\affiliation{School of Earth and Planetary Sciences, National Institute of Science Education and Research, Jatni 752050, Odisha, India}
\affiliation{Homi Bhabha National Institute, Training School Complex, Anushaktinagar, Mumbai 400094, India}
\affiliation{Universitäts-Sternwarte, Fakultät für Physik, Ludwig-Maximilians-Universität München, Scheinerstr. 1, D-81679 München, Germany}
\affiliation{Exzellenzcluster `Origins’, Boltzmannstr. 2, D-85748 Garching, Germany}

\author[0000-0001-7031-8039]{Liton Majumdar}
\affiliation{School of Earth and Planetary Sciences, National Institute of Science Education and Research, Jatni 752050, Odisha, India}
\affiliation{Homi Bhabha National Institute, Training School Complex, Anushaktinagar, Mumbai 400094, India}

\begin{abstract}

In the era of state-of-the-art space-borne telescopes, high-resolution ground-based observation has emerged as a crucial method for characterizing exoplanets, providing essential insights into their atmospheric compositions. In the optical and NIR regions, high-resolution spectroscopy has been powerful for hot Jupiters (HJ) and ultra-hot Jupiters (UHJ) during their primary transits, as it can probe molecules with better sensitivity. Here, we focus on a comparative simulation study of \textbf{WASP-76 b} (UHJ) and \textbf{WASP-77 A b} (HJ) for different number of transits, utilizing three ground-based spectrographs \textbf{(GIANO-B (TNG), CARMENES (CAHA), and ANDES (E-ELT))} with varying instrumental parameters, spectral coverages, and resolutions. We aim to evaluate the feasibility of the upcoming ground-based European Extremely Large Telescope (E-ELT) in probing molecules from planet atmospheres and how it surpasses other ground-based observatories in terms of detectability. With the 1-D model, petitCODE, we have self-consistently simulated the atmospheric pressure-temperature profiles, which are subsequently integrated into the 1-D chemical kinetics model, VULCAN, to evolve the atmospheric chemistry. High-resolution spectra are obtained by performing line-by-line radiative transfer using petitRADTRANS. Finally, we use the resulting spectra to assess the detectability ($\mathrm{\sigma_{det}}$) of molecular bands, employing the ground-based noise simulator SPECTR. Utilizing cross-correlation spectroscopy, we have successfully demonstrated the robust consistency between our simulation study and real-time observations for both planets. ANDES excels overall in molecular detection due to its enhanced instrumental architecture, reinforcing E-ELT's importance for studying exoplanet atmospheres. \textbf{Additionally, our theoretical simulations predict the detection of $\mathrm{CO}$, $\mathrm{NH_3}$, and $\mathrm{H_2S}$ on \textbf{WASP-76 b} atmosphere with a $\mathrm{\sigma_{det}}$ $>$ 3.}

\end{abstract}


\section{Introduction} \label{sec:intro}

Exoplanetary science has made remarkable advancements in recent years, driven by numerous discoveries that have significantly changed our understanding of planetary systems and their complex characteristics. In recent decades, spectroscopic measurements have become an essential window for understanding atmospheric compositions, its dynamics, and temperature profiles of exoplanets \citep{crossfield2015observations}. While space-based observations have historically been crucial in providing initial characterizations of atmospheres for transiting exoplanets, ground-based observations have increasingly taken on a significant role in undertaking in-depth investigations of their atmospheric chemistry \citep{snellen2010orbital, wyttenbach2015spectrally, brogi2016rotation, brogi2018exoplanet, alonso2019multiple,sanchez2019water, giacobbe2021five}. In contrast to low-resolution spectroscopy, high-resolution spectroscopy offers the clear differentiation of molecular fingerprints originating from various species, hence facilitating more accurate detection. \textbf{The initial breakthrough in ground-based observations occurred with the detection of Na in the atmospheres of HD 189733 b \citep{redfield2008sodium} and HD 209458 b \citep{snellen2008ground}. Since then, a large number of molecules have been identified in $\mathrm{H_2}$-dominated atmospheres of gas giants, primarily in the near-infrared spectral range, leveraging high-resolution spectroscopy: $\mathrm{H_2O}$ \citep{birkby2013detection,alonso2019multiple}, CO \citep{snellen2010orbital,brogi2012signature}, $\mathrm{CH_4}$ \citep{guilluy2019exoplanet}, $\mathrm{C_2H_2}$, $\mathrm{NH_3}$ \citep{giacobbe2021five,carleo2022gaps}, HCN \citep{hawker2018evidence, cabot2019robustness}, etc.}

Ground-based observatories, particularly the inbuilt high-resolution spectrographs, can resolve the molecular features into an ensemble of distinctive lines. In spectroscopy, the signal-to-noise ratio improves as the square root of the number of lines observed ($\mathrm{N_{lines}}$) in a planetary spectrum, hence making high-resolution spectroscopy more suitable for broadband absorbers of comparable line strengths \citep{birkby2018exoplanet}. \textbf{For high-resolution spectroscopy, this can be understood from the simplified first-order signal-to-noise ratio equation from \cite{birkby2018exoplanet} in the photon-limited region:}

\begin{equation*}
    \mathrm{SNR_{planet} = \Big(\frac{S_p}{S_*}\Big) SNR_{star}\sqrt{N_{lines}}}
\end{equation*}

\noindent \textbf{Here, $\mathrm{\frac{S_p}{S_*}}$ is the planet-to-star signal ratio and $\mathrm{SNR_{star}}$ is the square root of the photon-limited, total signal-to-noise of the host star.} However, this method encounters the inherent challenge of disentangling the planetary signals from strong stellar and telluric backgrounds \citep{sanchez2019water}. This challenge can be effectively addressed through the detrending process. It incorporates Doppler shifts into the planet spectra over an extended observation period during the transit and, consequently, separates the planet signal successfully from the complex backgrounds of stellar and telluric lines. The detrended spectra are then subjected to cross-correlation with molecular templates, thereby quantifying the detection significance associated with different molecules \citep{snellen2010orbital, brogi2012signature, birkby2013monthly}. 

The increased spectral resolution is particularly significant when studying the expanded absorption and emission characteristics observed in hot Jupiters (HJs) and ultra-hot Jupiters (UHJs), as they often display pronounced thermal inversions and exhibit unique chemical compositions \citep{fortney2008unified,madhusudhan2010inference,line2016no,parmentier2018thermal,arcangeli2018h,bell2018increased}. These gas giants are of great importance in the realm of exoplanetary science since they can challenge our current understanding of the planetary formation, migration, and atmospheric evolution \citep{wright2012frequency, Dash_2022}. While HJs are recognized for their elevated temperatures, UHJs, positioned in even closer proximity to their host stars, experience substantial irradiation, which causes them to have a day-side temperature $\gtrsim$ 2200 K \citep{parmentier2018thermal}. These extreme climatic conditions facilitate the development of unique and often exotic atmospheric chemical processes such as the thermal dissociation of molecules followed by recombination of them on the cooler night side \citep{parmentier2018thermal,arcangeli2018h,bell2018increased}. Furthermore, they represent the most extensively observed category of exoplanets to date and, consequently, the most preferred targets for ground-based observations. 

\textbf{WASP-77 A b} and \textbf{WASP-76 b}, our targeted HJ and UHJ, are crucial in this regard as they show distinct non-inverted and inverted behaviors, respectively, in the pressure-temperature profile (PT profile) \citep{Edwards_2020,line2021solar,changeat2022five,edwards2024measuring} and thus have different implications on their atmospheres.  While several atomic and ionic species have been detected in the atmospheres of UHJs \citep{hoeijmakers2019spectral,hoeijmakers2020hot,casasayas2019atmospheric,tabernero2021espresso}, it is equally expected to have diatomic and triatomic molecular species as a crucial part of their atmospheric compositions \citep{landman2021detection}. The recent detections of OH, $\mathrm{H_2O}$, and HCN in the atmosphere of \textbf{WASP-76 b} using CAREMENS spectrograph serve as compelling evidence supporting this fact \citep{sanchez2022searching,landman2021detection}. Conversely, CO and $\mathrm{H_2O}$ line detections on \textbf{WASP-77 A b} using IGRINS spectrometer paves the way for conducting a simultaneous comprehensive analysis to study HJs too using ground-based observation \citep{line2021solar}. 

\begin{figure*}
\centering
	\includegraphics[width=2\columnwidth]{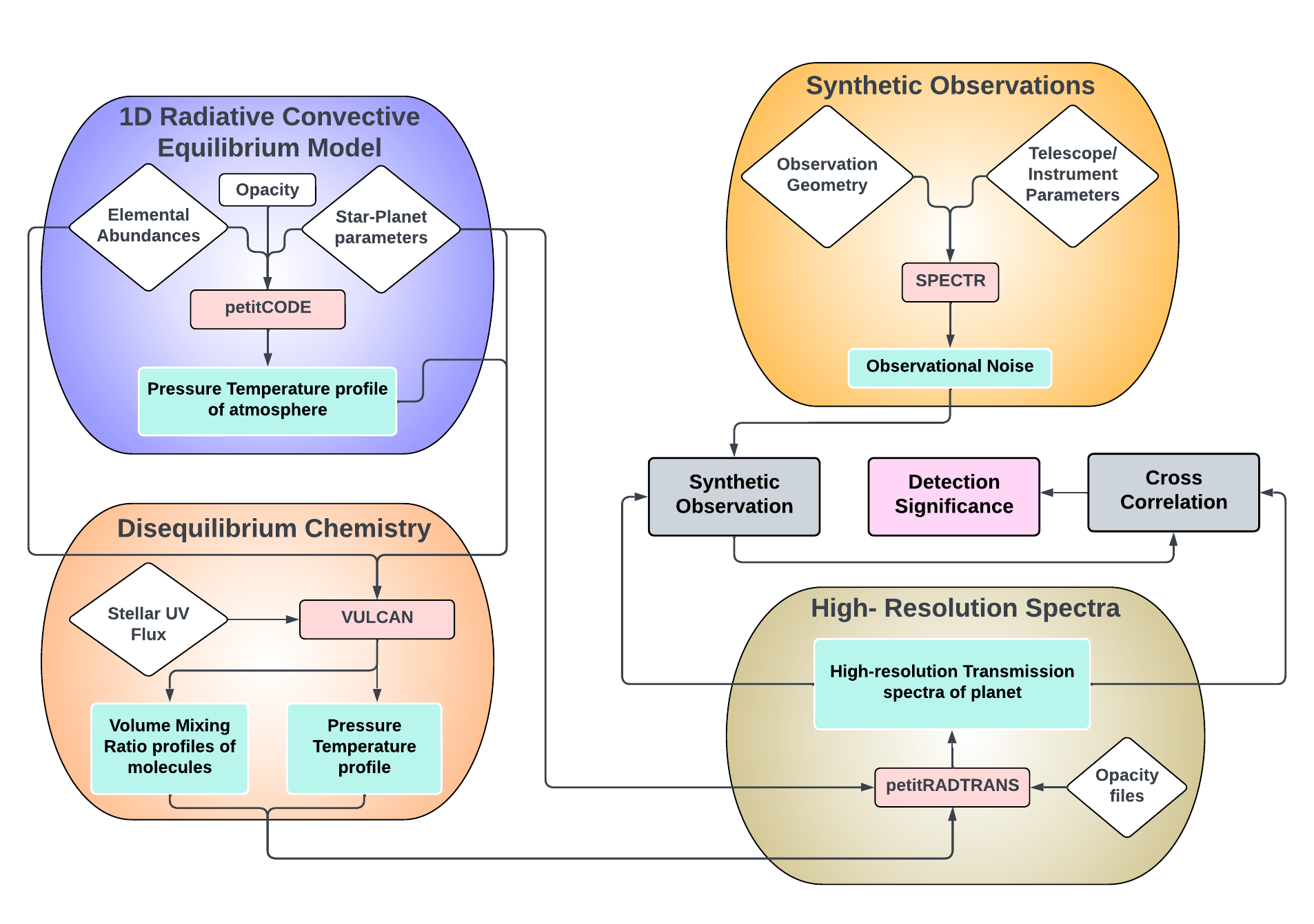}
%\plotone{Figures/Flowchart.pdf}
\caption{Flowchart representing our methodical pipeline: The big boxes represent the four different pillars (radiative-convective equilibrium calculation, disequilibrium chemistry, high-resolution spectra, and synthetic observations), which are connected with each other, of our analysis. The cyan rectangles represent the outputs from each model, while the white shapes are the user-specified inputs for the respective models. Gray rectangles are the outcomes of the whole pipeline, which lead to the calculation of the detection significance of molecular bands.}
\label{fig:flowchart}
\end{figure*}

In this work, we present a complementary analysis of \textbf{WASP-77 A b} and \textbf{WASP-76 b} atmospheres using three high-resolution ground-based spectrographs: GIANO-B (Telescopio Nazionale Galileo: TNG)\footnote{\href{https://www.tng.iac.es/instruments/giano-b/}{https://www.tng.iac.es/instruments/giano-b/}} \citep{oliva2012giano}, CARMENES (Centro Astronomico Hispano Alemán: CAHA)\footnote{\href{https://www.caha.es/telescope-3-5m/carmenes}{https://www.caha.es/telescope-3-5m/carmenes}} \citep{quirrenbach2014carmenes,quirrenbach2018carmenes}, and ANDES (European Extremely Large Telescope: E-ELT; From this point forward, throughout this paper, the E-ELT will be referred to as the ELT.)\footnote{\href{https://elt.eso.org/instrument/ANDES/}{https://elt.eso.org/instrument/ANDES/}} \citep{di2022andes}. Our main motivation lies in understanding the spectrographs' ability to probe different molecular bands using cross-correlation spectroscopy. In section \ref{sec:methods}, we have outlined our assumptions and methods for simulating the planet's atmosphere using the 1-D radiative-convective equilibrium model, petitCODE \citep{molliere2015model,molliere2017modeling}, and 1-D chemical kinetics model, VULCAN \citep{Tsai_2017,Tsai_2021}. In addition, we have briefed our method of calculating synthetic observation using the 1-D radiative transfer model, petitRADTRANS \citep{refId0, molliere2020}, and the ground-based noise simulator, Spectral Planetary ELT Calculator (SPECTR) \citep{currie2023there}. The next section, i.e., Section \ref{sec:results}, presents our results from the atmospheric models and the detection significance of molecular bands. Finally, we end up with our findings in Section \ref{sec:conclusion}.

%--------------------------------------------------------------------
\begin{table*}
\centering
%\captionsetup{justification=centering} % Center the caption
\caption{Relevant parameters of the \textbf{WASP-77 A b} and \textbf{WASP-76 b} systems}
\begin{tabular*}{2\columnwidth}{@{\extracolsep{\fill}}lcccccc@{}}

%\begin{tabular*}{cccccccc}
\hline
\hline
\noalign{\smallskip}
&\multicolumn{4}{c}{\textbf{\hspace{3.5cm}Star Parameters}}  & \\
\noalign{\smallskip}
\hline
\noalign{\smallskip}
& \multicolumn{2}{c}{\textbf{WASP-77 A}} & \multicolumn{3}{c}{\hspace{1cm}\textbf{WASP-76}} \\
\noalign{\smallskip}
\hlineB{3.5}
\noalign{\smallskip}
\multicolumn{1}{l|}{Parameters} & Values & \hspace{1cm}References & \multicolumn{1}{|c}{} & \hspace{-0.5cm}Values & \hspace{1cm}References \\
\noalign{\smallskip}
 \hline
\noalign{\smallskip}
    \multicolumn{1}{l|}{$M_*$/$M_{\odot}$} & 1.002 & \hspace{1cm}\cite{maxted2012wasp} & \multicolumn{1}{|c}{} & \hspace{-0.5cm}1.460 & \hspace{1cm}\cite{west2016three}\\
\multicolumn{1}{l|}{$R_*$/$R_{\odot}$} & 1.12 & \hspace{1cm}\cite{maxted2012wasp} & \multicolumn{1}{|c}{} & \hspace{-0.5cm}1.730 & \hspace{1cm}\cite{west2016three}\\
\multicolumn{1}{l|}{$[$Fe/H$]$} & 0 & \hspace{1cm}\cite{maxted2012wasp} & \multicolumn{1}{|c}{} & \hspace{-0.5cm}0.23 & \hspace{1cm}\cite{west2016three}\\
\multicolumn{1}{l|}{$T_*$ (K)} & 5500  & \hspace{1cm}\cite{maxted2012wasp} & \multicolumn{1}{|c}{} & \hspace{-0.5cm}6250  & \hspace{1cm}\cite{west2016three}\\
\noalign{\smallskip}
\hline
\hline
\noalign{\smallskip}\noalign{\smallskip}\noalign{\smallskip}
& \multicolumn{4}{c}{\textbf{\hspace{3.55cm}Planet Parameters}} &  \\
\noalign{\smallskip}
\hline
\noalign{\smallskip}
& \multicolumn{2}{c}{\textbf{WASP-77 A b}} & \multicolumn{3}{c}{\hspace{1cm}\textbf{WASP-76 b}} \\
\noalign{\smallskip}
\hlineB{3.5}
\noalign{\smallskip}
\multicolumn{1}{l|}{$M\mathrm{_p}$/$M\mathrm{_{Jup}}$} & 1.759 & \hspace{1cm}\cite{maxted2012wasp} & \multicolumn{1}{|c}{} & \hspace{-0.5cm}0.922 & \hspace{1cm}\cite{west2016three}\\
\multicolumn{1}{l|}{$R\mathrm{_p}$/$R\mathrm{_{Jup}}$} & 1.210 & \hspace{1cm}\cite{maxted2012wasp} & \multicolumn{1}{|c}{} & \hspace{-0.5cm}1.877 & \hspace{1cm}\cite{west2016three}\\
\multicolumn{1}{l|}{\textit{log}(g)} & 3.477 & \hspace{1cm}\cite{southworth2007method} & \multicolumn{1}{|c}{} & \hspace{-0.5cm}2.8 & \hspace{1cm}\cite{southworth2007method}\\
\multicolumn{1}{l|}{$T\mathrm{_{eq}}$ (K)} & 1715 & \hspace{1cm}\cite{cortes2020tramos} & \multicolumn{1}{|c}{} & \hspace{-0.5cm}2160 & \hspace{1cm}\cite{west2016three}\\
\multicolumn{1}{l|}{C/O} & 0.8 & \hspace{1cm}\cite{hoch2023assessing} & \multicolumn{1}{|c}{} & \hspace{-0.5cm}1.2 & \hspace{1cm}\cite{hoch2023assessing}\\
\multicolumn{1}{l|}{$a$ (AU)} & 0.023 & \hspace{1cm}\cite{cortes2020tramos} & \multicolumn{1}{|c}{} & \hspace{-0.5cm}0.033 & \hspace{1cm}\cite{west2016three}\\
\multicolumn{1}{l|}{Period (days)} & 1.36 & \hspace{1cm}\cite{ivshina2022tess} & \multicolumn{1}{|c}{} & \hspace{-0.5cm}1.81 & \hspace{1cm}\cite{ivshina2022tess}\\
\multicolumn{1}{l|}{Distance (parsec)} & 93 & \hspace{1cm}\cite{maxted2012wasp} & \multicolumn{1}{|c}{} & \hspace{-0.5cm}120 & \hspace{1cm}\cite{west2016three}\\
\noalign{\smallskip}
\noalign{\smallskip}
\hline
\hline
\label{tab:system}
\end{tabular*}
\end{table*}

%%%%%%%%%%%%%%%%%%%%%%%%%%%%%%%%%%%%%%%%%%%%%%%%%%%%%%%%%%%%%%%%%%%%%%%%%%%%%%%%%%%%%%%%%%%%%%%%%%%%%%%%%%%%%%%%%%%%%%%
%%%%%%%%%%%%%%%%%%%%%%%%%%%%%%%%%%%%%%%%%%%%%%%%%%%%%%%%%%%%%%%%%%%%%%%%%%%%%%%%%%%%%%%%%%%%%%%%%%%%%%
\section{Methods}
\label{sec:methods}
Our methodical pipeline for assessing the detectability of molecular spectral features using the cross-correlation analysis is outlined in Figure \ref{fig:flowchart}. In this current section, we discuss a detailed description of the methods followed.

\subsection{Atmospheric structures of \textbf{WASP-76 b} and \textbf{WASP-77 A b}}
\label{sec:atmo_structure}
\begin{figure}
\centering
	\includegraphics[width=\columnwidth]{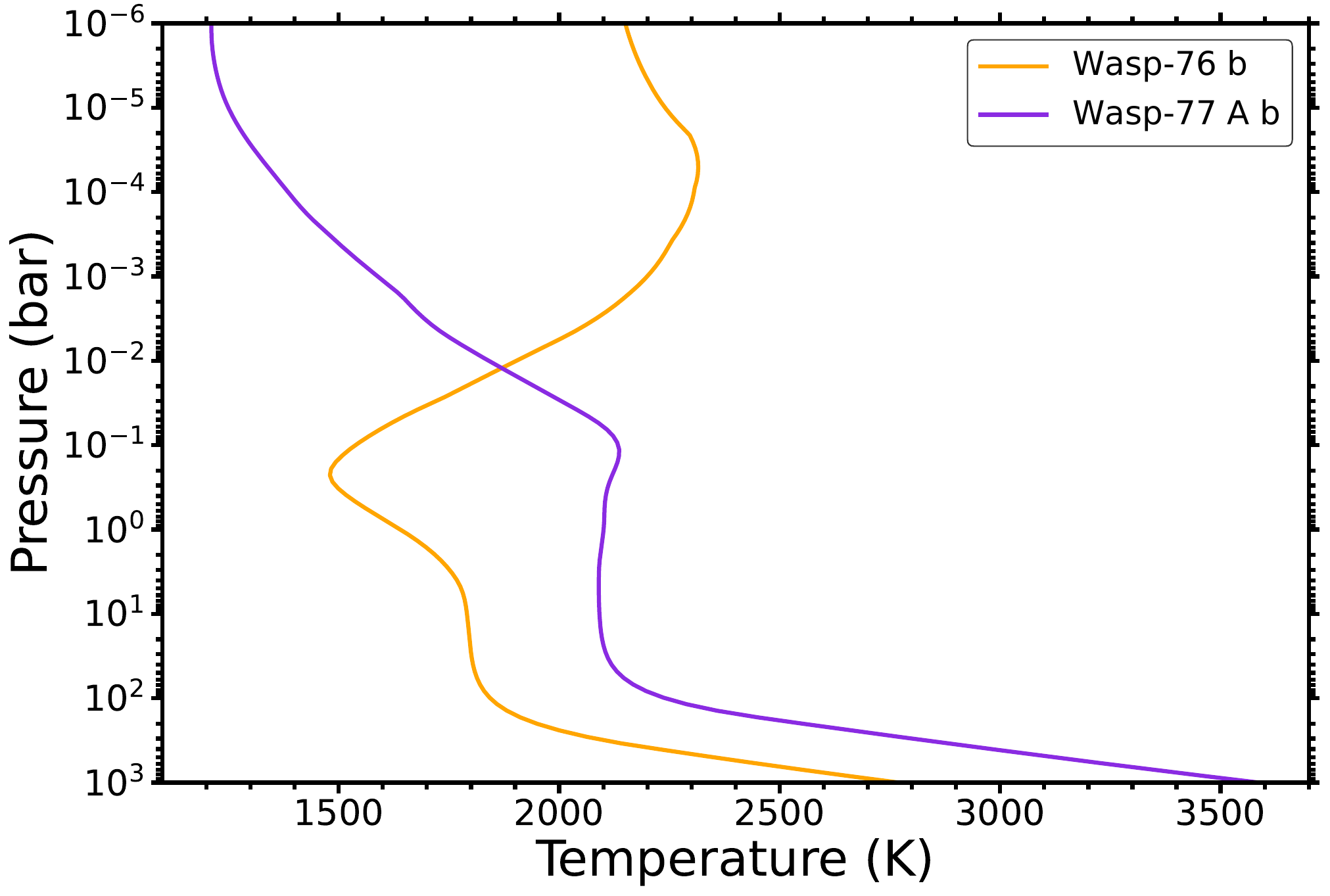}
  \caption{Self-consistent PT profile for both planets,  simulated with the 1-D radiative-convective equilibrium model, petitCODE, using the system parameters and elemental abundances listed in Table \ref{tab:system}. The orange line shows the inverted PT profile for \textbf{WASP-76 b}, whereas the violet one represents the non-inverted PT profile for \textbf{WASP-77 A b}.}
   \label{fig:TP_profile}
\end{figure}

The physical characteristics of a planet's atmosphere exert a direct influence on its spectral properties and molecular abundance profiles. We self-consistently construct the PT profiles for these two planets using 1-D forward model petitCODE (see Figure \ref{fig:TP_profile}). petitCODE employs a radiative-convective and thermochemical equilibrium assumption to simulate planetary atmospheres numerically. Based on the input parameters (star and planet parameters for both systems are listed in Table \ref{tab:system}), petitCODE iterates the atmospheric structure of the planet till it achieves an equilibrium between the chemical abundances, their opacities and radiation fields with the PT profile of the atmosphere. It solves the radiative transfer using the Feautrier method and implements the Accelerated Lambda Iteration \citep{olson1986rapidly} and Ng acceleration \citep{ng1974hypernetted} methods to expedite convergence.

%%%%%%%%%%%%%%%%%%%%%%

We have used petitCODE's default chemical parameters for the calculation and considered $\mathrm{CH_4}$, $\mathrm{H_2O}$, $\mathrm{CO_2}$, $\mathrm{HCN}$, $\mathrm{CO}$, $\mathrm{H_2S}$, $\mathrm{NH_3}$, $\mathrm{C_2H_2}$, $\mathrm{Na}$, $\mathrm{K}$, $\mathrm{TiO}$, and $\mathrm{VO}$ as opacity sources in the atmosphere. Additionally, $\mathrm{H_2}$-$\mathrm{H_2}$ and $\mathrm{H_2}$-$\mathrm{He}$ opacities were used as collision induced absorption (CIA) species. For defining the carbon-to-oxygen ratio (C/O), we followed the same prescription as \cite{madhusudhan2012c}, \cite{molliere2015model}, \cite{woitke2018equilibrium}, and \cite{molaverdikhani2019cold}, where the oxygen elemental abundance was altered, but the carbon elemental abundance was maintained at a constant value. %This approach allowed us to emulate the dynamic accumulation of variable water content on young accreting planets.

We considered globally averaged insolation treatment for radiative transfer calculation. For cloud treatment, we opted for Ackerman \& Marley \citep{ackerman2001precipitating} prescription: providing the settling factor ($f\mathrm{_{sed}}$) = 2 and lognormal particle distribution width ($\sigma\mathrm{_g}$) = 1.05.

%%%%%%%%%%%%%%%%%%%%%%%%%%%%%%%%%%%%%%%%%%%%%%%%%%%%%%%%%%%%%%%%%%%%%%%%%%%%%%%%%%%%%%%%%%%%%%%%%%%%%%
\subsection{Chemical evolution: vertical mixing and photochemistry in action}
\label{sec:diseqm}

In our study, we have taken the PT profiles from petitCODE as input for the 1-D chemical kinetics model VULCAN\footnote{\href{https://github.com/exoclime/VULCAN}{https://github.com/exoclime/VULCAN}} and then evolved the atmospheric chemistry for both planets. We utilize the original `\texttt{SNCHO photo network}' from the repository, which consists of 1150 reactions connecting 96 different species. Disequilibrium chemistry is effectively incorporated into planet atmospheres by implementing eddy diffusion, molecular diffusion, and photochemical processes. While chemical reactions govern molecular diffusion and photochemistry, eddy diffusion is regulated by the Eddy diffusion coefficient ($K_{\mathrm{zz}}$), which is kept constant = $\mathrm{10^{10}}$ $\mathrm{cm^2/s}$ (following the most common value from \cite{Tsai_2017}) for both planets. Apart from the star-planet parameters, planet-specific elemental abundances (O/H, N/H, C/H, S/H, and He/H) are also necessary for VULCAN to initiate the equilibrium chemistry calculation using its coupled equilibrium chemistry solver, Fastchem\footnote{\href{https://github.com/exoclime/FastChem}{https://github.com/exoclime/FastChem}} \citep{stock2018fastchem}. We keep the He/H abundance ratio constant at 0.085 and change the remaining abundance values accordingly, following the C/O ratio and metallicity of planets (see Table \ref{tab:system}). \textbf{The C/O ratio influences the elemental abundances of carbon and oxygen specifically (refer to Section \ref{sec:atmo_structure}). In contrast, metallicity affects the elemental abundances of all elements. The notation [Fe/H] is used to represent metallicity on a logarithmic scale. Here, [Fe/H] = 0 denotes solar metallicity, [Fe/H] = 1 indicates an elemental abundance ten times greater than solar, and [Fe/H] = -1 signifies an elemental abundance ten times lower than solar.}

%%%%%%%%%%%%%%%%%%%%%
\begin{table*}
\centering
\begin{threeparttable}
%\captionsetup{justification=centering} % Center the caption
\caption{Instrumental configuration for three ground-based observatories}
\begin{tabular*}{2\columnwidth}{@{\extracolsep{\fill}}lcccccc@{}}

%\begin{tabular*}{cccccccc}
\hline
\hline
\noalign{\smallskip}
Telescope/ & Wavelength & Resolution & Diameter (m) & Read Noise & Dark Current  & $\mathrm{T_{opt}}$\\
Instrument & ($\mu$m)& & (m) & ($\mathrm{e^{-}}$) & ($\mathrm{e^{-}}$/pixel/s) & (K)\\
\noalign{\smallskip}
\hlineB{3.5}
\noalign{\smallskip}
TNG (GIANO-B) & 0.95-2.45 & 50000 & 3.58 & 5 & 0.05 & 50\\
CAHA (CARMENES) & 0.52-1.71 & 80000 & 3.5 & 13.52 & 0.002 & 140 \\
%ARIES (TANSPEC) & 0.55-2.54 & 2750 & 3.6 & 20.7 & 0.060 & 75 \\
ELT (ANDES) & 0.4-1.8 & 100000 & 39.3 & 3 & 0.00056 (optical) & 143 \\
& & & & & 0.0011 (NIR) &\\
\noalign{\smallskip}
\hline
\hline
\label{tab:instruments}
\end{tabular*}
\begin{tablenotes}[flushleft]
        \item[a] Note: The instrumental parameters are taken from the respective websites of the high-resolution spectrographs. GIANO-B: \href{https://www.tng.iac.es/instruments/giano-b/}{https://www.tng.iac.es/instruments/giano-b/}. CARMENES: \href{https://www.caha.es/telescope-3-5m/carmenes}{https://www.caha.es/telescope-3-5m/carmenes}. ANDES: \href{https://elt.eso.org/instrument/ANDES/}{https://elt.eso.org/instrument/ANDES/}.
        %\item[b] Another footnote.
    \end{tablenotes}
    \end{threeparttable}
%\footnotesize{Note: The instrumental parameters are taken from the respective websites of the high-resolution spectrographs. GIANO-B: \href{https://www.tng.iac.es/instruments/giano-b/}{https://www.tng.iac.es/instruments/giano-b/}. CARMENES: \href{https://www.caha.es/telescope-3-5m/carmenes}{https://www.caha.es/telescope-3-5m/carmenes}. ANDES: \href{https://elt.eso.org/instrument/ANDES/}{https://elt.eso.org/instrument/ANDES/}.}
\end{table*}

%%%%%%%%%%%%%%%%%%%%%%%%%%%%%%%%%%%%%%%%%%%%%%%%%%%%%%%%%%%%%%%%%%%%%%%%%%%%%%%%%%%%%%%%%%%%%%%%%%%%%%
\subsection{Transmission spectra simulation from planet atmospheres}
\label{sec:spectra}

We couple the time-evolved chemistry from VULCAN to the 1-D radiative transfer model, petitRADTRANS\footnote{\href{https://petitradtrans.readthedocs.io/en/latest/}{https://petitradtrans.readthedocs.io/en/latest/}}, to simulate high-resolution line-by-line synthetic transmission spectra for both planets at a reference pressure level of $P_0$ = 0.01 bar. We utilize a customized Python3 wrapper for this purpose. We have considered OH, $\mathrm{H_2O}$, CO, $\mathrm{CO_2}$ \citep{rothman2010hitemp}, $\mathrm{CH_4}$, $\mathrm{HCN}$, $\mathrm{NH_3}$, $\mathrm{H_2S}$ \citep{chubb2021exomolop} (for \textbf{WASP-76 b}) and OH, $\mathrm{H_2O}$, CO, $\mathrm{CO_2}$, $\mathrm{NH_3}$, $\mathrm{H_2S}$ (for \textbf{WASP-77 A b}) as opacity sources. The selection of chemical species is determined by their average molecular volume mixing ratios (VMRs) $>$ $\mathrm{10^{-8}}$ at the photospheric level. Molecules with a lower VMR than $\mathrm{10^{-8}}$ would not have a significant impact on the planet's spectra. Hence, considering other molecules does not add any additional change to the planet's spectra and hence, to the analysis. The photosphere is the region in the atmosphere that contributes more to the planet's spectra and lies in the nexus of 1 bar-0.1 mbar region in the atmosphere. In addition, we use CIA opacities for $\mathrm{H_2}$-$\mathrm{H_2}$ and $\mathrm{H_2}$-$\mathrm{He}$ and $\mathrm{H_2}$ and He as Rayleigh scattering species. The wavelength range and resolution for different spectra are instrument-specific (see Table \ref{tab:instruments}). The default resolution for line-by-line opacities is R = $\mathrm{10^{6}}$ in petitRADTRANS. To avoid numerical inefficiency, we downsample the high-resolution molecular opacity files to the spectrograph resolutions before the radiative transfer calculation. For different instrumental resolutions, the rebinning of the opacity files is done using the inbuilt opacity sampling method (by providing the resolution downsampling factor) of petitRADTRANS.

%%%%%%%%%%%%%%%%%%%%%%%%%%%%%%%%%%%%%%%%%%%%%%%%%%%%%%%%%%%%%%%%%%%%%%%%%%%%%%%%%%%%%%%%%%%%%%%%%%%%%%
\subsection{Noise calculation and observation simulation}
\label{sec:noise}

Ground-based observations pose significant challenges, primarily due to the presence of Earth's atmosphere. The prominent challenge stems from the interference of telluric lines in the observed spectra, which introduces a caveat when attempting to estimate the molecular fingerprints within the planetary spectrum.  To address this issue, we have employed a cutting-edge, coronagraph-free, ground-based noise simulator pipeline, SPECTR\footnote{\href{https://github.com/curriem/spectr}{https://github.com/curriem/spectr}}, to simulate the observations for both planets using different ground-based instruments. It quantifies the planetary photon count as the signal while considering starlight, zodiacal light, exo-zodiacal light, moonlight, airglow, instrumental thermal emission, and Earth's atmosphere (telluric) photon counts as noise sources.

\begin{figure}[h]
    \centering  
            \includegraphics[width=\columnwidth]{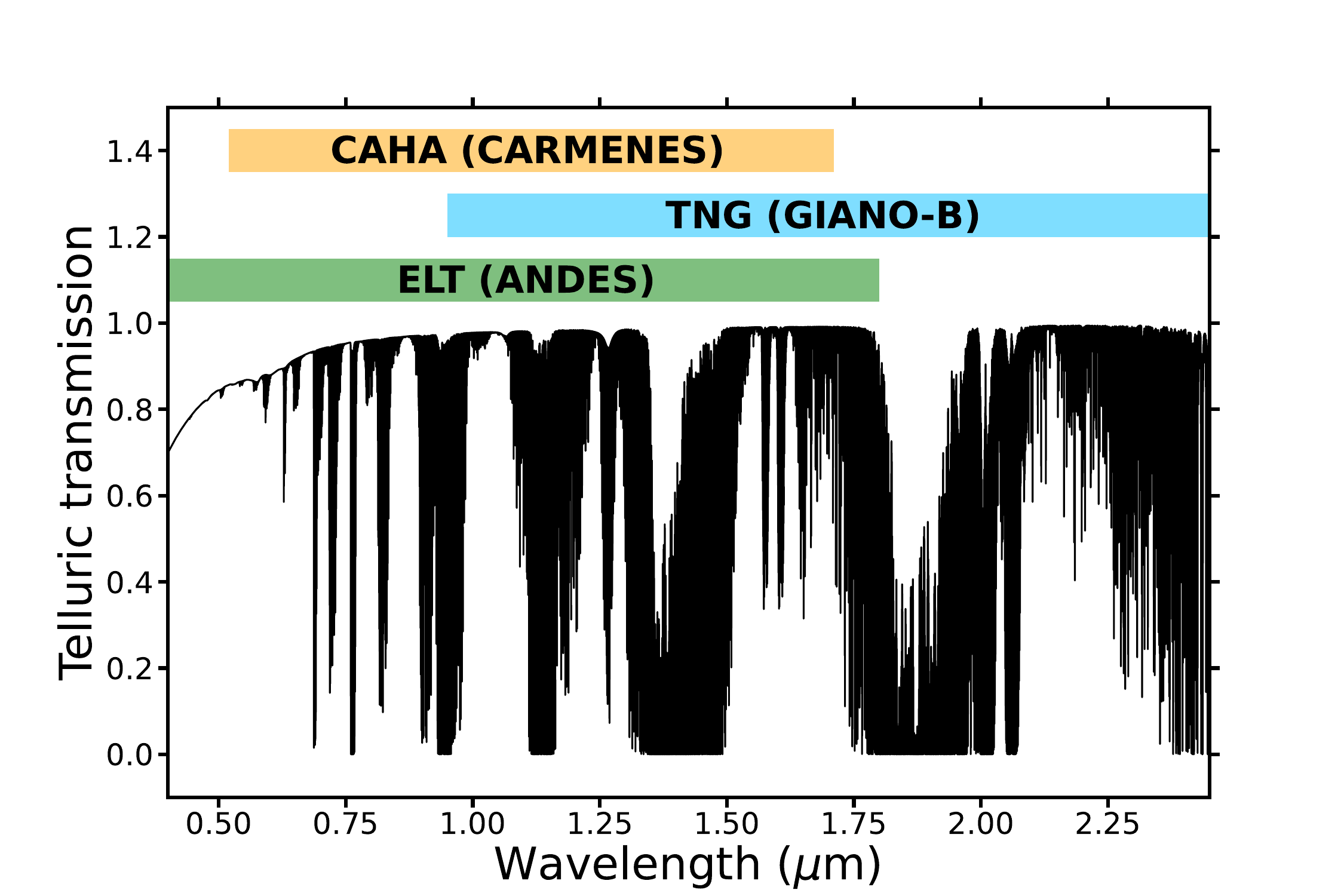}
\caption{Telluric transmission from SkyCalc for the total wavelength coverage of all the instruments: blue band is for TNG (0.95-2.45 $\mu$m), orange band for CAHA (0.52-1.71 $\mu$m), and green band for ELT (0.4-1.8 $\mu$m). The telluric lines were calculated for the resolution R = 80000. \textbf{It shows the differential effect of telluric lines on the instrumental wavelength coverages. While similar telluric lines impact CARMENES and ANDES observations, GIANO-B has some influence at the longer wavelengths.}}
\label{fig:telluric}
\end{figure}

SPECTR is integrated with the state-of-the-art atmospheric telluric model, Cerro Paranal Advanced Sky Model (SkyCalc)\footnote{\href{https://www.eso.org/observing/etc/doc/skycalc/helpskycalccli.html}{https://www.eso.org/observing/etc/doc/skycalc/helpskycalccli.html}} \citep{noll2012atmospheric}, from the European Southern Observatory for simulating wavelength-dependent background and telluric lines at the resolution specific to the instrument. In line with the approach outlined in the \citep{currie2023there}, we have selected Paranal as the observatory location and assumed a precipitable water vapor level of 3.5 mm for calculating Earth's atmospheric telluric lines. Observed telluric transmission at the site of observation is shown in Figure \ref{fig:telluric}. The read noise and dark current values for different instruments are listed in Table \ref{tab:instruments}. We have set the exposure duration equal to 488 seconds for each target, following the \textbf{WASP-76 b} observations by \cite{landman2021detection} and \cite{sanchez2022searching} using CARMENES spectrograph. We have kept the same exposure time for all the spectrographs to estimate a comparative understanding of the instrumental capability of molecular detection.

\begin{figure}
    \centering  
            \includegraphics[width=\columnwidth]{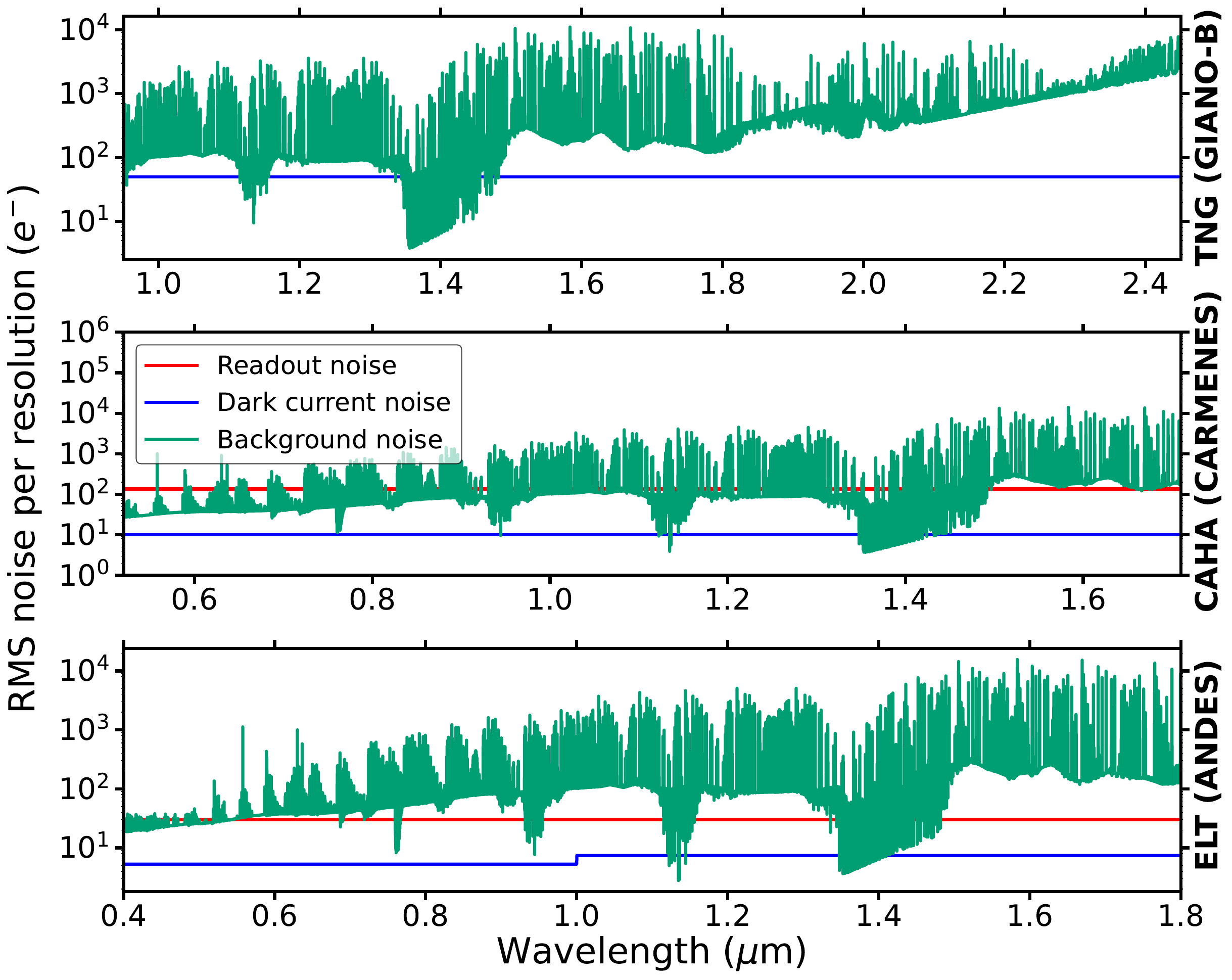}
\caption{Different noise sources in the visible and NIR region for different instruments: 1. Background noise (telluric emission + thermal emission of instrument + zodiacal light + airglow) (bluish-green) 2. Readout noise (red), and 3. Dark current noise (blue). The noises are measured \textbf{for an exposure duration of 488 seconds} at native resolutions specific to instruments (see Table \ref{tab:instruments}). \textbf{It shows the variability of background noise over a wide wavelength span, the different instrumental read-out noise, and the dark current values. Overall, it presents the differential influence of noise profiles on different telescopic observations.} The step function observed in the Dark current noise for the ELT arises from the distinct dark current counts in the optical and NIR regions. \textbf{For TNG, the values for readout noise and dark current noise are the same.}}
\label{fig:noise_components}
\end{figure}

It is important to note that we have assumed a consistent throughput of 21\% for GIANO-B \citep{oliva2006giano}, 7.9\% for CARMENES \citep{seifert2016carmenes}, and 10\% for ANDES \citep{currie2023there}. Throughputs are usually wavelength-dependent. However, the measurements of instrumental throughputs are limited to certain wavelength points. To overcome this issue, we have considered the best throughput value for the spectrographs. Hence, our assumption serves as an upper limit for molecular detection from the perspective of the instrumental throughput parameter. For telescope mirror temperature, we have followed \citep{currie2023there} and maintained a consistent temperature of 273 K across all observatories. But the detector temperatures ($\mathrm{T_{opt}}$) are set according to the specifications outlined in Table \ref{tab:instruments}. Different sources of noise components are shown in Figure \ref{fig:noise_components}. Furthermore, for uniformity, we have designated 100 detector pixels per resolution for all the detectors and carried out the observation for 1 and 3 transits for both planets (following the observations on hot gas giant planets by \cite{sanchez2020discriminating}, \cite{landman2021detection}, \cite{sanchez2022searching}, \cite{carleo2022gaps}, \cite{borsato2024small}, and \cite{prinoth2024atlas}).

The pipeline Doppler shifts the planet's spectrum and multiplies the telluric lines at each exposure to it. The systematic radial velocity shifts relative to Earth are taken to be -1.07 km $\mathrm{s^{-1}}$ for \textbf{WASP-76 b} \citep{west2016three} and 1.68 km $\mathrm{s^{-1}}$ for \textbf{WASP-77 A b} \citep{maxted2012wasp}. The pipeline systematically includes out-of-transit observations alongside each in-transit observation to facilitate the precise isolation of the planet signal and noise within observed transmission spectra. To enhance data quality, the pipeline employs an outlier mask and high-pass filters to eliminate data points with exceptionally low signal-to-noise ratios and mitigate low-frequency variations. 

\begin{figure*}
    \centering  
        \begin{minipage}[b]{\columnwidth}
            \includegraphics[width=\columnwidth]{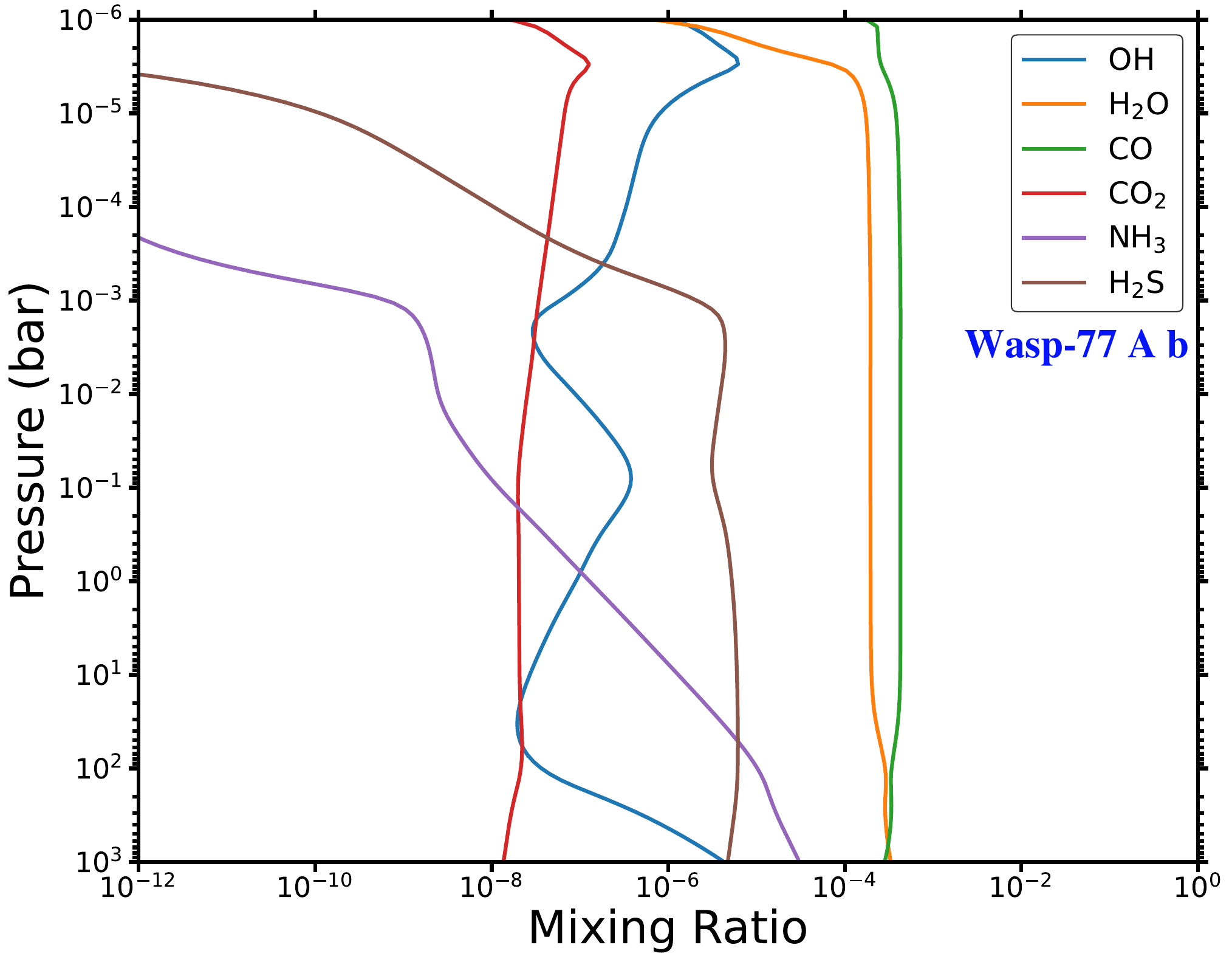}
        \end{minipage}
        %\columnbreak
         \begin{minipage}[b]{\columnwidth}
            \includegraphics[width=\columnwidth]{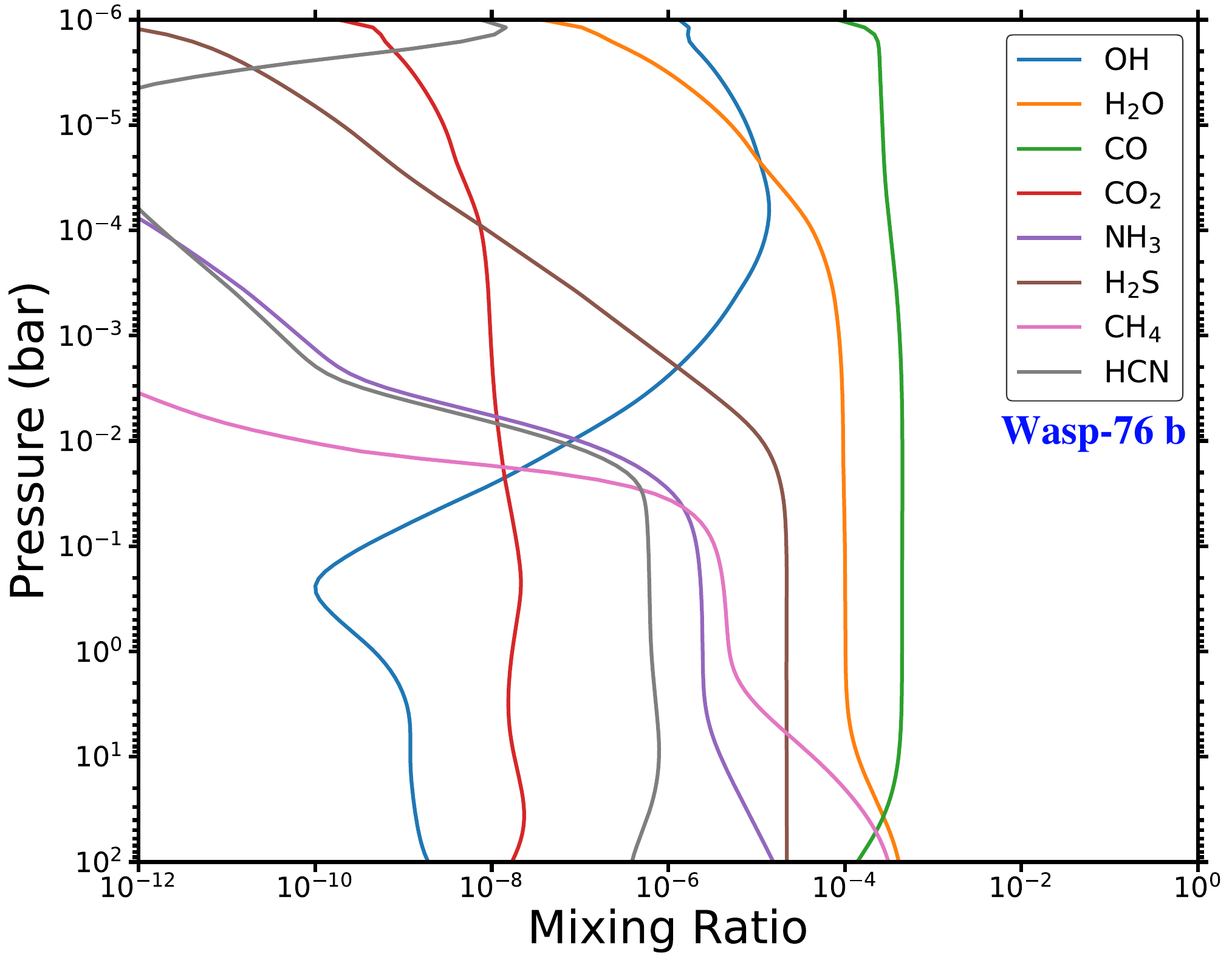}
        \end{minipage}
\caption{Mixing ratio profiles from VULCAN for specific molecules in the atmosphere of both planets. \textbf{WASP-77 A b} (\textit{left}): OH, $\mathrm{H_2O}$, CO, $\mathrm{CO_2}$, $\mathrm{NH_3}$, $\mathrm{H_2S}$, and \textbf{WASP-76 b} (\textit{right}): OH, $\mathrm{H_2O}$, CO, $\mathrm{CO_2}$, $\mathrm{NH_3}$, $\mathrm{H_2S}$, $\mathrm{CH_4}$, HCN. Molecules are selected considering the average mixing ratios $>$ $\mathrm{10^{-8}}$ between the 1 bar-0.1 mbar pressure layers (photosphere). }
\label{fig:VMR}
\end{figure*}

It computes the detection significance ($\mathrm{\sigma_{det}}$) of molecular bands using the cross-correlation spectroscopy technique as outlined in \citep{brogi2016rotation}. This method exhibits a high degree of sensitivity to line features and demonstrates robustness even in the presence of minimal perturbations in radial velocity. Our approach involves cross-correlating the simulated spectra ($\mathrm{f(\lambda)}$) for both planets with Doppler-shifted template spectra ($\mathrm{f(\lambda - \lambda')}$) representing various molecular bands and, thus, calculating the detectability of these molecules across a range of ground-based facilities operating at different resolutions and different wavelength regimes. Here, $\mathrm{\lambda}$ denotes the bin number, while $\mathrm{\lambda'}$ signifies the bin shift attributable to relative velocity. To generate velocity-shifted template spectra, we opt for an extensive grid of Doppler velocities, spanning from -150 km $\mathrm{s^{-1}}$ to +150 km $\mathrm{s^{-1}}$, with 101 intervals. Here, we will brief the numerical recipe to compute the $\mathrm{\sigma_{det}}$ following \citep{currie2023there} (for more details see \cite{brogi2019retrieving}):
\begin{equation}\label{eq1}\
    \mathrm{s_f^2 = \frac{1}{N} \sum_{\lambda} f^2(\lambda),} \text{ $\mathrm{s_f^2}$ = variance of the observed spectrum}
\end{equation}

\begin{equation}\label{eq2}
    \mathrm{s_g^2 = \frac{1}{N} \sum_{\lambda} f^2(\lambda - \lambda'),} \text{ $\mathrm{s_g^2}$ = variance of the template spectrum}
\end{equation}

\noindent where N is the total number of pixels. Hence, the cross-covariance can be defined as
\begin{equation}\label{eq3}
    \mathrm{R(\lambda) = \frac{1}{N} \sum_{\lambda} f(\lambda)f(\lambda - \lambda')} 
\end{equation}

\noindent Now, Eqs. \ref{eq1},\ref{eq2}, and \ref{eq3} can be coupled together to estimate the cross-correlation function (CCF) ($\mathrm{C(\lambda)}$), which  peaks at the relative velocity of zero for a detectable planetary signal 

\begin{equation}\label{eq4}
    \mathrm{C(\lambda) = \frac{R(\lambda)}{\sqrt{\mathrm{s_f^2s_g^2}}}}
\end{equation}

\noindent Thereafter, a $\mathrm{\chi^2}$ analysis is followed between estimated CCF and a flat line to test for a signal \citep{brogi2016rotation}

\begin{equation}\label{eq5}
    \mathrm{\chi^2_{CCF}= \sum_{\lambda} \frac{[C(|\lambda|< \lambda_0) - 0]^2}{\sigma^2_{C(|\lambda|> \lambda_0)}}}
\end{equation}

\noindent In Eq. \ref{eq5}, the $\mathrm{\chi^2_{CCF}}$ is summed over the velocity shifts ($\mathrm{|\lambda|}$) less than $\mathrm{\lambda_0}$ = 5 km $\mathrm{s^{-1}}$. These values only represent the planet signal in the calculation and the typical width of the CCF peak. The denominator, on the other hand, signifies the variance, excluding any potential planetary signals. Utilizing the cumulative $\mathrm{\chi^2}$ distribution function and inverse survival function of a normal distribution, $\mathrm{\chi^2_{CCF}}$ is converted to \textit{p}-value and later on, to a sigma interval. Finally, $\mathrm{\sigma_{det}}$ is estimated as a measure of the deviation of the CCF from a Gaussian distribution.

\begin{figure*}[b]
    \centering  
        \begin{minipage}[b]{\columnwidth}
            \includegraphics[width=\columnwidth]{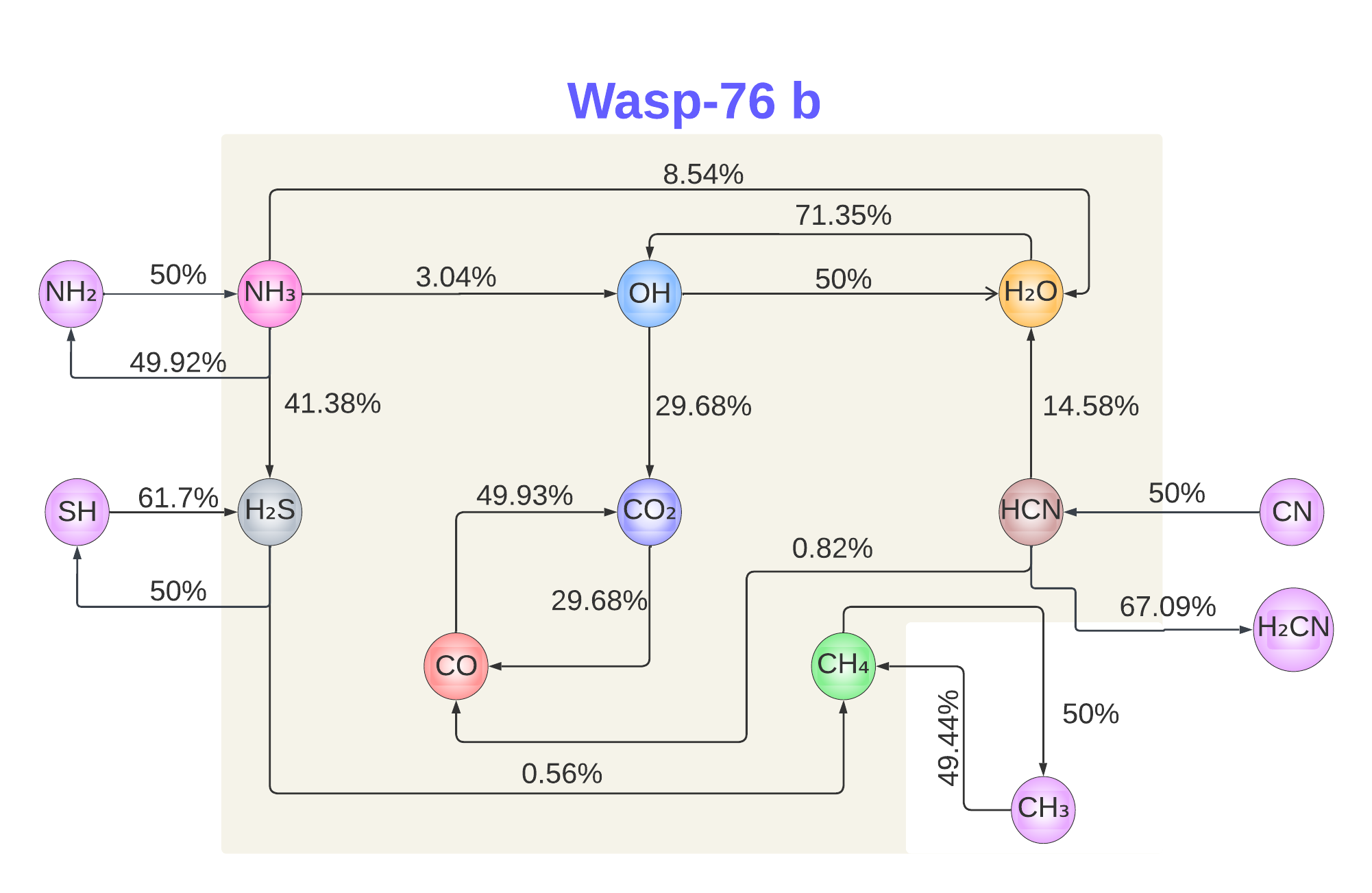}
        \end{minipage}
        %\columnbreak
         \begin{minipage}[b]{\columnwidth}
            \includegraphics[width=\columnwidth]{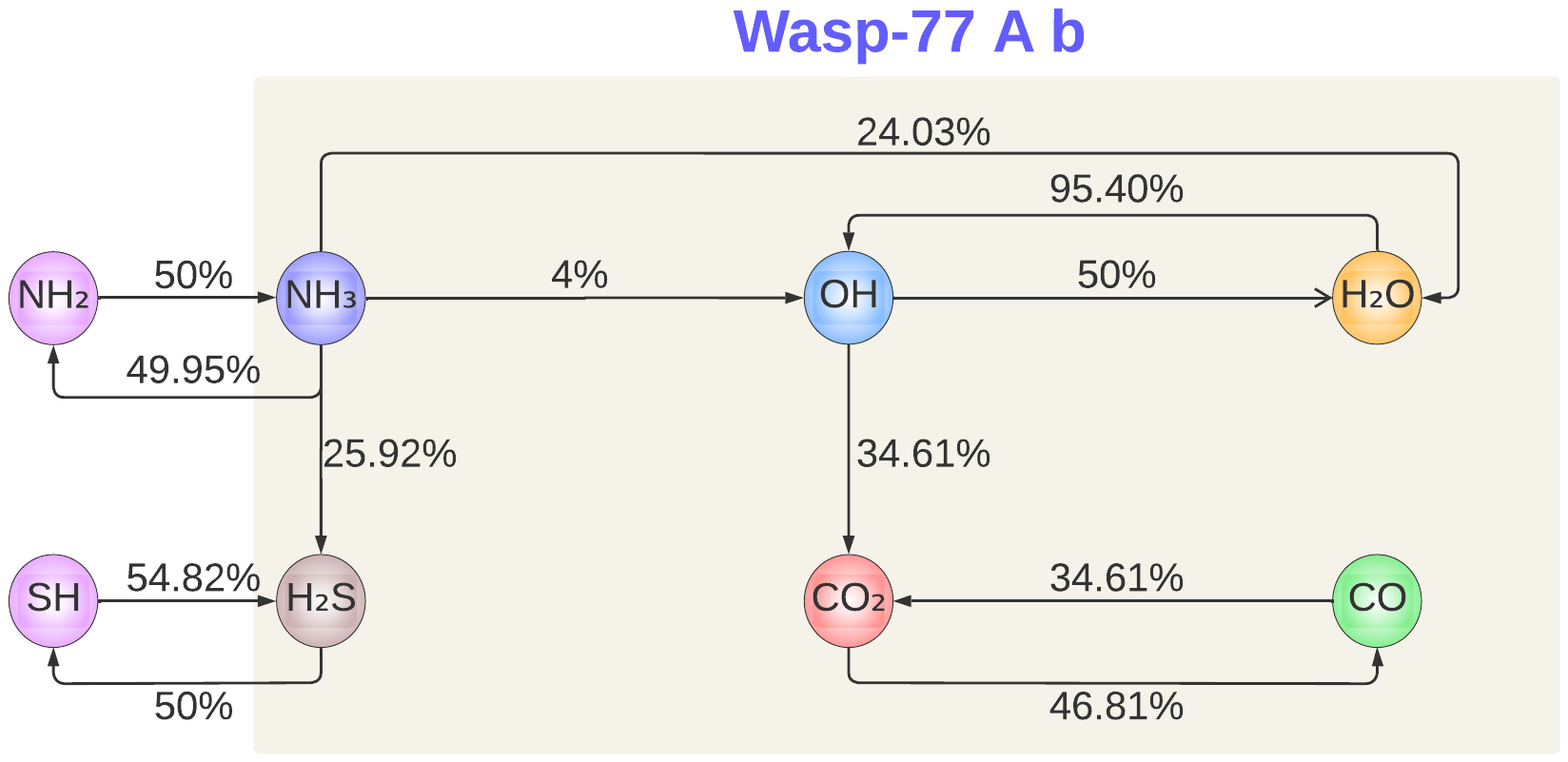}
        \end{minipage}
\caption{Illustration of the principal chemical pathways and their conversion rates at a pressure level of 0.01 bar (consistent with the photosphere) for specific molecules. \textit{Left panel}: \textbf{WASP-76 b}. \textit{Right panel}: \textbf{WASP-77 A b}. The ``tan" shaded areas represent molecules with an average VMR  $>$ $\mathrm{10^{-8}}$ between 1 bar-0.1 mbar (photosphere). Molecules outside these `tan' regions, indicated by `magenta' circles, are non-significant in terms of VMR, but they serve as crucial sources and sinks for some molecules. For detailed model-specific analyses of reaction pathways, including the incorporation of all secondary reactants, see Figures \ref{fig:shortest_wasp76b}, \ref{fig:shortest_wasp77ab} in the Appendix \ref{appendix_1}.}
\label{fig:chemisrty}
\end{figure*}

%%%%%%%%%%%%%%%%%%%%%%%%%%%%%%%%%%%%%%%%%%%%%%%%%%%%%%%%%%%%%%%%%%%%%%%%%%%%%%%%%%%%%%%%%%%%%%%%%%%%%%%
%%%%%%%%%%%%%%%%%%%%%%%%%%%%%%%%%%%%%%%%%%%%%%%%%%%%%%%%%%%%%%%%%%%%%%%%%%%%%%%%%%%%%%%%%%%%%%%%%%%%%%%
\section{Results}
\label{sec:results}

%\subsection{Trends in molecular VMR for \textbf{WASP-76 b} and \textbf{WASP-77 A b}}
%\label{result:chemistry_trend}

%%%%%%%%%%%%%%%%%%%%%%%%%%%%%%%%%%%%%%%%%%%%%%%%%%%%%%%%%%%%%%%%%%%%%%%%%%%%%%%%%%%%%%%%%%%%%%%%%%%%%%%%%%%%%%%%%%%%%%%%%%%%%%%%%%
\subsection{High-resolution transmission spectra: linking molecular VMRs to their spectral signatures}
\label{result:spectra}

As discussed in Section \ref{sec:spectra}, the determination of molecules for simulating synthetic high-resolution spectra depends on their average VMRs $>$ $\mathrm{10^{-8}}$ in the planetary photosphere. The criterion leads us to the inclusion of OH, $\mathrm{H_2O}$, $\mathrm{CH_4}$,  CO, $\mathrm{CO_2}$, $\mathrm{HCN}$, $\mathrm{NH_3}$, $\mathrm{H_2S}$ in the atmosphere of \textbf{WASP-76 b} and OH, $\mathrm{H_2O}$, CO, $\mathrm{CO_2}$, $\mathrm{NH_3}$, $\mathrm{H_2S}$ in the atmosphere of \textbf{WASP-77 A b}. The molecular VMRs for both planets have been showcased in Figure \ref{fig:VMR}. The atmospheric compositions of the planets are significantly influenced by CO and $\mathrm{H_2O}$, with $\mathrm{H_2S}$ also showing a considerable presence in both of them. It is worth mentioning that at a pressure level of $\sim$ 0.01 bar, the VMR of OH starts rising for both planets, and this can be attributed to the reduction in levels of $\mathrm{H_2O}$ and, in particular, $\mathrm{NH_3}$ (see Figure \ref{fig:chemisrty}). They both contribute $\sim$ 74\% (for \textbf{WASP-76 b}) and $\sim$ 99\% (for \textbf{WASP-77 A b}) to the net OH formation at the specific pressure level. While $\mathrm{H_2O}$ undergoes a single reaction: $\mathrm{H_2O}$ + H $\rightarrow$ OH + $\mathrm{H_2}$ for the conversion, $\mathrm{NH_3}$ has different routes to form OH. For \textbf{WASP-76 b}, the fastest interconversion follows the path

\begin{table}[h]
    \centering
    \begin{tabular}{c}
        \noindent $\mathrm{NH_3}$ + H $\rightarrow$ $\mathrm{NH_2}$ + $\mathrm{H_2}$\\
        \noindent $\mathrm{NH_2}$ + H $\rightarrow$ $\mathrm{H_2}$ + $\mathrm{NH}$\\
        \noindent \textcolor{red}{$\mathrm{NH}$ + S $\rightarrow$ $\mathrm{H}$ + $\mathrm{NS}$}\\
        \noindent $\mathrm{NS}$ + H $\rightarrow$ $\mathrm{SH}$ + $\mathrm{N}$\\
        \noindent $\mathrm{H_2O}$ + SH $\rightarrow$ $\mathrm{H_2S}$ + $\mathrm{OH}$\\
        \noalign{\smallskip}
        \hline
        \noalign{\smallskip}
        \noindent net: $\mathbf{NH_3}$ + $\mathrm{H_2O}$ + 2H + S $\rightarrow$ \textbf{OH} + $\mathrm{H_2S}$ + N + $\mathrm{2H_2}$
    \end{tabular}
    %\caption{Caption}
    \label{tab:nh3_h2s_77ab}
\end{table}

\noindent whereas, for \textbf{WASP-77 A b} it undergoes

\begin{table}[h]
    \centering
    \begin{tabular}{c}
        \noindent\textcolor{red}{$\mathrm{NH_3}$ + SH $\rightarrow$ $\mathrm{NH_2}$ + $\mathrm{H_2}S$}\\
        \noindent $\mathrm{H_2S}$ + H $\rightarrow$ $\mathrm{H_2}$ + $\mathrm{SH}$\\
        \noindent $\mathrm{SH}$ + $\mathrm{H_2O}$ $\rightarrow$ $\mathrm{H_2S}$ + $\mathrm{OH}$\\
        \noalign{\smallskip}
        \hline
        \noalign{\smallskip}
        \noindent net: \textbf{$\mathbf{NH_3}$} + $\mathrm{H_2O}$ + SH + H $\rightarrow$ \textbf{OH} + $\mathrm{H_2S}$ + $\mathrm{NH_2}$ + $\mathrm{H_2}$
    \end{tabular}
    %\caption{Caption}
    \label{tab:nh3_h2s_77ab}
\end{table}

\begin{figure*}[b]
    \centering  
        \begin{minipage}[b]{\columnwidth}
            \includegraphics[width=\columnwidth]{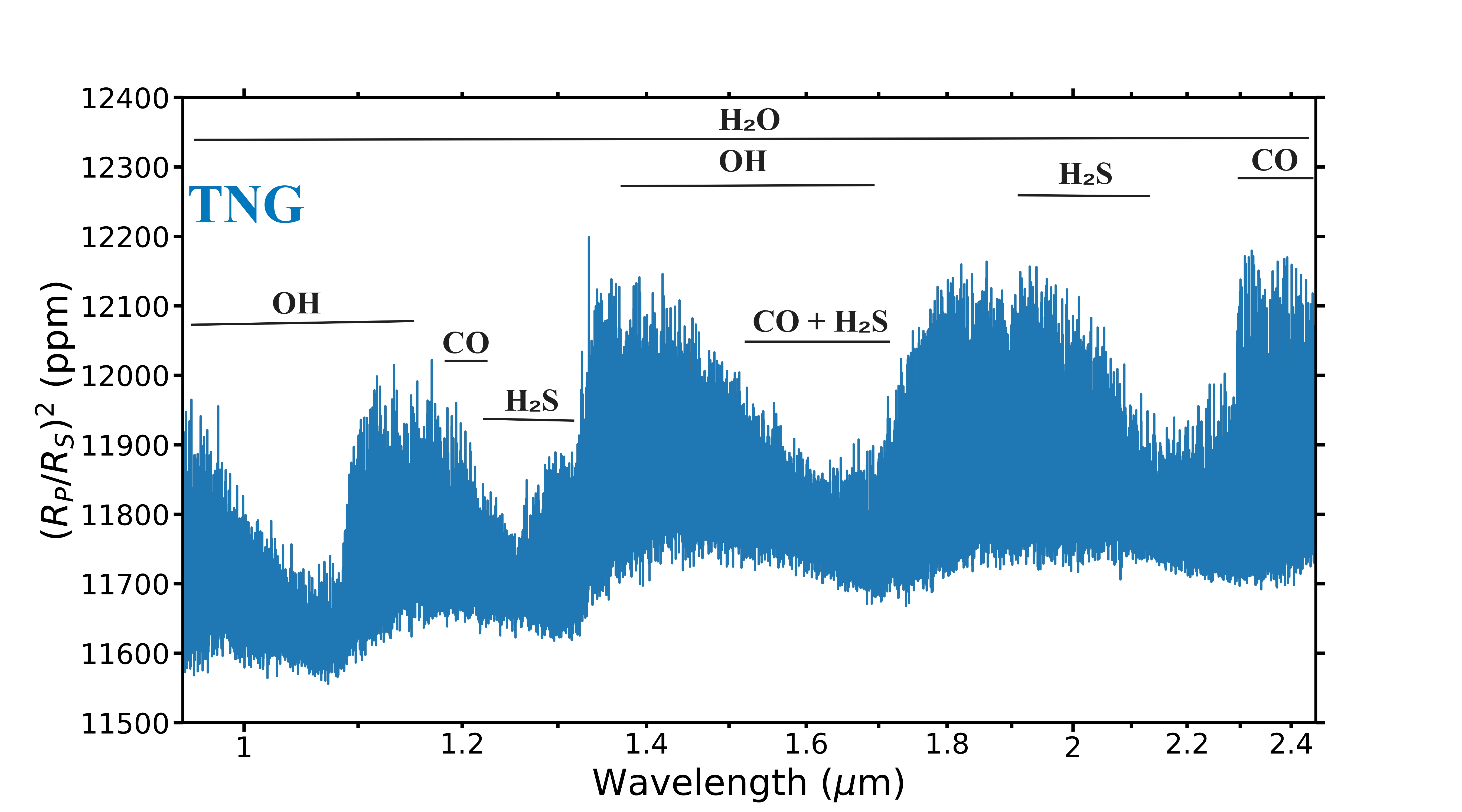}
        \end{minipage}
        %\columnbreak
         \begin{minipage}[b]{\columnwidth}
            \includegraphics[width=\columnwidth]{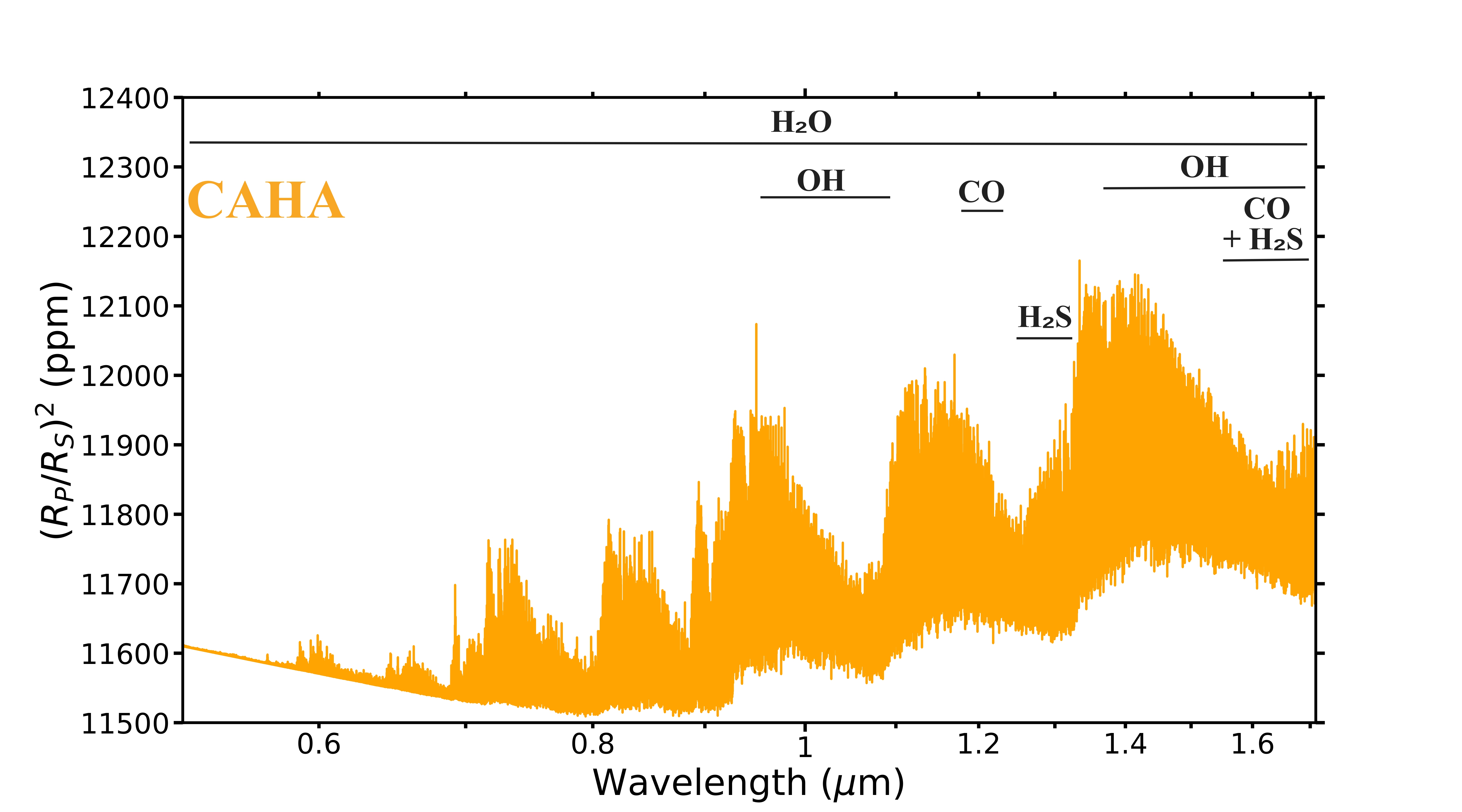}
        \end{minipage}
         \begin{minipage}[b]{\columnwidth}
            \includegraphics[width=\columnwidth]{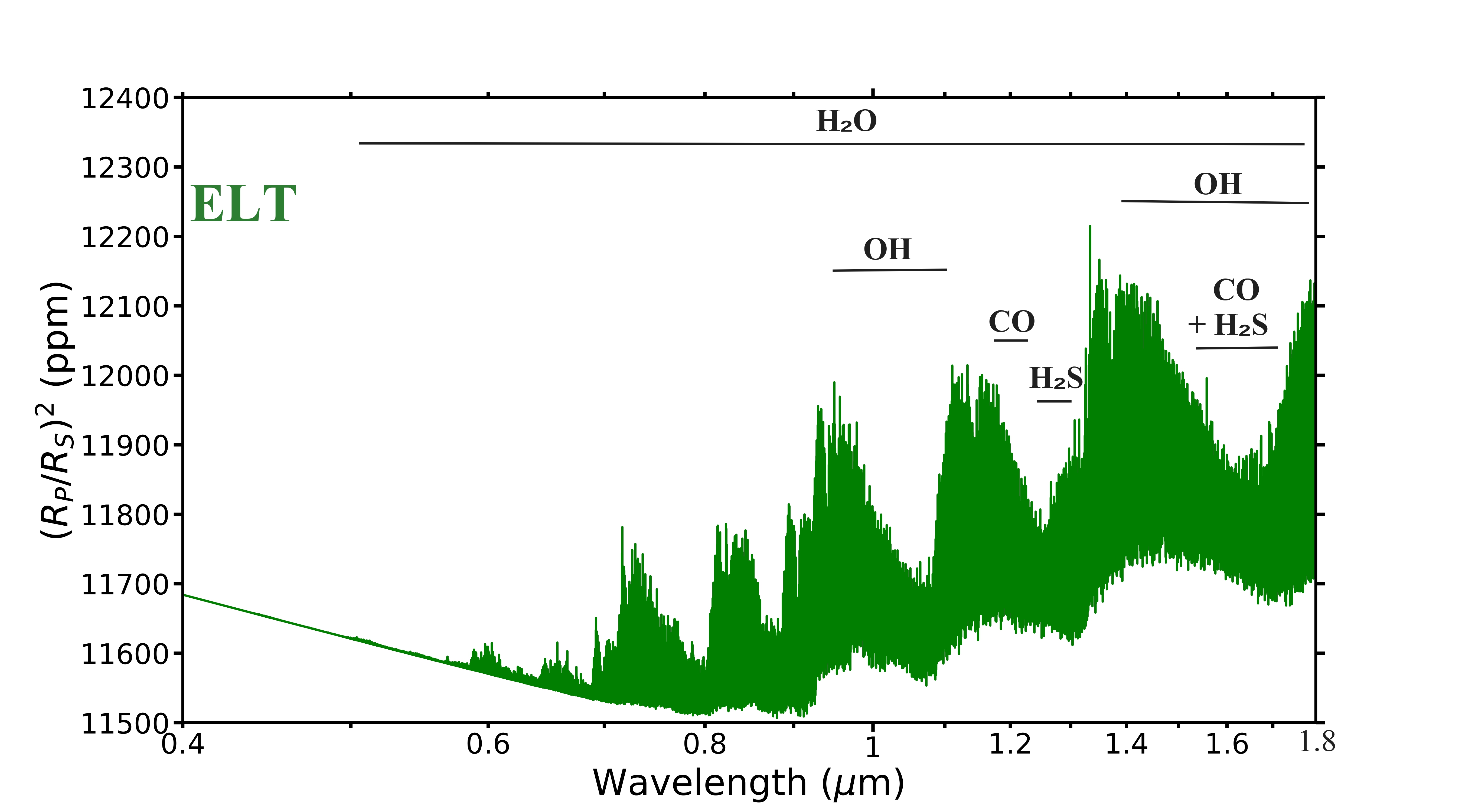}
        \end{minipage}
\caption{High-resolution transmission spectra of \textbf{WASP-77 A b} for different instruments (see instrumental parameters from Table \ref{tab:instruments}. \textit{Blue}: GIANO-B (TNG). \textit{Orange}: CARMENES (CAHA). \textit{Green}: ANDES (ELT). \textbf{The figure illustrates the contributions from different molecular bands to the net planet spectra for all cases.} Despite having a significant average photospheric VMR, discernible $\mathrm{CO_2}$ and $\mathrm{NH_3}$ features are not present in any of the synthetic spectra.}
\label{fig:Transmission_Wasp77ab}
\end{figure*}

\begin{figure*}
    \centering  
        \begin{minipage}[b]{\columnwidth}
            \includegraphics[width=\columnwidth]{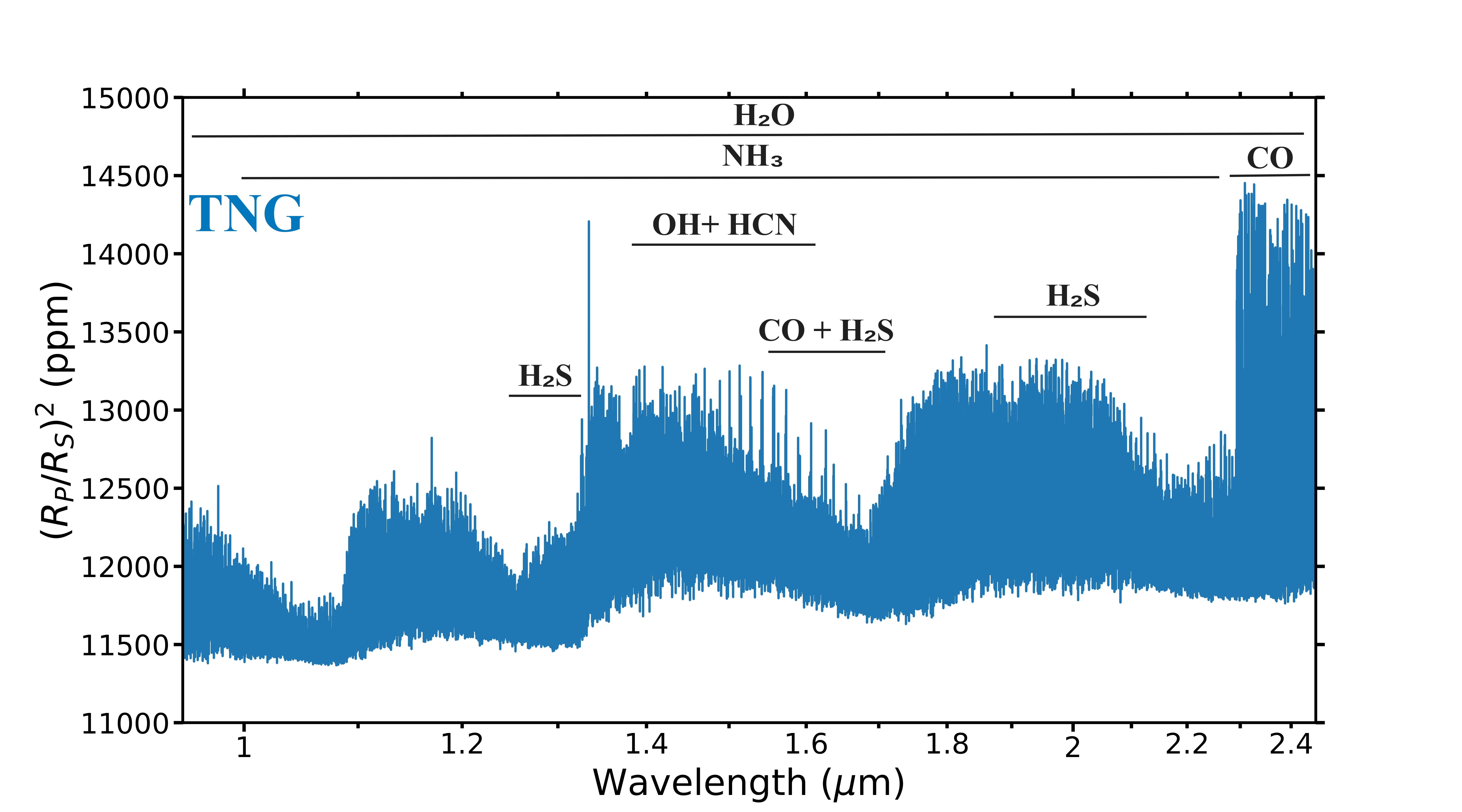}
        \end{minipage}
        %\columnbreak
         \begin{minipage}[b]{\columnwidth}
            \includegraphics[width=\columnwidth]{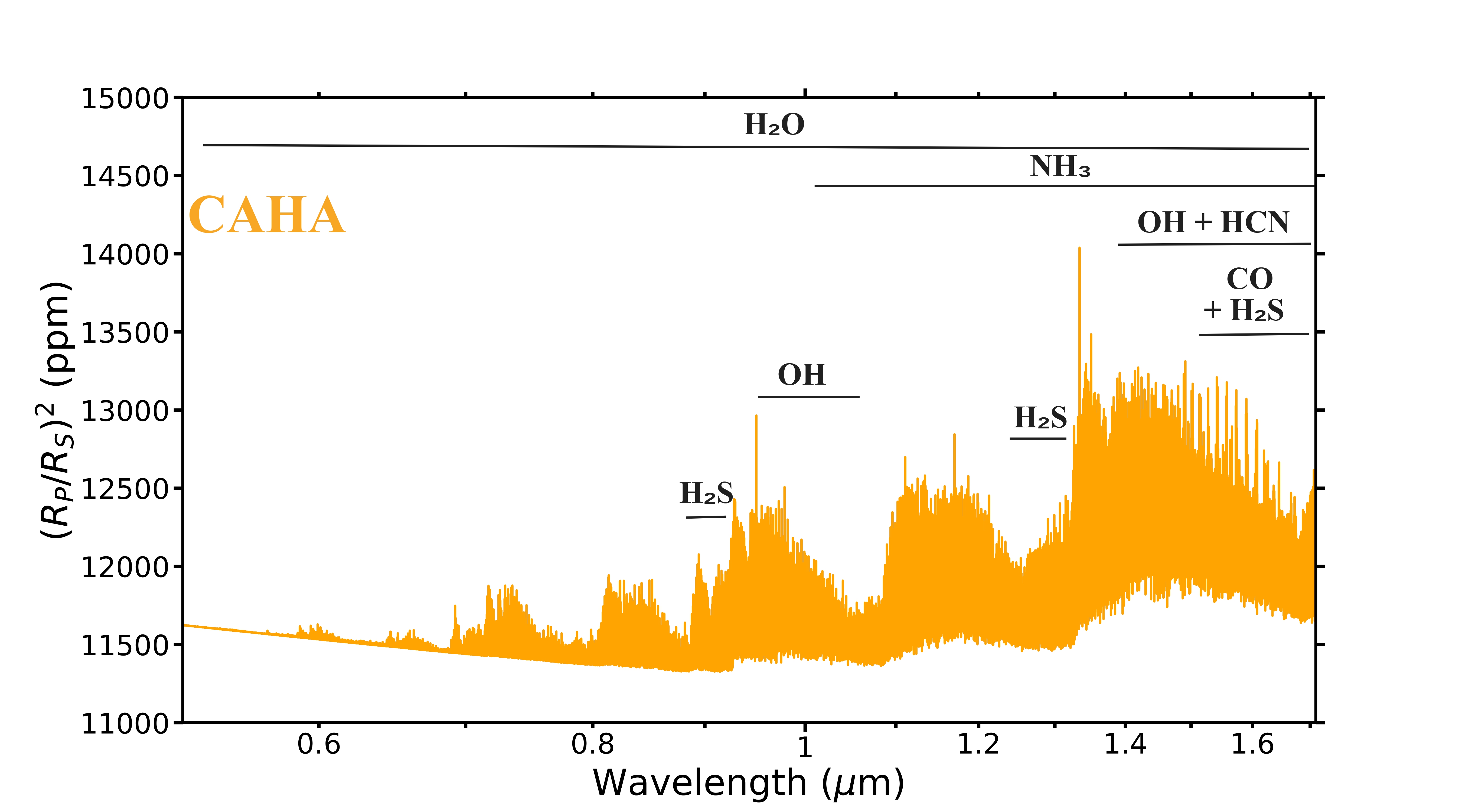}
        \end{minipage}
         \begin{minipage}[b]{\columnwidth}
            \includegraphics[width=\columnwidth]{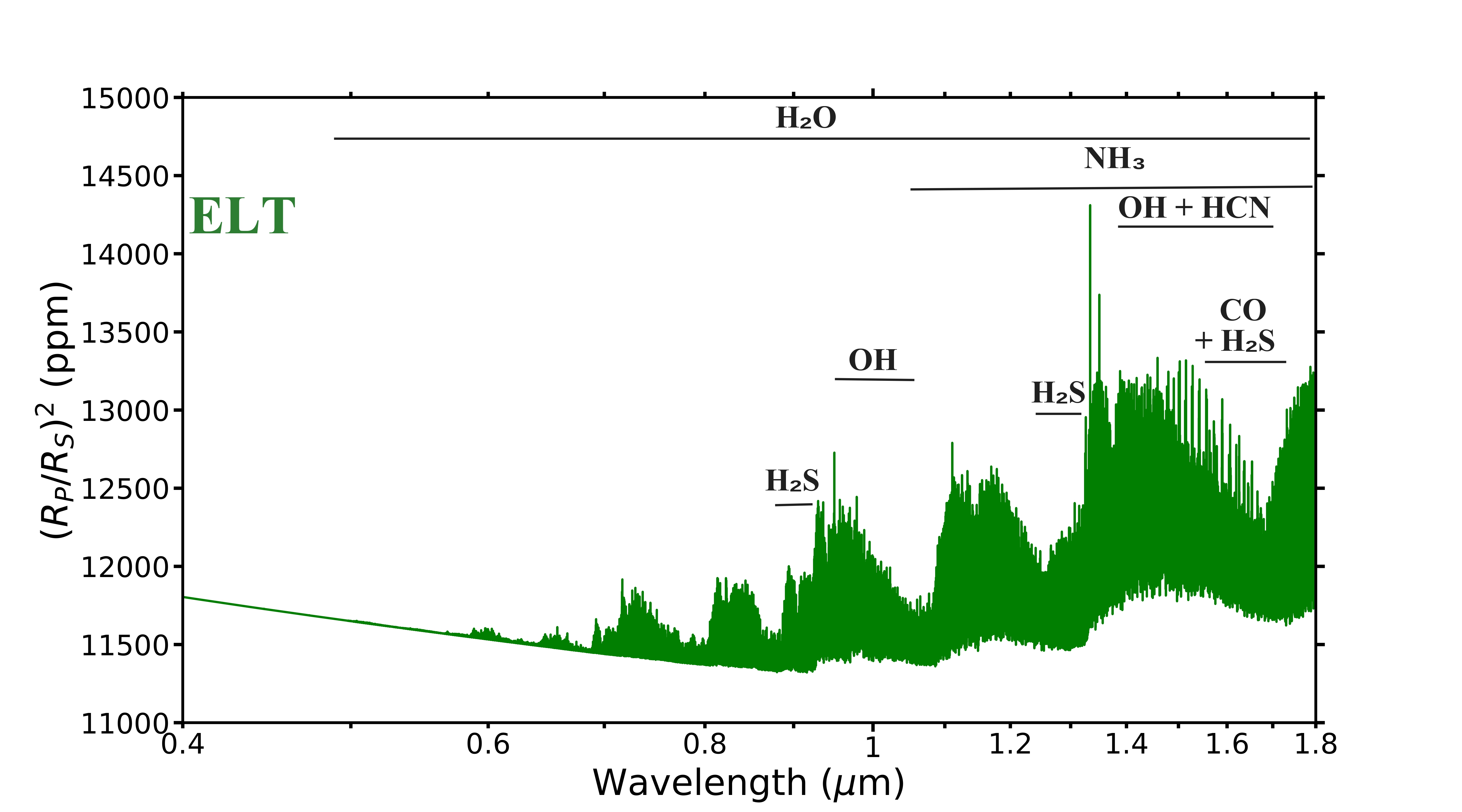}
        \end{minipage}
\caption{High-resolution transmission spectra of \textbf{WASP-76 b} for different instruments (see instrumental parameters from Table \ref{tab:instruments}. \textit{Blue}: GIANO-B (TNG). \textit{Orange}: CARMENES (CAHA). \textit{Green}: ANDES (ELT). \textbf{The figure illustrates the contributions from different molecular bands to the net planet spectra for all cases.} Despite having significant average photospheric VMRs, discernible $\mathrm{CH_4}$ and $\mathrm{CO_2}$ features are not present in any of the synthetic spectra.}
\label{fig:Transmission_Wasp76b}
\end{figure*}

\noindent The reactions in red colors indicate the rate-limiting steps for the whole reaction pathways. On the other hand, a large amount of $\mathrm{NH_3}$ gets converted to $\mathrm{H_2S}$ and $\mathrm{H_2O}$, leading to a further reduction in $\mathrm{NH_3}$ concentration on both planets. A similar story for HCN has been observed too on \textbf{WASP-76 b}, producing $\mathrm{H_2O}$ and $\mathrm{H_2CN}$ significantly. Figure \ref{fig:chemisrty}, which provides a robust visual representation of chemical networks for both planets, emphasizes this fact further. The figure highlights the molecular interconversions, with their net percentage contribution to the formation and destruction of other molecules. Types of reaction intermediates that participate in the fastest molecular conversion pathway are shown in Figures \ref{fig:shortest_wasp76b} and \ref{fig:shortest_wasp77ab}. The shortest route for molecular interconversions is built upon the principle of Dijkstra's algorithm (the principle of finding the least distance between two nodes in a complex network with other nodes present in between) \citep{dijkstra2022note} and is adopted from \cite{Tsai_2018}. While not being present significantly in the atmospheres, NH, SH, NS, and $\mathrm{NH_2}$ play very crucial roles in the interconversions (see Appendix \ref{appendix_1}).

Figures \ref{fig:Transmission_Wasp77ab} and \ref{fig:Transmission_Wasp76b} present synthetic spectra for \textbf{WASP-77 A b} and \textbf{WASP-76 b}, respectively, showcasing their spectral characteristics. These models are generated at the resolutions and wavelength ranges corresponding to the TNG (GIANO-B), CAHA (CARMENES), and ELT (ANDES) instruments. The figures also illustrate the molecular contributors responsible for various spectral features. It is noteworthy that \textbf{WASP-77 A b} lacks discernible  $\mathrm{NH_3}$ and $\mathrm{CO_2}$ features, while \textbf{WASP-76 b} exhibits an absence of $\mathrm{CH_4}$ and $\mathrm{CO_2}$ features in their respective spectra.  This is due to the prominent features of other molecules, coupled with the lower abundance of the respective molecules.

%%%%%%%%%%%%%%%%%%%%%%%%%%%%%%%%%%%%%%%%%%%%%%%%%%%%%%%%%%%%%%%%%%%%%%

\begin{figure*}
    \centering  
        \begin{minipage}[b]{1.95\columnwidth}
            \includegraphics[width=\columnwidth]{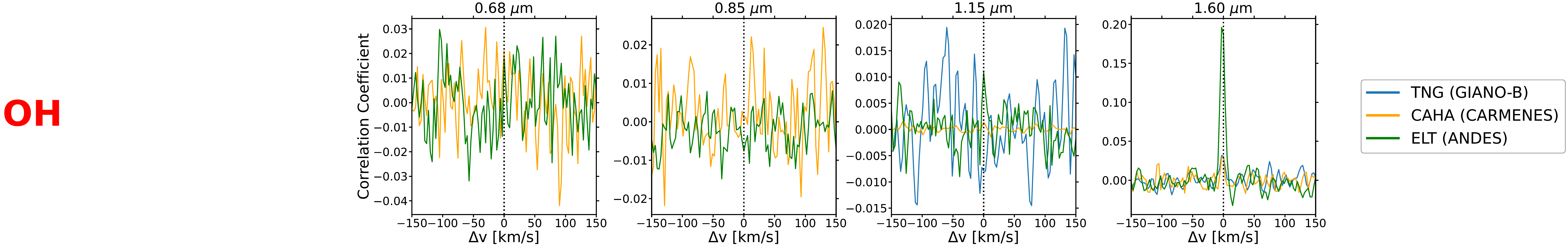}
        \end{minipage}
        %\columnbreak
         \begin{minipage}[b]{1.95\columnwidth}
            \includegraphics[width=\columnwidth]{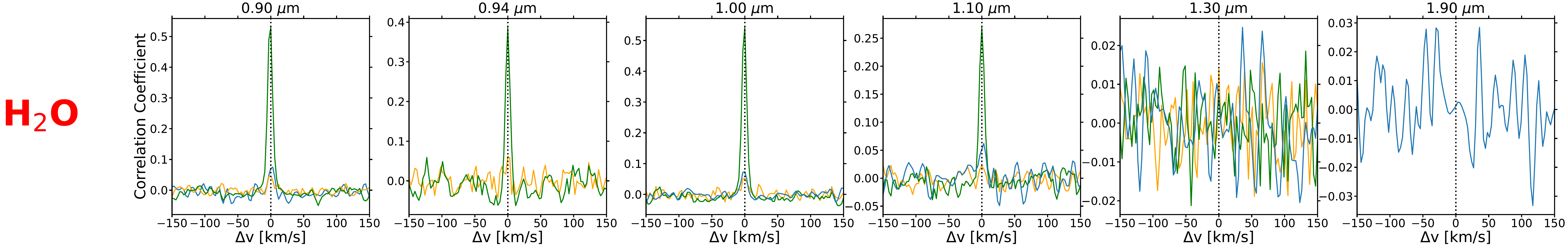}
        \end{minipage}
        \begin{minipage}[b]{1.95\columnwidth}
            \includegraphics[width=\columnwidth]{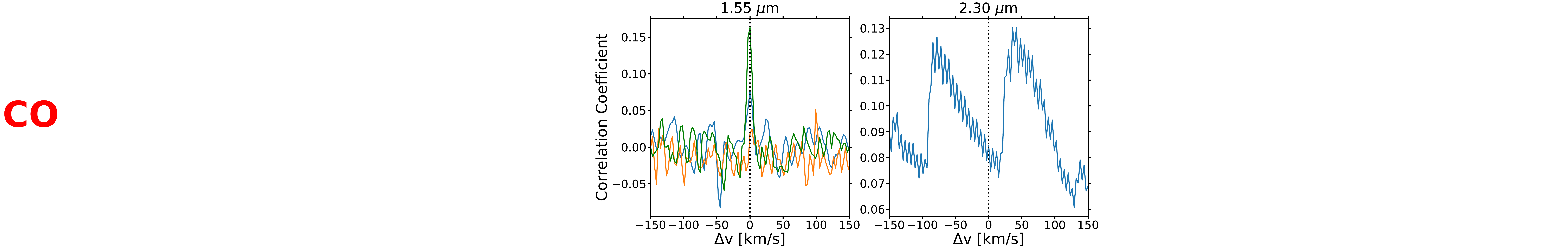}
        \end{minipage}
        \begin{minipage}[b]{1.95\columnwidth}
            \includegraphics[width=\columnwidth]{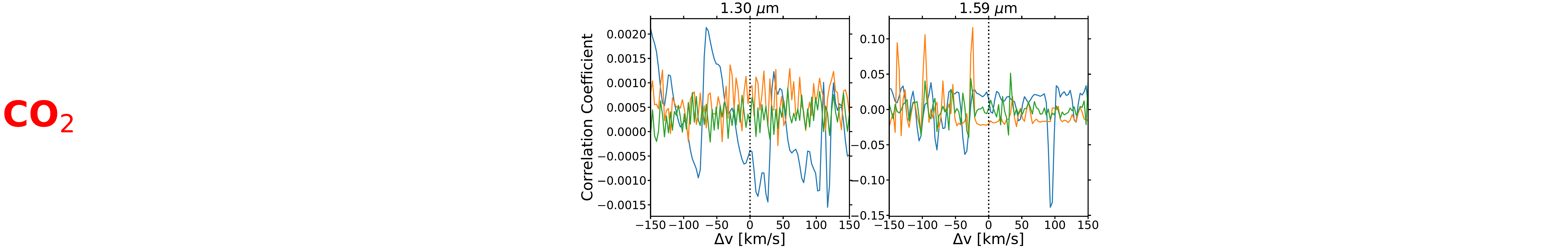}
        \end{minipage}
        \begin{minipage}[b]{1.95\columnwidth}
            \includegraphics[width=\columnwidth]{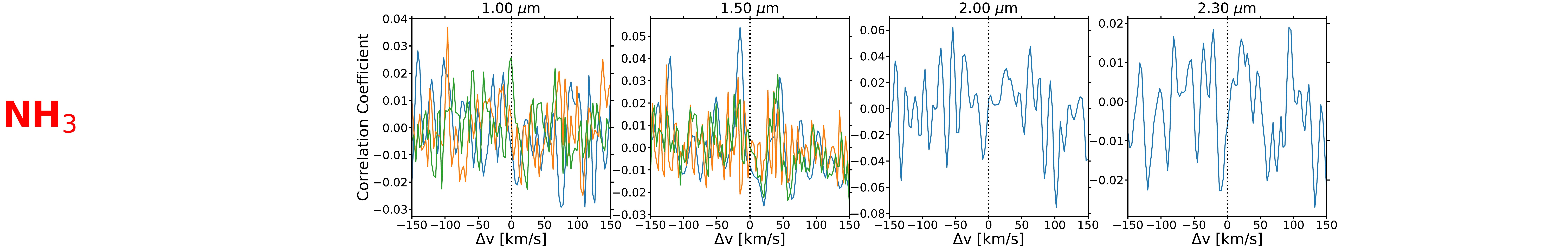}
        \end{minipage}
        \begin{minipage}[b]{1.95\columnwidth}
            \includegraphics[width=\columnwidth]{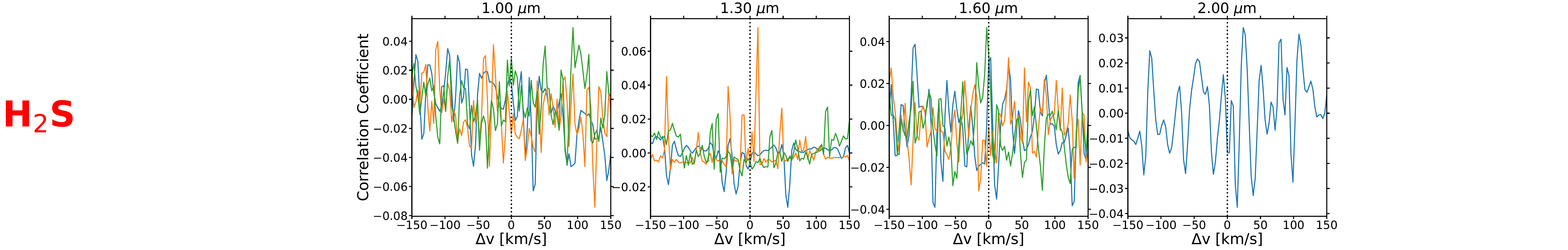}
        \end{minipage}
         \begin{minipage}[b]{1.95\columnwidth}
            \includegraphics[width=\columnwidth]{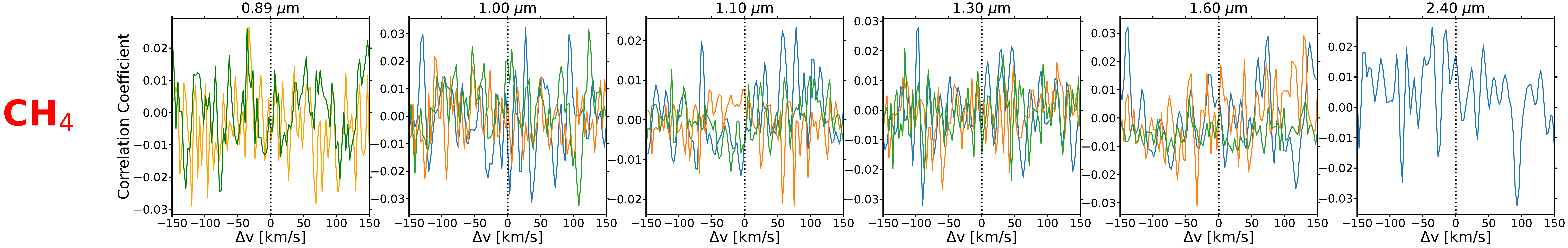}
        \end{minipage}
        \begin{minipage}[b]{1.95\columnwidth}
            \includegraphics[width=\columnwidth]{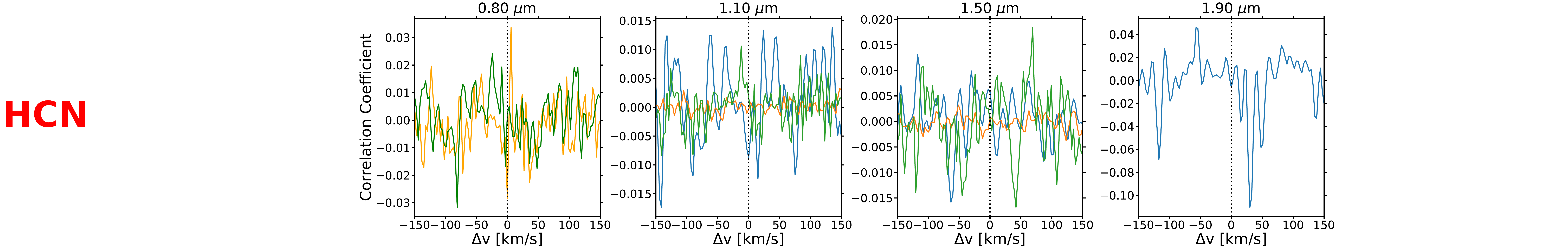}
        \end{minipage}
\caption{Correlation coefficient vs. velocity shift map for different molecular bands for 3 transits of \textbf{WASP-76 b}. \textit{Blue}: GIANO-B (TNG). \textit{Orange}: CARMENES (CAHA). \textit{Green}: ANDES (ELT). Each row corresponds to a specific molecule (\textit{from top to bottom}: OH, $\mathrm{H_2O}$, $\mathrm{CH_4}$, CO, $\mathrm{CO_2}$, $\mathrm{HCN}$, $\mathrm{NH_3}$, $\mathrm{H_2S}$). Molecular bands are listed explicitly in Table \ref{tab:detetcion_Wasp76b}.}
\label{fig:CCF_Wasp76b}
\end{figure*}

%%%%%%%%%%%%%%%%%%%%%%%%%%%%%%%%%%%%%%%%%%%%%%%%%%%%%%%%%%%
\begin{table*}
\centering
%\captionsetup{justification=centering} % Center the caption
\caption{Table summarizing the $\mathrm{\sigma_{det}}$ achieved for different molecular bands for different number of transits of \textbf{WASP-76 b} from different ground-based instruments (bands with $\mathrm{\sigma_{det}}$ $>$ 3 are shown in bold)}
\begin{tabular*}{\textwidth}{@{\extracolsep{\fill}}lcc|ccc|ccc@{}}

%\begin{tabular*}{cccccccc}
\hline
\hline
\noalign{\smallskip}
&  & &   & \textbf{No. of transit = 1} &  &   & \textbf{No. of transit = 3} & \\
\noalign{\smallskip}
\hlineB{3.5}
\noalign{\smallskip}
Molecules & Bands & Wavelength range &  TNG & CAHA & E-ELT &  TNG & CAHA & E-ELT\\
 &  ($\mu$m) & ($\mu$m)&  (GIANO-B) & (CARMENES) & (ANDES) &  (GIANO-B) & (CARMENES) & (ANDES)  \\
\noalign{\smallskip}
\hlineB{3.5}
\noalign{\smallskip}
OH & 0.68 & 0.66 - 0.70 &-- & No detection & No detection &-- & No detection & No detection\\
& 0.85 & 0.77 - 0.93 & -- & No detection & No detection &-- & No detection & No detection\\
& 1.15 & 1.0 - 1.3 &  No detection & No detection & 2.36 & No detection & \textbf{3.06} & \textbf{3.31}\\
& 1.60 & 1.5 - 1.7 &  3.98 & 3.23 & \textbf{12.87} & \textbf{4.66} & \textbf{4.41} & \textbf{14.75}\\
\noalign{\smallskip}
\hline
\noalign{\smallskip}
$\mathrm{H_2O}$ & 0.90 & 0.89 - 0.99 & \textbf{6.35} & \textbf{5.10} & \textbf{19} & \textbf{6.68} & \textbf{5.64} & \textbf{19.12}\\
& 0.94 & 0.93 - 0.95 & -- & 1.96 & \textbf{13.17} & \textbf{--} & \textbf{3.27} & \textbf{14.45}\\
& 1.00 &0.95 - 1.05 & \textbf{3.58} & 1.21 & \textbf{18.57} & \textbf{8.64} & \textbf{4.54} & \textbf{19.94}\\
& 1.10 & 1.1 - 1.2 & \textbf{3.01} & No detection & \textbf{15.87} & \textbf{3.34} & No detection & \textbf{17.63}\\
& 1.30 & 1.3 - 1.55 & No detection & No detection & No detection & No detection & No detection & No detection\\
& 1.90 & 1.8 - 2.1 & No detection & -- & -- & No detection & -- & --\\
\noalign{\smallskip}
\hline
\noalign{\smallskip}
CO & 1.55 & 1.555 - 1.60 &  2.36 & 1.24 & \textbf{9.38} &  \textbf{3.15} & 2.10 & \textbf{10.34}\\
& 2.30 & 2.319 - 2.3315 & No detection & -- & -- & No detection & -- & --\\
\noalign{\smallskip}
\hline
\noalign{\smallskip}
$\mathrm{CO_2}$ & 1.30 & 1.2 - 1.4 &  No detection & No detection & No detection &  No detection & No detection & No detection\\
& 1.59 & 1.45 - 1.66 &  No detection & No detection & No detection &  No detection & No detection & No detection\\
\noalign{\smallskip}
\hline
\noalign{\smallskip}
$\mathrm{NH_3}$ & 1.00 & 1.0 - 1.1 & No detection & No detection & \textbf{3.83} & 1.97 & 1.71 & \textbf{4.29}\\
& 1.50 & 1.45 - 1.65 &  No detection & No detection & No detection &  No detection & No detection & No detection\\
& 2.00 & 2.0 - 2.1  & No detection & -- & -- & No detection & -- & --\\
& 2.30 & 2.2 - 2.4 & No detection & -- & -- & No detection & -- & --\\
\noalign{\smallskip}
\hline
\noalign{\smallskip}
$\mathrm{H_2S}$ & 1.00 & 1.0 - 1.03 & No detection & No detection & No detection  & No detection & No detection & No detection\\
& 1.30 & 1.27 - 1.37 &No detection & No detection & No detection & No detection & No detection & No detection\\
& 1.60 & 1.55 - 1.66 &1.85 & No detection & \textbf{3.46}  & \textbf{3.18} & No detection & \textbf{4.35}\\
& 2.00 & 1.9 - 2.1 & No detection & -- & -- & No detection & -- & --\\
\noalign{\smallskip}
\hline
\noalign{\smallskip}
$\mathrm{CH_4}$ & 0.89 & 0.875 - 0.91 & -- & No detection & No detection & -- & No detection & No detection\\
& 1.00 & 0.98 - 1.05 & No detection & No detection & 1.41 & 1.78 & No detection & 2.26\\
& 1.10 & 1.1 - 1.2 & No detection & No detection & No detection & No detection & No detection & No detection \\
& 1.30 & 1.3 - 1.55 & No detection & No detection & No detection & No detection & No detection & No detection\\
& 1.60 & 1.6 - 1.8 & No detection & No detection & 1.16 & No detection & No detection & 2.70\\
& 2.40 & 2.2 - 2.5 &No detection & -- & -- & No detection & -- & --\\
\noalign{\smallskip}
\hline
\noalign{\smallskip}
HCN & 0.80 & 0.8 - 0.9 &  -- & No detection &  1.07 &  -- & 2.42 &  2.65\\
& 1.10 & 1.0 - 1.2 &  No detection & No detection & No detection &  No detection & No detection & No detection\\
& 1.50 & 1.47 - 1.56 &  No detection & No detection & No detection &  No detection & No detection & No detection\\
& 1.90 & 1.85 - 1.95 &  No detection & -- & -- &  No detection & -- & --\\
\noalign{\smallskip}
\hline
\hline
\label{tab:detetcion_Wasp76b}
\end{tabular*}
\end{table*}

%%%%%%%%%%%%%%%%%%%%%%%%%%%%%%%%%%%%%%%%%%%%%%%%%%%%%%%%%%%%%%%%%%%%%%%%%%%%%%%%%%%
\subsection{Cross-correlation spectroscopy: estimation of $\mathrm{\sigma_{det}}$ for molecular bands}
\label{result:cross-correlation}

In our analysis, assessing the detectability of molecular signatures presents a unique challenge, primarily arising from the complex spectral profiles characterized by more spectral lines present in the spectrum. Unlike the conventional signal-to-noise ratio (S/N) estimation, where a non-detection typically results in an S/N of zero, our approach yields a non-detection consistent with a significance level of $\mathrm{\sigma_{det}}$ $\sim$  1 \citep{currie2023there}. This deviation from the traditional methodology is driven by the intricacies of our detection scheme, which accounts for the complex nature of the spectral lines and their interactions, making our scheme a more appropriate means of determining detectability in this context.

\begin{figure*}
    \centering  
        \begin{minipage}[b]{2\columnwidth}
            \includegraphics[width=\columnwidth]{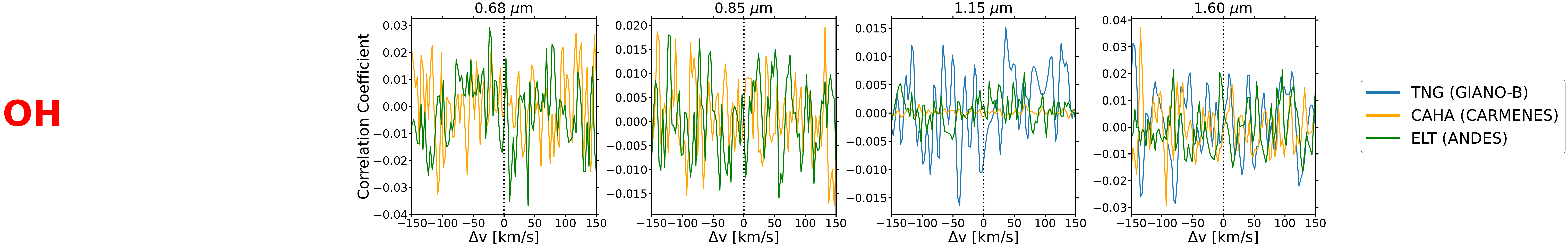}
        \end{minipage}
        %\columnbreak
         \begin{minipage}[b]{2\columnwidth}
            \includegraphics[width=\columnwidth]{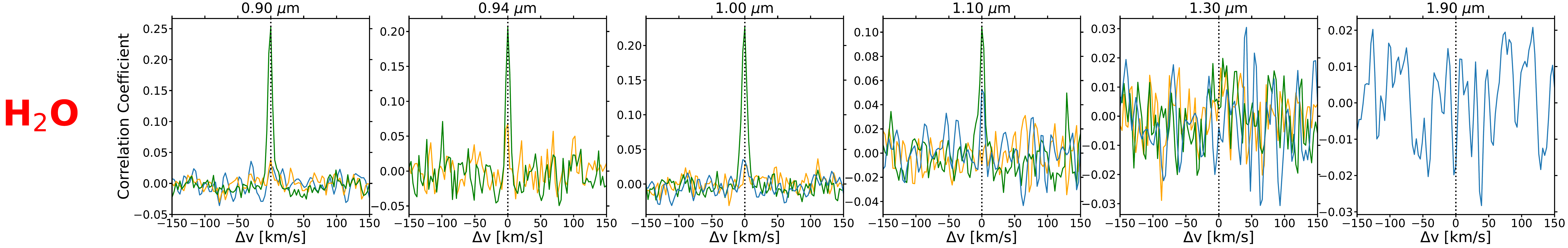}
        \end{minipage}
         \begin{minipage}[b]{2\columnwidth}
            \includegraphics[width=\columnwidth]{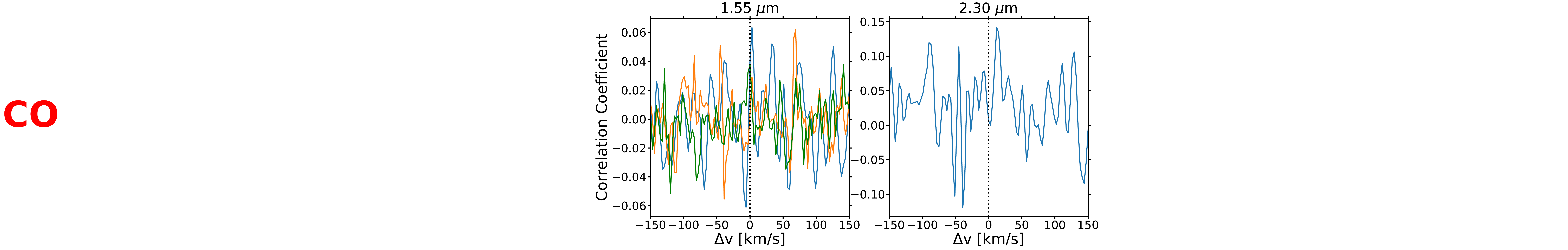}
        \end{minipage}
        \begin{minipage}[b]{2\columnwidth}
            \includegraphics[width=\columnwidth]{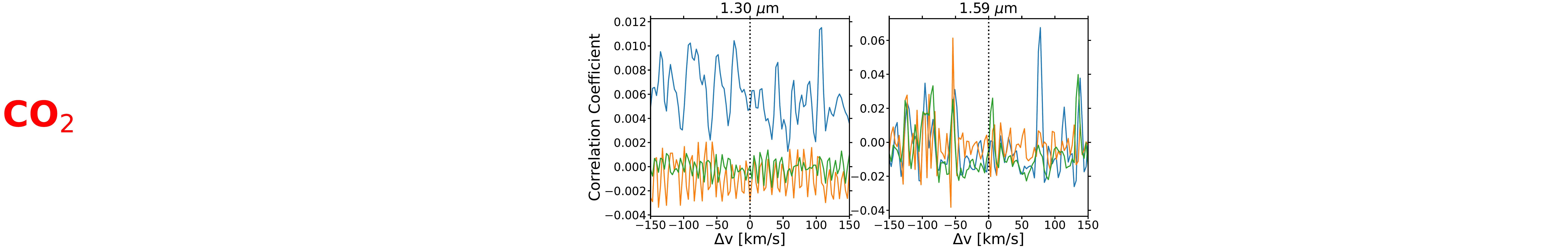}
        \end{minipage}
        \begin{minipage}[b]{2\columnwidth}
            \includegraphics[width=\columnwidth]{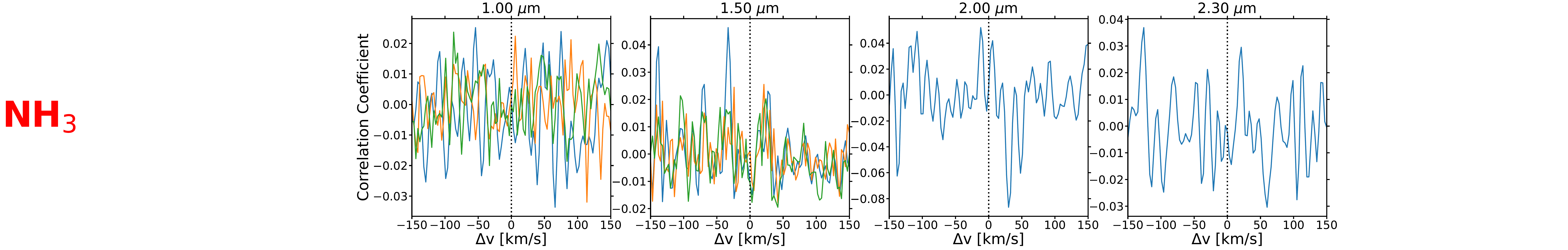}
        \end{minipage}
        \begin{minipage}[b]{2\columnwidth}
            \includegraphics[width=\columnwidth]{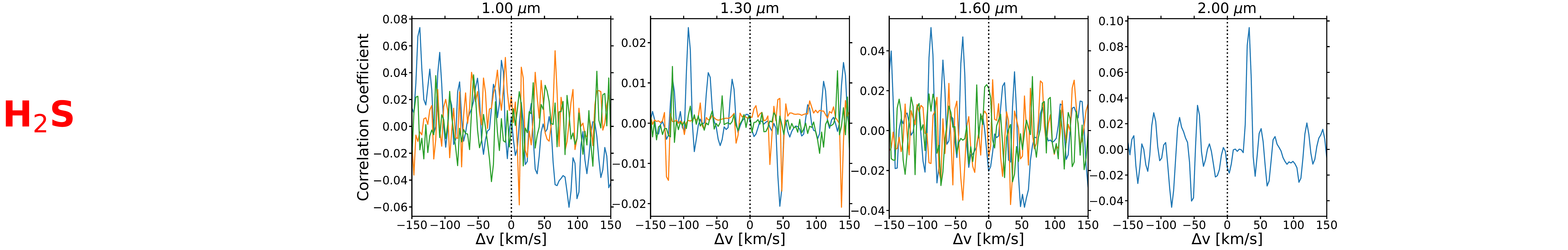}
        \end{minipage}
\caption{Correlation coefficient vs. velocity shift map for different molecular bands for 3 transits of \textbf{WASP-77 A b}. \textit{Blue}: GIANO-B (TNG). \textit{Orange}: CARMENES (CAHA). \textit{Green}: ANDES (ELT). Each row corresponds to a specific molecule (\textit{from top to bottom}: OH, $\mathrm{H_2O}$, CO, $\mathrm{CO_2}$, $\mathrm{NH_3}$, $\mathrm{H_2S}$). Molecular bands are listed explicitly in Table \ref{tab:detetcion_Wasp77ab}.}
\label{fig:CCF_Wasp77ab}
\end{figure*}

Therefore, our approach involves a more detailed analysis rather than employing the conventional S/N estimator, which provides a detectability estimate for the entire wavelength range. We have defined specific molecular bands corresponding to each molecule of interest (see Table \ref{tab:detetcion_Wasp76b} and \ref{tab:detetcion_Wasp77ab}) and provided an overview in terms of $\mathrm{\sigma_{det}}$ for each of the molecular bands using three different spectrographs for 1 and 3 transits, respectively. This approach allows us to assess the detectability of individual molecular features within the broader spectrum, providing a more detailed and accurate characterization of their presence in a better way. In the subsequent sections, we have discussed a comprehensive analysis of detecting various molecular bands on both planets using all the instruments. Transit depths for all bands are shown in Figures \ref{fig:band_spectrum_TNG_76b}, 
 \ref{fig:band_spectrum_CAHA_76b}, \ref{fig:band_spectrum_ELT_76b} (for \textbf{WASP-76 b}) and Figures \ref{fig:band_spectrum_TNG_77ab}, 
 \ref{fig:band_spectrum_CAHA_77ab}, \ref{fig:band_spectrum_ELT_77ab} (for \textbf{WASP-77 A b}). The cross-correlation spectroscopy for molecular bands has been depicted in Figures \ref{fig:CCF_Wasp76b} and \ref{fig:CCF_Wasp77ab}, where a clear detection is indicated by a pronounced peak at zero velocity shift in the correlation coefficient vs. velocity shift map.

%%%%%%%%%%%%%%%%%%%%%%%%%%%%%%%%%%%%%%%%%%%%%%%%%%%%%%%%%

%%%%%%%%%%%%%%%%%%%%%%%%%%%%%%%%%%%%%%%%%%%%%%%%%%%%%%%%%%%%%%%%%%%%%%%%

\begin{table*}
\centering
%\captionsetup{justification=centering} % Center the caption
\caption{Table summarizing the $\mathrm{\sigma_{det}}$ achieved for different molecular bands for different number of transits of \textbf{WASP-77 A b} from different ground-based instruments (bands with $\mathrm{\sigma_{det}}$ $>$ 3 are shown in bold)}
\begin{tabular*}{\textwidth}{@{\extracolsep{\fill}}lcc|ccc|ccc@{}}

%\begin{tabular*}{cccccccc}
\hline
\hline
\noalign{\smallskip}
&  & &   & \textbf{No. of transit = 1} &  &   & \textbf{No. of transit = 3} & \\
\noalign{\smallskip}
\hlineB{3.5}
\noalign{\smallskip}
Molecules & Bands & Wavelength range &  TNG & CAHA & E-ELT &  TNG & CAHA & E-ELT\\
 &  ($\mu$m) & ($\mu$m)&  (GIANO-B) & (CARMENES) & (ANDES) &  (GIANO-B) & (CARMENES) & (ANDES)  \\
\noalign{\smallskip}
\hlineB{3.5}
\noalign{\smallskip}
OH & 0.68 & 0.66 - 0.70 & -- & No detection & No detection &-- & No detection & No detection\\
& 0.85 & 0.77 - 0.93 & -- & No detection & No detection &-- & No detection & No detection\\
& 1.15 & 1.0 - 1.3 &  No detection & No detection & No detection & No detection & No detection & 2.02\\
& 1.60 & 1.5 - 1.7 &  No detection & No detection & No detection & No detection & No detection & 2.76\\
\noalign{\smallskip}
\hline
\noalign{\smallskip}
$\mathrm{H_2O}$ & 0.90 & 0.89 - 0.99 & 2.49 & 2.10 & \textbf{13.04} & \textbf{3.44} & \textbf{3.30} & \textbf{18.77}\\
& 0.94 & 0.93 - 0.95 & -- & 1.74 & \textbf{5.76} & -- & \textbf{4.53} & \textbf{10.52}\\
& 1.00 &0.95 - 1.05 & 1.56 & 1.50 & \textbf{12.85} & \textbf{4.56} & \textbf{3.70} & \textbf{15.52}\\
& 1.10 & 1.1 - 1.2 & 2.56 & No detection & \textbf{5.45} & \textbf{4.87} & No detection & \textbf{7.89}\\
& 1.30 & 1.3 - 1.55 & No detection & No detection & No detection & No detection & No detection & No detection\\
& 1.90 & 1.8 - 2.1 & 1.63 & -- & -- & 2.98 & -- & --\\
\noalign{\smallskip}
\hline
\noalign{\smallskip}
CO & 1.55 & 1.555 - 1.60 &  1.55 & 1.16 & 1.92 &  2.75 & 1.93 & \textbf{3.49}\\
& 2.30 & 2.319 - 2.3315 & No detection & -- & -- & No detection & -- & --\\
\noalign{\smallskip}
\hline
\noalign{\smallskip}
$\mathrm{CO_2}$ & 1.30 & 1.2 - 1.4 &  No detection & No detection & No detection &  No detection & No detection & No detection\\
& 1.59 & 1.45 - 1.66 &  No detection & No detection & No detection &  No detection & No detection & No detection\\
\noalign{\smallskip}
\hline
\noalign{\smallskip}
$\mathrm{NH_3}$ & 1.00 & 1.0 - 1.1 & No detection & No detection & No detection & No detection & No detection & 2.61\\
& 1.50 & 1.45 - 1.65 &  No detection & No detection & No detection &  No detection & No detection & No detection\\
& 2.00 & 2.0 - 2.1  & No detection & -- & -- & No detection & -- & --\\
& 2.30 & 2.2 - 2.4 & No detection & -- & -- & No detection & -- & --\\
\noalign{\smallskip}
\hline
\noalign{\smallskip}
$\mathrm{H_2S}$ & 1.00 & 1.0 - 1.03 & No detection & No detection & No detection  & No detection & No detection & No detection\\
& 1.30 & 1.27 - 1.37 & No detection & No detection & No detection & No detection & No detection & No detection\\
& 1.60 & 1.55 - 1.66 & No detection & No detection & 1.72  & No detection & No detection & 2.11\\
& 2.00 & 1.9 - 2.1 & No detection & -- & -- & No detection & -- & --\\
\noalign{\smallskip}
\hline
\hline
\label{tab:detetcion_Wasp77ab}
\end{tabular*}
\end{table*}
%%%%%%%%%%%%%%%%%%%%%%%%%%%%%%%%%%%%%%%%%%%%%%%%%%%%%%%%%%%%%%%%%%%%%%

\subsubsection{$\mathrm{OH}$}
\label{sec:OH}

OH is one of the major byproducts that have formed during the atmospheric dissociation of $\mathrm{H_2O}$ \citep{landman2021detection}. For HJs and UHJs, the enrichment of OH at the top of the atmosphere can be attributed to the photolysis of water molecules. It was proven during the detection of OH (with a 5.5$\sigma$ significance) on Wasp-33 b, where the insignificant detection of $\mathrm{H_2O}$ lends support to the hypothesis that OH formation is facilitated by the thermal dissociation of $\mathrm{H_2O}$ molecules \citep{nugroho2021first}. However, from Figure \ref{fig:chemisrty}, it is evident that $\mathrm{H_2O}$ does play a substantial role in the generation of OH radicals through the following reaction: $\mathrm{H_2O}$ + H $\rightarrow$ OH + $\mathrm{H_2}$ (see Figure \ref{fig:shortest_wasp76b} and \ref{fig:shortest_wasp77ab}), but not with direct photolysis. A more detailed understanding of this necessitates further exploration, which can be achieved through the utilization of more intricate chemical networks.

For \textbf{WASP-76 b}, all spectrographs demonstrate superior sensitivity in probing the 1.60 $\mu$m band. While achieving a better $\mathrm{\sigma_{det}}$ requires more transits for GIANO-B and CARMENES, ANDES is capable of reaching a commendable $\mathrm{\sigma_{det}}$ within 1 transit. Additionally, CARMENES and ANDES exhibit significant precision in detecting the 1.15 $\mu$m band with a higher number of transits. Similarly, in the case of \textbf{WASP-77 A b}, ANDES exhibits reasonable precision in detecting the 1.15 $\mu$m and 1.60 $\mu$m features with an increased number of transits. However, it's worth noting that ANDES exhibits a lower level of confidence for detecting the 1.60 $\mu$m band compared to the observations of \textbf{WASP-76 b}.

\subsubsection{$\mathrm{H_2O}$}
\label{sec:H2O}

From the planet spectra, it is clear that both planets are water-dominated. Among all six bands, the 0.90, 1.00, and 1.10 $\mu$m bands in the optical and NIR spectral regions exhibit notably higher precision across three instruments for a higher number of transits. It is important to note that ANDES demonstrates superior capability in containing these features even with a single transit. Both CARMENES and ANDES exhibit favorable detectability for the 0.94 $\mu$m band too. All over, CARMENES shows a lower level of sensitivity towards detecting different bands. This is attributed to the higher read noise value of the spectrograph and the lower collecting area of the mirror.

In our analysis, the 0.90, 0.94, 1.00, and 1.10 $\mu$m bands have proven to be more dependable in terms of detection. Although the 1.30 $\mu$m band exhibits pronounced features, they are susceptible to overlap with the prominent telluric transmission features, resulting in reduced significance during detection. A similar effect is also observed for the 1.10 $\mu$m band, albeit to a lesser extent.

%%%%%%%%%%%%%%%%%%%%%%%%%%%%%%%%%%%%%%%%%%%%%%%%%%%%%%%%%%%%

%%%%%%%%%%%%%%%%%%%%%%%%%%%%%%%%%%%%%%%%%%%%%%%%%%%%%%%%%%%%%

\subsubsection{$\mathrm{CO}$}
\label{sec:CO}

CO emerges as the predominant molecule in terms of VMR on both planets (see Figure \ref{fig:VMR}). It is produced mainly by the direct reduction $\mathrm{CO_2}$ molecule in the atmospheres: $\mathrm{CO_2}$ + H $\rightarrow$ CO + OH (see Figure \ref{fig:chemisrty} and Appendix \ref{appendix_1}). GIANO-B and ANDES exhibit potential in detecting the 1.55 $\mu$m band with $\mathrm{\sigma_{det}}$ $>$ 3 in the atmosphere of \textbf{WASP-76 b} (GIANO-B: with 3 transits and ANDES: with a single transit). \textbf{Moreover, the detection of CO on \textbf{WASP-77 A b} is insignificant with GIANO-B and CARMENES with a $\mathrm{\sigma_{det}}$ $<$ 3 even with 3 transits.} ANDES can achieve a significance level of $\mathrm{\sigma_{det}}$ $>$ 3 for the 1.55 $\mu$m CO band when observing for 3 transits. The detectability increases in the sequence from CARMENES to GIANO-B to ANDES.

\subsubsection{$\mathrm{CO_2}$}
\label{sec:CO2}

With a threshold of average photospheric VMR $>$ $\mathrm{10^{-8}}$, it is apparent that $\mathrm{CO_2}$ is the least abundant species on both planets. Unfortunately, neither of its bands is detectable for all instruments.

\subsubsection{$\mathrm{NH_3}$}
\label{sec:NH3}

As previously discussed in Section \ref{result:spectra}, the contribution of $\mathrm{NH_3}$ to the planetary spectra is not visually distinguishable for \textbf{WASP-77 A b}, indicating a minimal likelihood of detection. ANDES exhibits reasonable detection significance for \textbf{WASP-77 A b} for the 1.00 $\mu$m band. Nevertheless, it demands more transits to achieve a significant detection. In contrast, for \textbf{WASP-76 b}, the ANDES instrument is capable of resolving the 1.00 $\mu$m band with higher precision ($\mathrm{\sigma_{det}}$ $>$ 3) even with a single transit.

\subsubsection{$\mathrm{H_2S}$}
\label{sec:H2S}

On both studied planets, sulfur is exclusively found in the form of $\mathrm{H_2S}$. Despite being the third most prevalent compound in their atmospheres, no prior detection has been recorded for this molecule. Our analysis, however, indicates a substantial detection of the 1.60 $\mu$m band on \textbf{WASP-76 b} and a moderate level of detection on \textbf{WASP-77 A b}, using the ANDES instrument. With additional transits, GIANO-B too shows prominence in detecting the 1.60 $\mu$m band on \textbf{WASP-76 b}. The significant formation of this molecule in the atmosphere can be attributed to the destruction $\mathrm{NH_3}$ and SH radical (see Figure \ref{fig:chemisrty} and Appendix \ref{appendix_1}).

\subsubsection{$\mathrm{CH_4}$}
\label{sec:CH4}

Detecting $\mathrm{CH_4}$ is particularly challenging across all three instruments due to its extremely low atmospheric abundance on \textbf{WASP-76 b}. Nevertheless, ANDES displays a little promising capability in detecting the 1.00 and 1.60 $\mu$m bands with 3 transits, each with a $\mathrm{\sigma_{det}}$ $>$ 2. However, none of these instruments exhibits a distinct, sharp feature in the correlation coefficient vs. velocity shift map.

\subsubsection{$\mathrm{HCN}$}
\label{sec:HCN}

Despite having a prominent detection on \textbf{WASP-76 b}, our model falls short of establishing a sufficient detection of HCN in the atmosphere, leading to the following conclusion: it is anticipated that HCN will originate in the lower temperature region of the planetary nightside and subsequently migrate to the dayside via atmospheric circulation. Given the higher detection significance of $\mathrm{H_2O}$, the presence of HCN is intriguing, as $\mathrm{H_2O}$ and HCN are not typically expected to coexist \citep{sanchez2022searching}. %A more comprehensive analysis of this phenomenon can be achieved by expanding the investigation from 1-D models to 3-D GCM study.
However, HCN can be detected to some extent with an increased number of transits on \textbf{WASP-76 b}. Notably, the CARMENES and ANDES instruments demonstrate a reasonable capability to resolve the 0.80 $\mu$m band with a detection significance $\mathrm{\sigma_{det}}$ $>$ 2.

%%%%%%%%%%%%%%%%%%%%%%%%%%%%%%%%%%%%%%%%%%%%%%%%%%%%%%%%%%%%%%%%%%%%%%%%%%%%%%%%%%%%%%%%%%%%%%%%%%%%%%%%%%%%%%%%
\subsection{Present observations and the future with ELT}
\label{sec:future_with_ELT}

Recent advancements in ground-based observatories have opened up new horizons in the field of exoplanetary science, particularly exploring planet atmospheres and comprehending the underlying atmospheric chemistry. By comparing present observations with simulation studies, we can anticipate future observations using more advanced instruments like ELT and gain insights into how ELT will excel the existing instruments in detecting molecular signatures in planetary atmospheres. Our model exhibits good coherence with the present observations for both planets. On \textbf{WASP-76 b}, \cite{landman2021detection} detect OH with a significance of 6.1, while \cite{sanchez2022searching} report the detection of $\mathrm{H_2O}$ (with a significance of 5.5) and HCN (with a significance of 5.2) using the CARMENES spectrograph during a single transit observation. We successfully retrieve $\mathrm{H_2O}$ signature from the 0.90 $\mu$m band with a $\mathrm{\sigma_{det}}$ of 5.10 and OH from the 1.60 $\mu$m band with a little lower $\mathrm{\sigma_{det}}$ of 3.23 for single transit. As discussed in Section \ref{sec:HCN}, the production of HCN is more favorable in the cooler regions of the planet. Due to the limitations imposed by 1D models, we are unable to attain a significant VMR of HCN on the planet and detect it solely through single transit observations. However, with 3 transits, CARMENES can detect the 0.80 $\mu$m band of HCN with a $\mathrm{\sigma_{det}}$ of 2.42.

On the other hand, observations conducted during pre and post-eclipse of \textbf{WASP-77 A b} using the IGRINS spectrograph indicate the presence of CO (with a significance $\sim$ 3.5) and $\mathrm{H_2O}$ (with a significance $\sim$ 9) in the atmosphere \citep{smith2024combined}. Our analysis utilizing GIANO-B and CARMENES also detects CO and $\mathrm{H_2O}$, albeit with slightly lower significance levels for three transits. This is due to the differential spectrograph architectures between GIANO-B, CARMENES, and IGRINS. Specifically, the IGRINS spectrograph has a larger mirror diameter of 8.1 meters compared to 3.58 meters for GIANO-B and 3.5 meters for CARMENES. This larger diameter translates to a doubling in collecting area, leading to a greater influx of photons and consequently enhancing detectability. Thus, the observed differences in detection significance values across the instruments are justified by these architectural distinctions. Given the accordance between our simulations and real-time observations, it is anticipated that the ELT will surpass other ground-based observatories, paving the way for a new era in atmospheric studies with heightened sensitivity for detecting molecular signatures in exoplanet atmospheres. With a larger \textbf{diameter} of 39.3 meters and a resolution of 100000, ELT can probe molecules that are hard to observe using other instruments and disentangle molecular features with better precision.
%%%%%%%%%%%%%%%%%%%%%%%%%%%%%%%%%%%%%%%%%%%%%%%%%%%%%%%%%%%%%%%%%%%%%%%%%%%%%%%%%%%%%%%%%%%%%%%%%%%%%%%%%%%%%%%%

\section{Conclusions}
\label{sec:conclusion}
 
HJs and UHJs present us with a unique opportunity to characterize their atmospheres and determine the molecular abundances of O- and C-bearing species \citep{guillot2022giant}, offering insights into the exotic chemical processes taking place within their atmospheres. The primary objective of this work was to investigate the atmospheres of \textbf{WASP-77 A b} (HJ) and \textbf{WASP-76 b} (UHJ) using three high-resolution ground-based spectrographs (GIANO-B, CARMENES, ANDES) and conduct a complementary analysis to assess the detectability of different molecular bands. Among all instruments, ANDES demonstrated significantly better precision in detecting molecular bands, thanks to its larger collecting area and higher instrumental resolution. The alignment between our simulation analyses and actual observations for both planets underscores the significance of utilizing the ELT. With its advanced capabilities, ELT holds the promise of significantly enhancing the ability to probe molecules within exoplanet atmospheres with better sensitivity, further advancing our understanding of these distant worlds.

Based on the diverse wavelength coverage and distinct instrumental parameters, spectrographs possess a sweet spot where they excel in detecting particular molecules with enhanced significance. Employing the cross-correlation spectroscopy method, we anticipate the detection of various molecules, including OH (1.15 $\mu$m: CARMENES, ANDES; 1.60 $\mu$m: GIANO-B, CARMENES, ANDES), $\mathrm{H_2O}$ (0.90, 1.00 $\mu$m: GIANO-B, CARMENES, ANDES; 0.94 $\mu$m: CARMENES, ANDES; 1.10 $\mu$m: GIANO-B, ANDES), $\mathrm{CH_4}$ (1.00, 1.60 $\mu$m: ANDES), $\mathrm{CO}$ (1.55 $\mu$m: GIANO-B, ANDES), HCN (0.80 $\mu$m: CARMENES, ANDES), $\mathrm{NH_3}$ (1.00 $\mu$m: ANDES), and $\mathrm{H_2S}$ (1.60 $\mu$m: GIANO-B, ANDES) for \textbf{WASP-76 b}. This indeed encompasses molecular detection from single transit to a maximum of 3 transits for the planet (see Table \ref{tab:detetcion_Wasp76b}). ANDES surpasses CARMENES and GIANO-B in terms of the instrumental capability for detecting most of the molecular bands with a single transit. More transits are usually favorable for detecting the molecules (e.g. OH, CO, and $\mathrm{H_2S}$) using GIANO-B and CARMENES spectrographs.

For \textbf{WASP-77 A b}, we have identified the presence of OH (1.15, 1.60 $\mu$m: ANDES), $\mathrm{H_2O}$ (0.90, 1.00 $\mu$m: GIANO-B, CARMENES, ANDES; 0.94 $\mu$m: CARMENES, ANDES; 1.10 $\mu$m: GIANO-B, ANDES), $\mathrm{CO}$ (1.55$\mu$m: ANDES), $\mathrm{NH_3}$ (1.00 $\mu$m: ANDES) and $\mathrm{H_2S}$ (1.60 $\mu$m: ANDES). For both planets, we have reached a $\mathrm{\sigma_{det}}$ $>$ 2 for above mentioned molecular bands for different numbers of transits (see Tables \ref{tab:detetcion_Wasp76b} and \ref{tab:detetcion_Wasp77ab}). The spectral bands of $\mathrm{H_2O}$, CO, OH, $\mathrm{NH_3}$, and $\mathrm{H_2S}$ are sufficiently robust for \textbf{WASP-76 b}, enabling their detection with a $\mathrm{\sigma_{det}}$ $>$ 3 within 3 transits. Conversely, the signals are a little fainter for \textbf{WASP-77 A b} except $\mathrm{H_2O}$. This prioritizes the need for more transits to significantly detect CO, OH, $\mathrm{NH_3}$, and $\mathrm{H_2S}$ in the atmosphere of \textbf{WASP-77 A b}.

By examining the interlinked chemical network and molecular conversion pathways, we have developed a detailed understanding of the atmospheric chemistry on both planets. It is noteworthy that, aside from photolysis, OH can be formed through alternative reaction mechanisms from water. Additionally, within the photosphere, the rise in the OH mixing ratio is related to the depletion of $\mathrm{H_2O}$ and $\mathrm{NH_3}$. Our efforts to identify HCN signatures in the atmosphere of \textbf{WASP-76 b} were inconclusive due to the extremely low VMR in our 1-D models. Nevertheless, we highlight the significance of nitrogen chemistry, particularly in the cooler night side of the planet \citep{sanchez2022searching}. The non-detection of $\mathrm{CO_2}$ in the atmosphere of \textbf{WASP-77 A b} is consistent with recent JWST observations \citep{august2023confirmation}. The detection of $\mathrm{H_2S}$, particularly on \textbf{WASP-76 b}, as the only sulfur-bearing molecular entity, serves as motivation for emphasizing the significance of sulfur chemistry within the atmospheres of irradiated gas giants.

\newpage
\section*{acknowledgements}L.M. acknowledges financial support from DAE and DST-SERB research grants (SRG/2021/002116 and MTR/2021/000864) of the Government of India. D.D. also acknowledges financial support from the DST-SERB research grant SRG/2021/002116, which supported this work. L.M. thanks Dr. Paul Mollière of MPIA Heidelberg, Germany, for providing access to petitCODE and for actively participating in discussions regarding its use and applications. We would like to thank the anonymous referee for constructive comments that helped improve the manuscript.

\bibliographystyle{aasjournal}
\bibliography{references.bib}
\clearpage





\appendix
\section{Reaction profile diagrams for molecular interconversions}
\label{appendix_1}
\begin{figure}[H]
    \centering  
        \begin{minipage}[c]{0.32\textwidth}
            \includegraphics[width=0.95\textwidth]{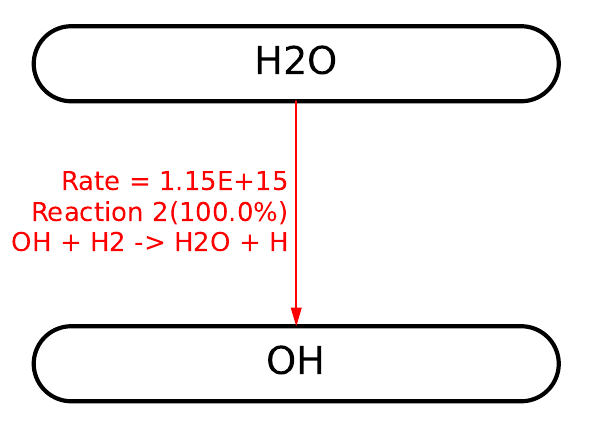}
        \end{minipage}
        %\columnbreak
         \begin{minipage}[c]{0.32\textwidth}
           \includegraphics[width=0.95\textwidth]{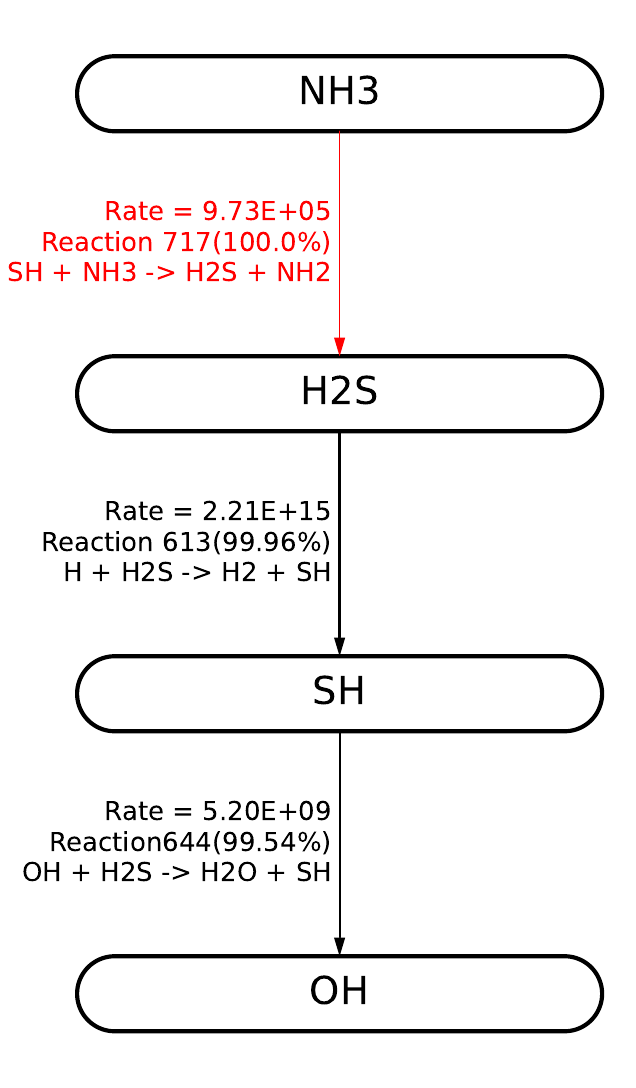}
        \end{minipage}
         \begin{minipage}[c]{0.32\textwidth}
            \includegraphics[width=0.95\textwidth]{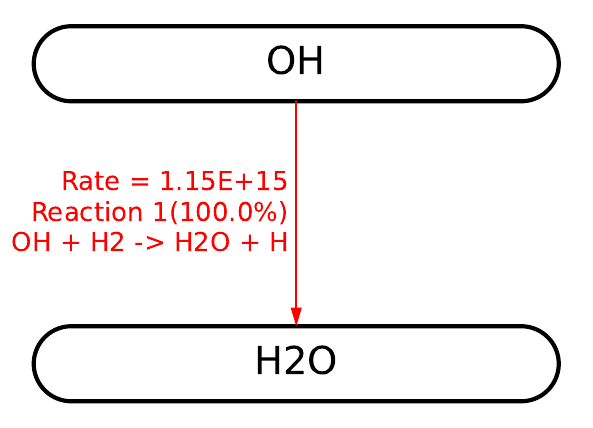}
        \end{minipage}
        \begin{minipage}[c]{0.32\textwidth}
         \includegraphics[width=0.95\textwidth]{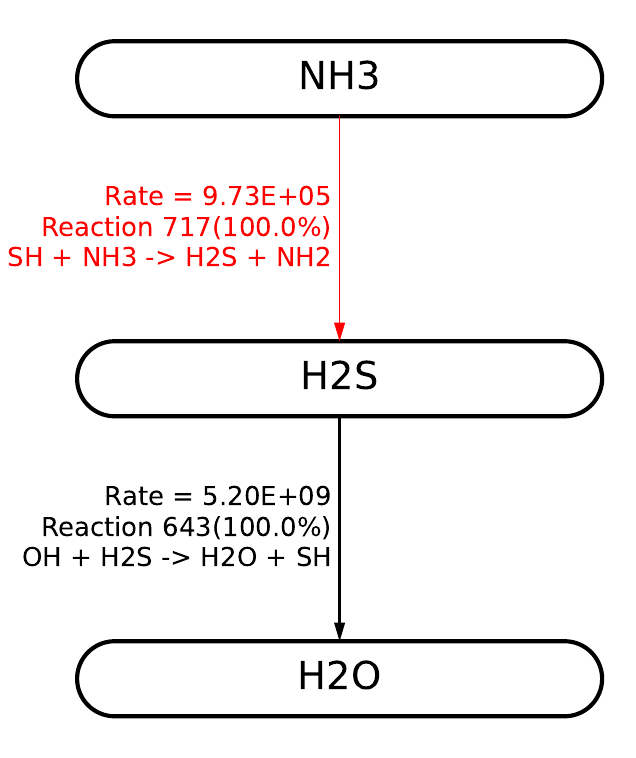}
        \end{minipage}
         \begin{minipage}[c]{0.32\textwidth}
          \includegraphics[width=0.95\textwidth]{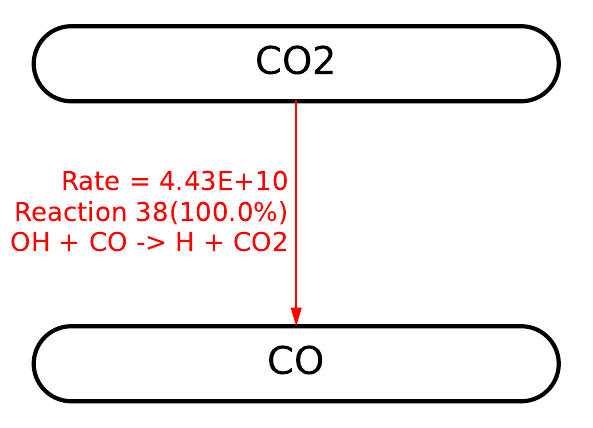}
        \end{minipage}
        \begin{minipage}[c]{0.32\textwidth}
          \includegraphics[width=0.95\textwidth]{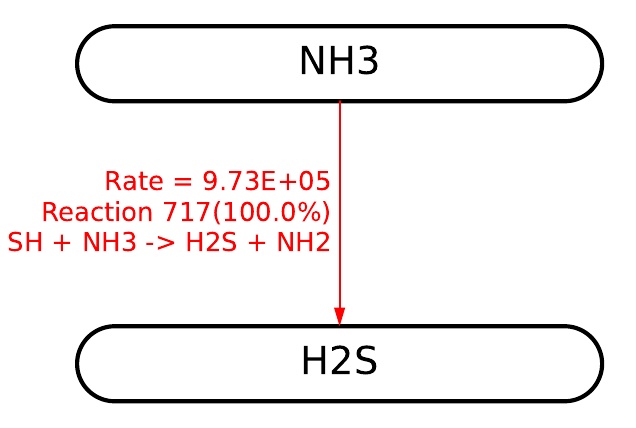}
        \end{minipage}
\caption{Shortest pathway analysis for molecular conversion using Dijkstra's algorithm at 0.01 bar. The conversion networks are shown based on Figure \ref{fig:chemisrty} for the atmosphere of \textbf{WASP-77 A b}. The texts besides the arrows explain the reactions involved (\textit{First row}: reaction rates in $\mathrm{cm^{-1} s^{-1}}$, \textit{Second row}: reaction number from the VULCAN network. The odd number is for the forward reaction and the even number is for the backward reaction. The contribution of that particular reaction to the interconversion has also been provided in brackets. \textit{Third row}: reaction involved in the interconversion.) Thicker black lines represent faster reactions, while the red lines represent the slowest reaction, i.e. the rate-limiting step for the molecular conversion pathway.}
\label{fig:shortest_wasp76b}
\end{figure}

\begin{figure*}
    \centering  
        \begin{minipage}[c]{0.19\textwidth}
            \includegraphics[width=0.95\textwidth]{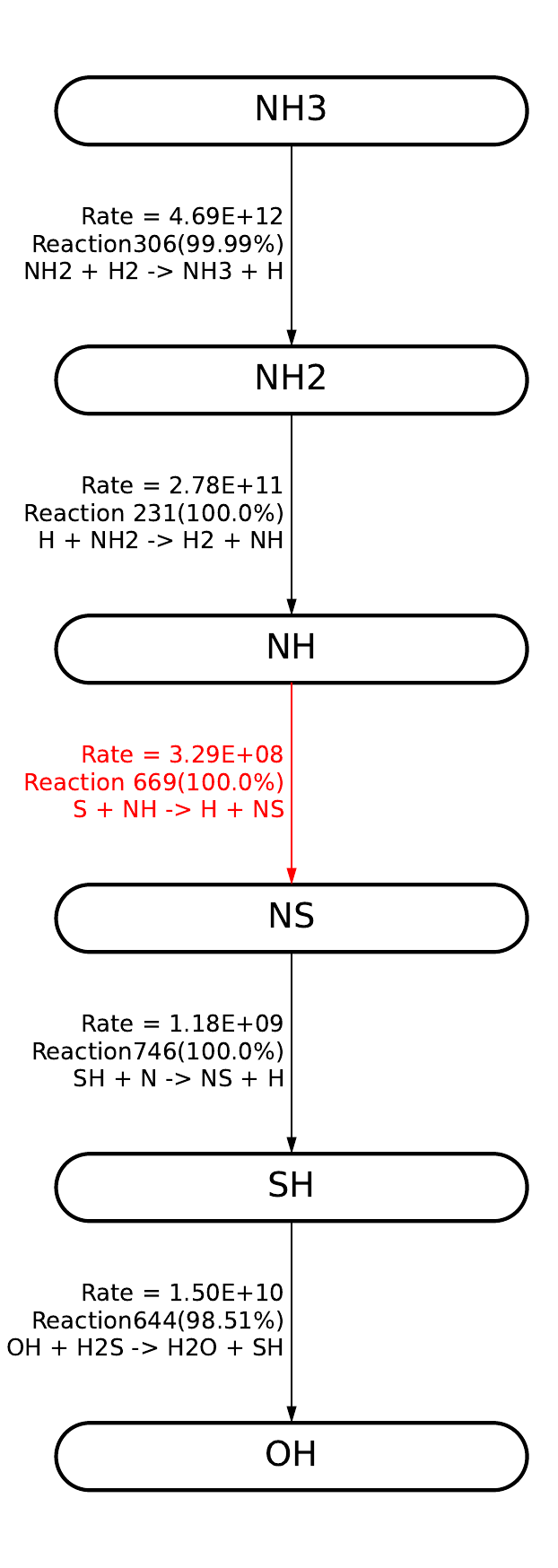}
        \end{minipage}
        %\columnbreak
         \begin{minipage}[c]{0.19\textwidth}
            \includegraphics[width=0.95\textwidth]{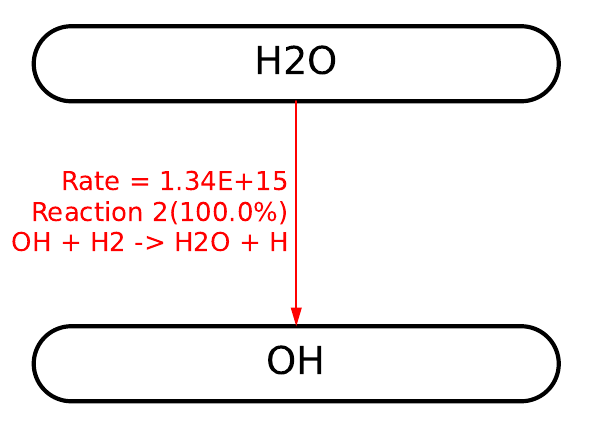}
        \end{minipage}
         \begin{minipage}[c]{0.19\textwidth}
            \includegraphics[width=0.95\textwidth]{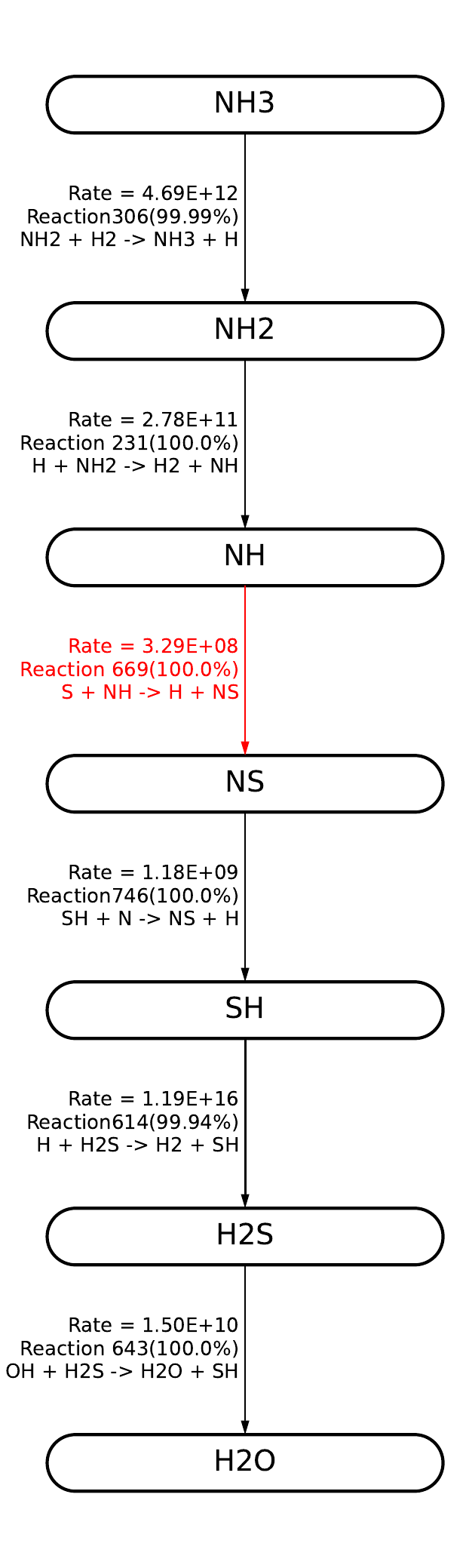}
        \end{minipage}
        \begin{minipage}[c]{0.19\textwidth}
          \includegraphics[width=0.95\textwidth]{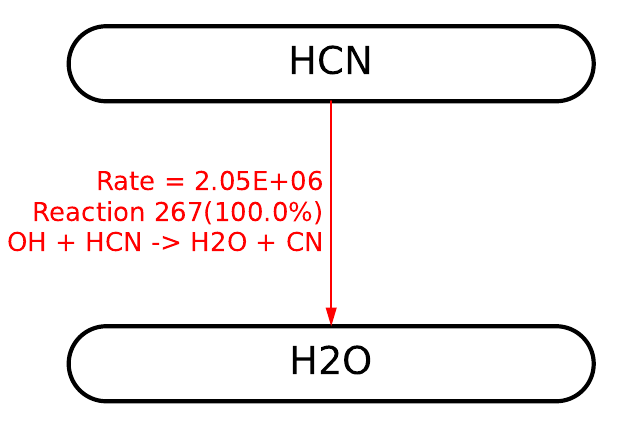}
        \end{minipage}
         \begin{minipage}[c]{0.19\textwidth}
          \includegraphics[width=0.95\textwidth]{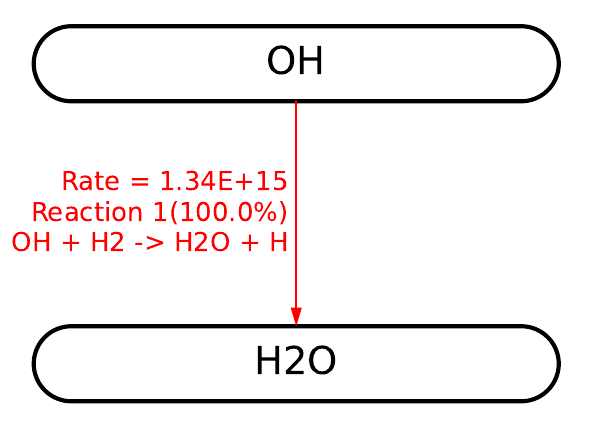}
        \end{minipage}
         \begin{minipage}[c]{0.24\textwidth}
          \includegraphics[width=0.85\textwidth]{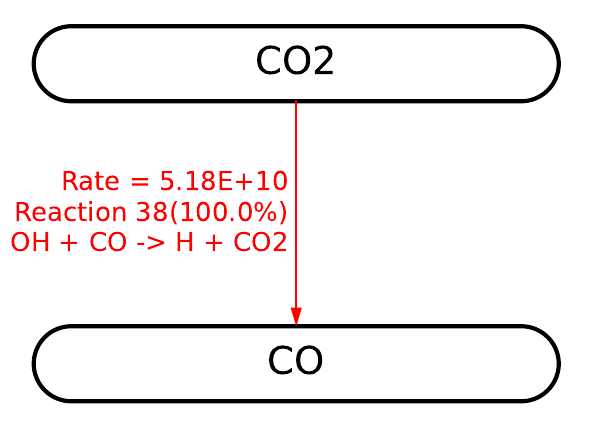}
        \end{minipage}
        \begin{minipage}[c]{0.24\textwidth}
           \includegraphics[width=0.85\textwidth]{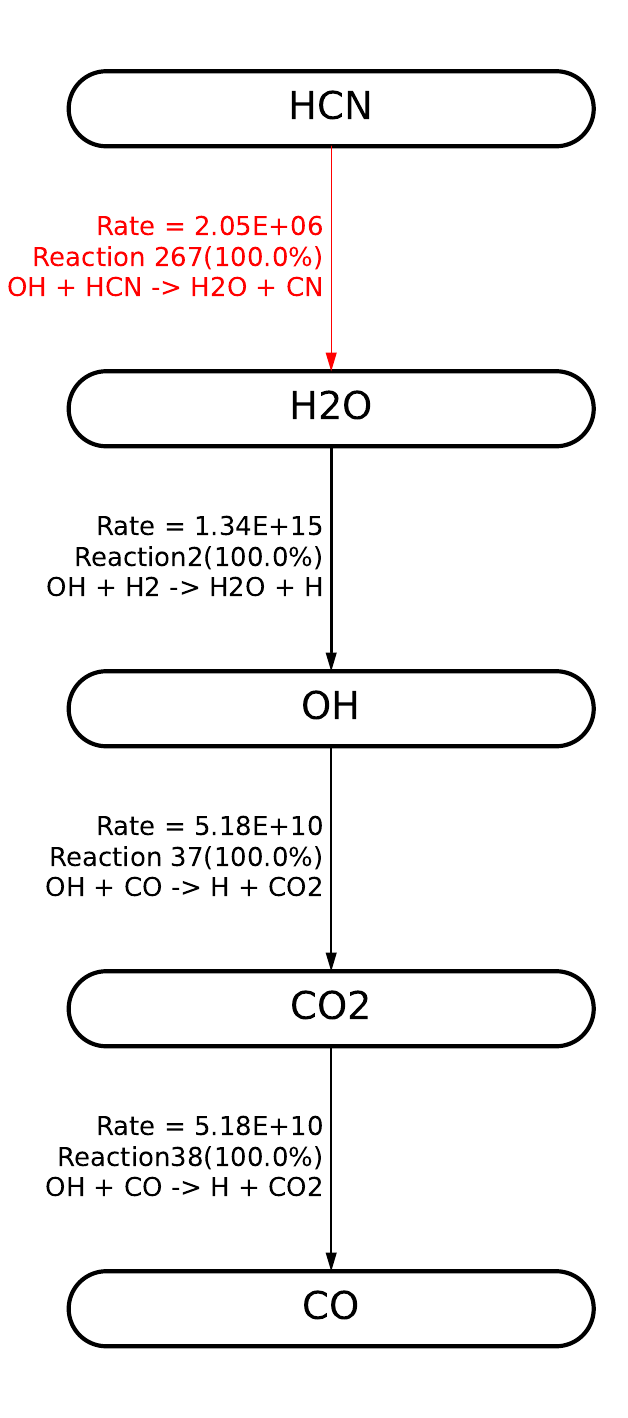}
        \end{minipage}
         \begin{minipage}[c]{0.24\textwidth}
            \includegraphics[width=0.85\textwidth]{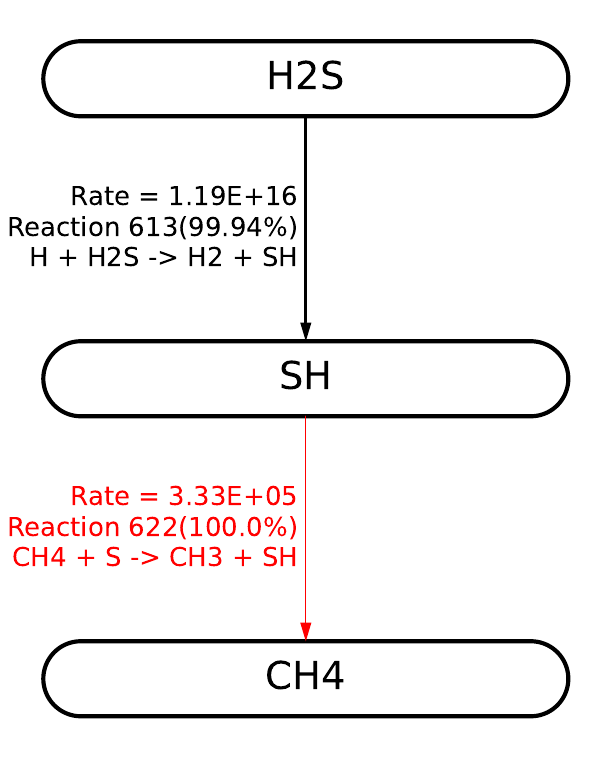}
        \end{minipage}
        \begin{minipage}[c]{0.24\textwidth}
           \includegraphics[width=0.85\textwidth]{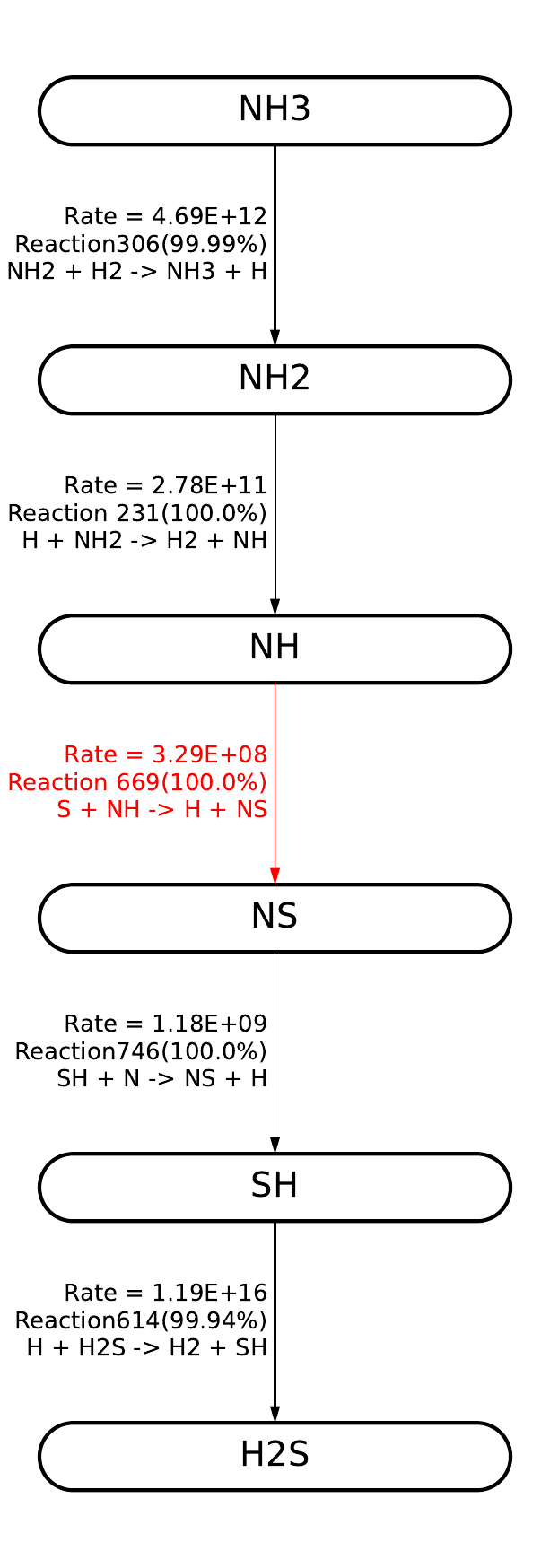}
        \end{minipage}
\caption{Same as Figure \ref{fig:shortest_wasp76b}, but for the \textbf{WASP-76 b} molecular network in Figure \ref{fig:chemisrty}.}
\label{fig:shortest_wasp77ab}
\end{figure*}

%%%%%%%%%%%%%%%%%%%%%%%%%%%%%%%%%%%%%%%%%%%%%%%%%%%%%%%%%%%%%%%%%%%%%%%%%%%%%%%%%%%%%%%%%
%%%%%%%%%%%%%%%%%%%%%%%%%%%%%%%%%%%%%%%%%%%%%%%%%%%%%%%%%%%%%%%%%%%%%%%%%%%%%%%%%%%%%%%%%

\clearpage
\section{Molecular bands of \textbf{WASP-76 b}}
\label{appendix_2}

\begin{figure}[H]

    \centering  
        \begin{minipage}[b]{0.98\columnwidth}
            \includegraphics[width=\columnwidth]{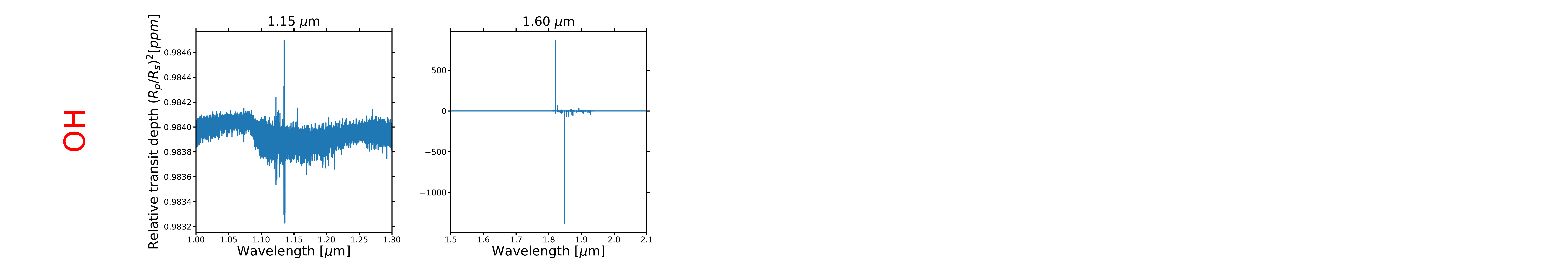}
        \end{minipage}
        %\columnbreak
         \begin{minipage}[b]{0.98\columnwidth}
            \includegraphics[width=\columnwidth]{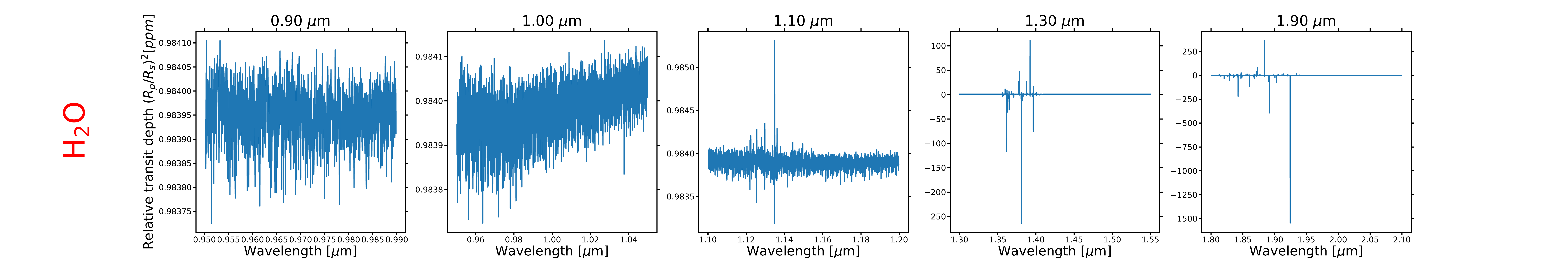}
        \end{minipage}
         \begin{minipage}[b]{0.98\columnwidth}
            \includegraphics[width=\columnwidth]{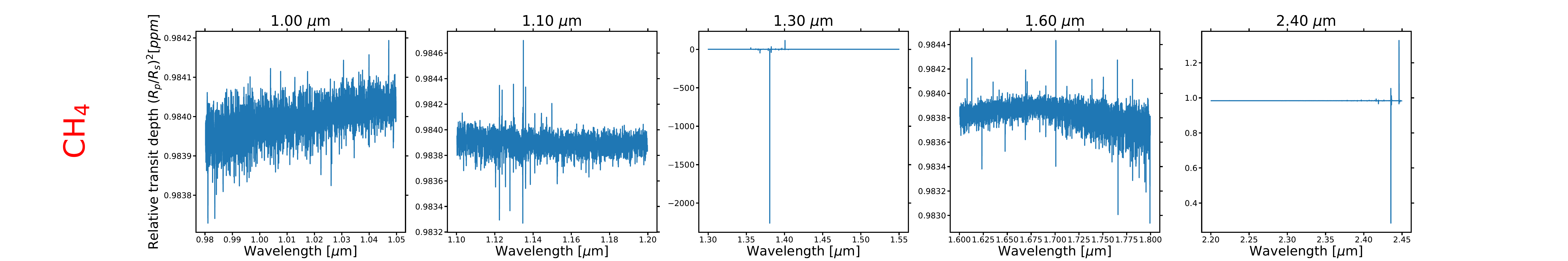}
        \end{minipage}
        \begin{minipage}[b]{0.98\columnwidth}
            \includegraphics[width=\columnwidth]{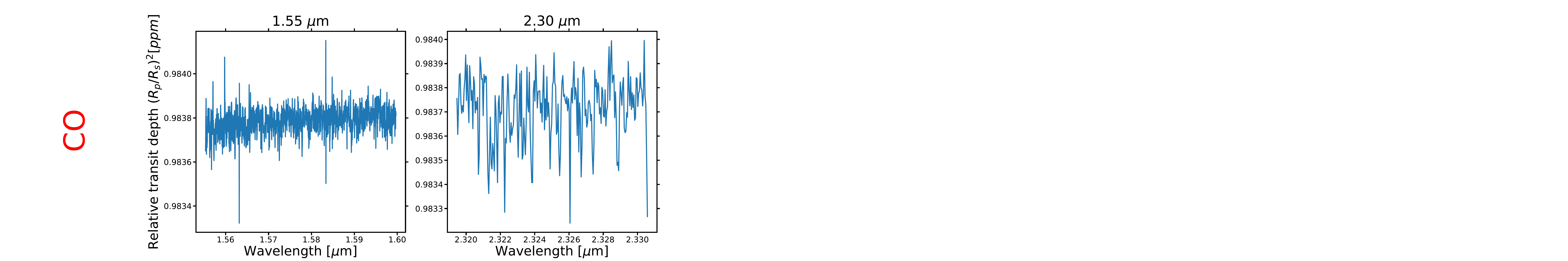}
        \end{minipage}
        \begin{minipage}[b]{0.98\columnwidth}
            \includegraphics[width=\columnwidth]{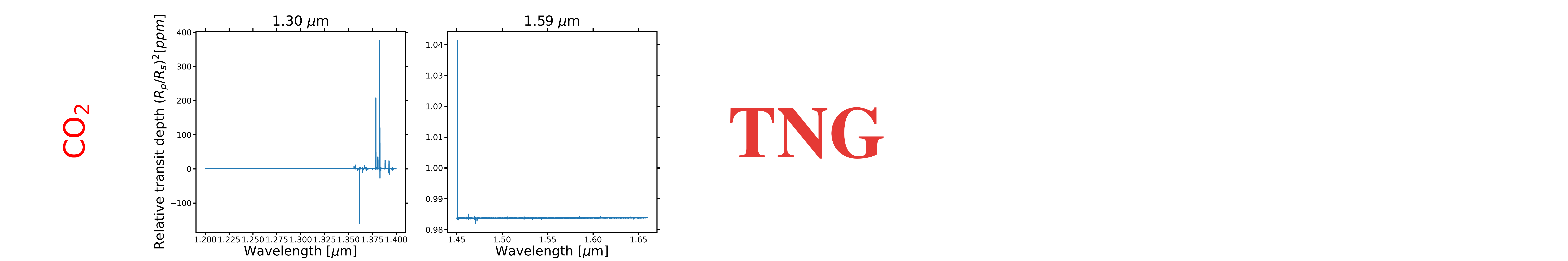}
        \end{minipage}
        \begin{minipage}[b]{0.98\columnwidth}
            \includegraphics[width=\columnwidth]{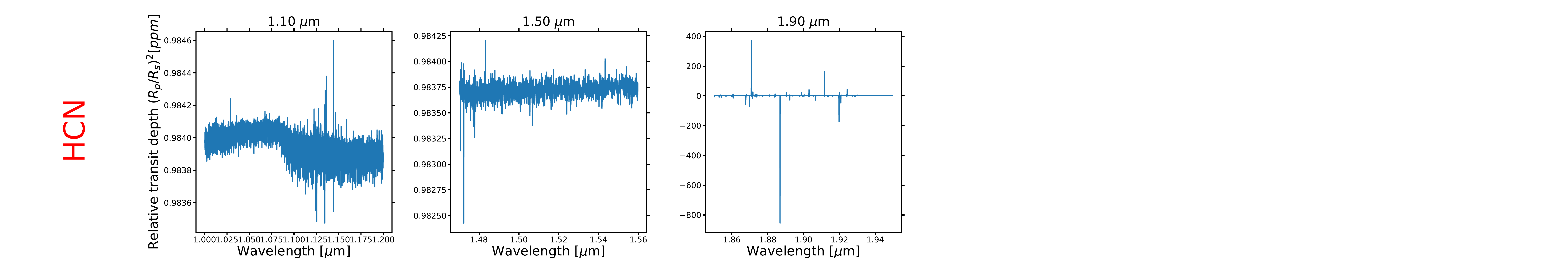}
        \end{minipage}
        \begin{minipage}[b]{0.98\columnwidth}
            \includegraphics[width=\columnwidth]{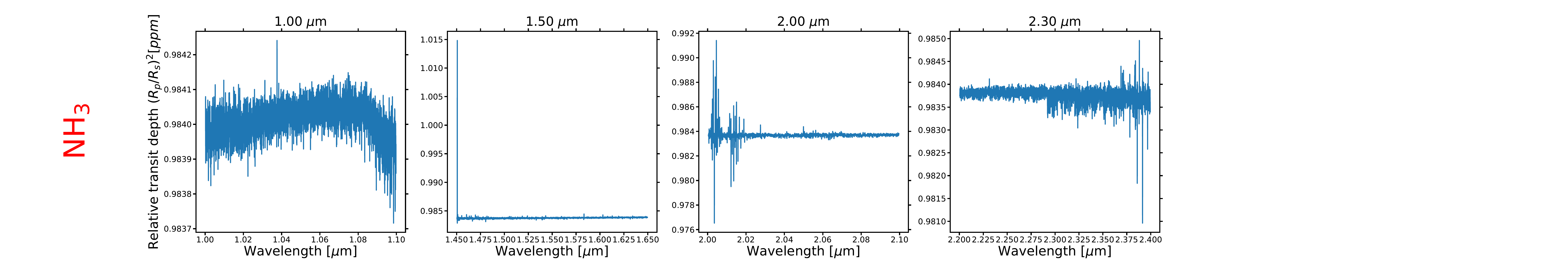}
        \end{minipage}
        \begin{minipage}[b]{0.98\columnwidth}
            \includegraphics[width=\columnwidth]{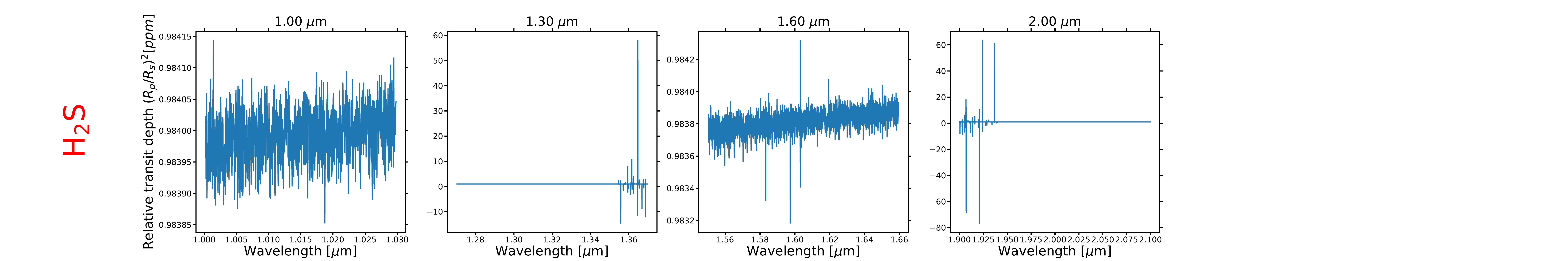}
        \end{minipage}      
\caption{Continuum subtracted \textbf{WASP-76 b} molecular bands using high pass filters, calculated for 3 transits for TNG.}
\label{fig:band_spectrum_TNG_76b}
\end{figure}

\begin{figure*}
    \centering  
        \begin{minipage}[b]{\columnwidth}
            \includegraphics[width=\columnwidth]{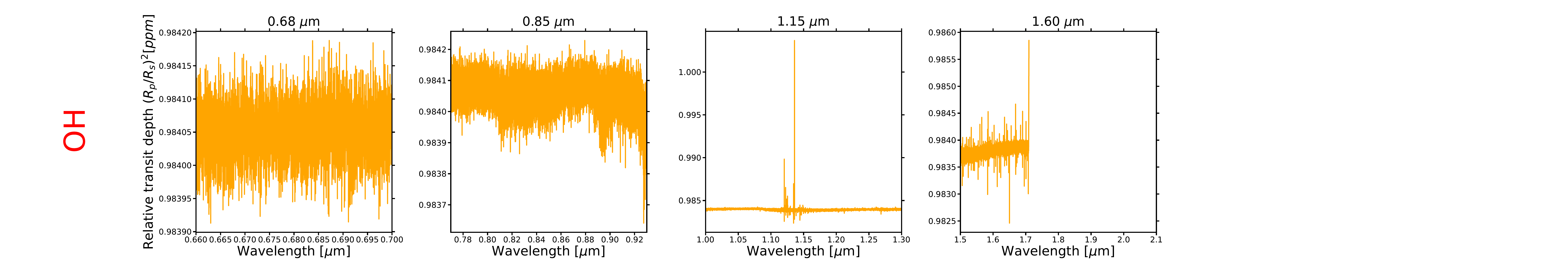}
        \end{minipage}
        %\columnbreak
         \begin{minipage}[b]{\columnwidth}
            \includegraphics[width=\columnwidth]{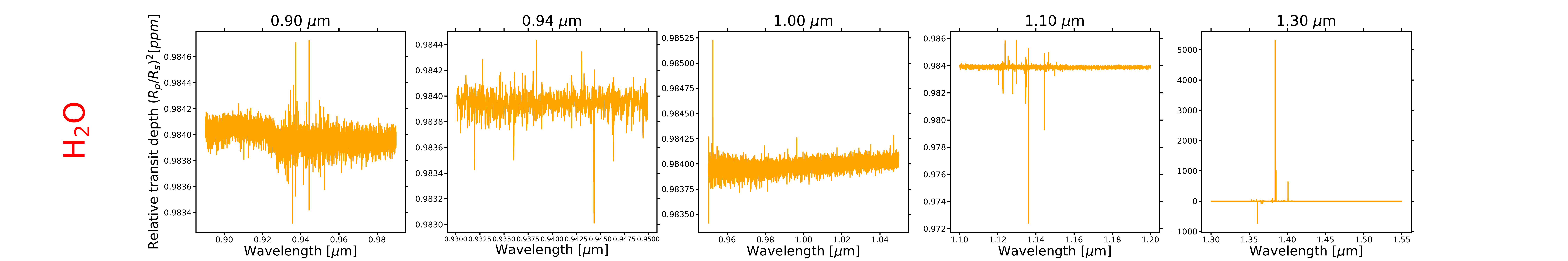}
        \end{minipage}
         \begin{minipage}[b]{\columnwidth}
            \includegraphics[width=\columnwidth]{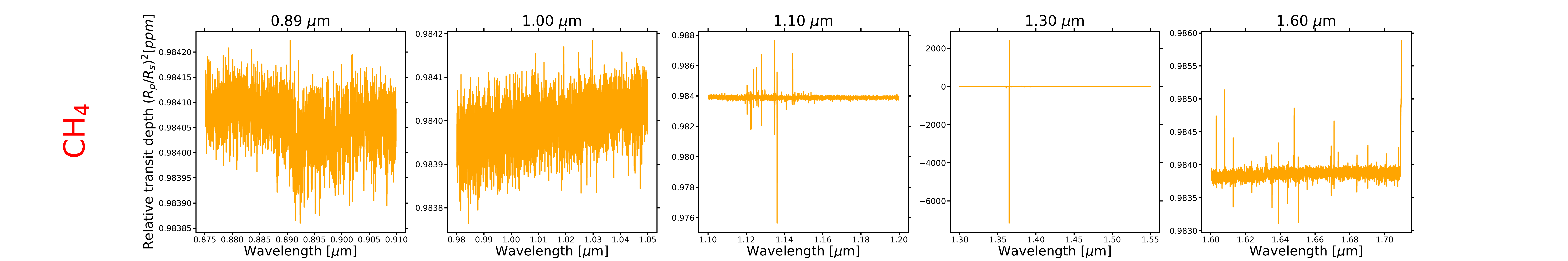}
        \end{minipage}
        \begin{minipage}[b]{\columnwidth}
            \includegraphics[width=\columnwidth]{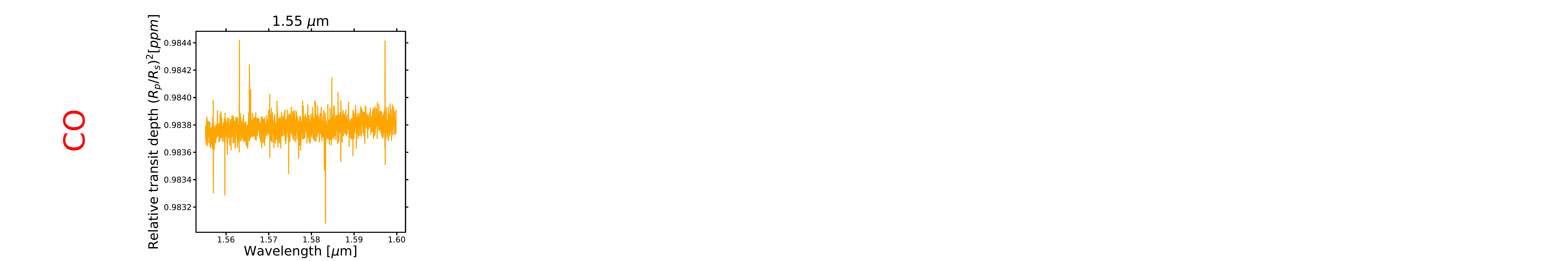}
        \end{minipage}
        \begin{minipage}[b]{\columnwidth}
            \includegraphics[width=\columnwidth]{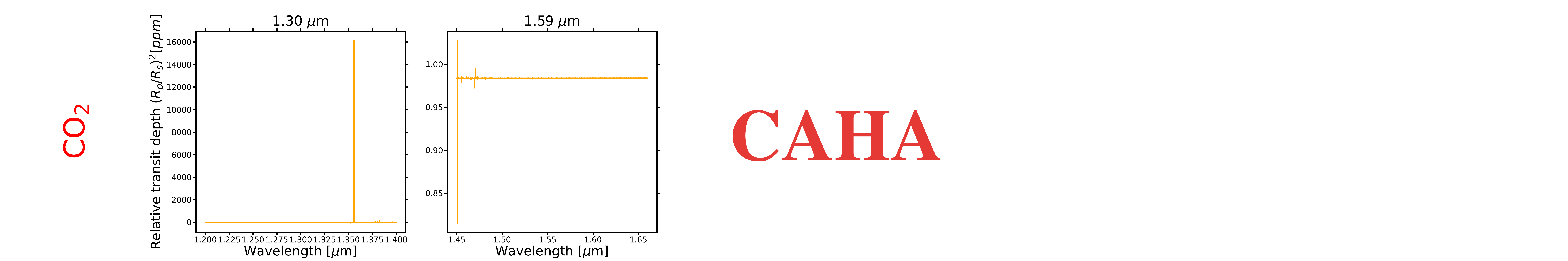}
        \end{minipage}
        \begin{minipage}[b]{\columnwidth}
            \includegraphics[width=\columnwidth]{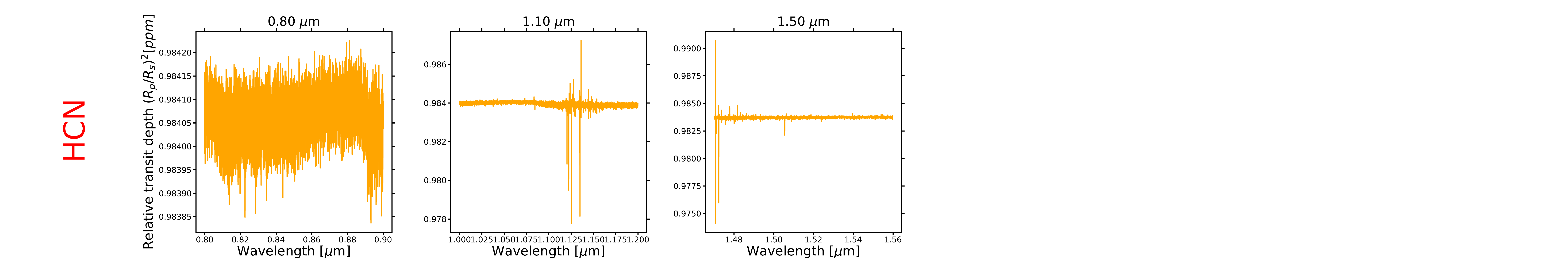}
        \end{minipage}
        \begin{minipage}[b]{\columnwidth}
            \includegraphics[width=\columnwidth]{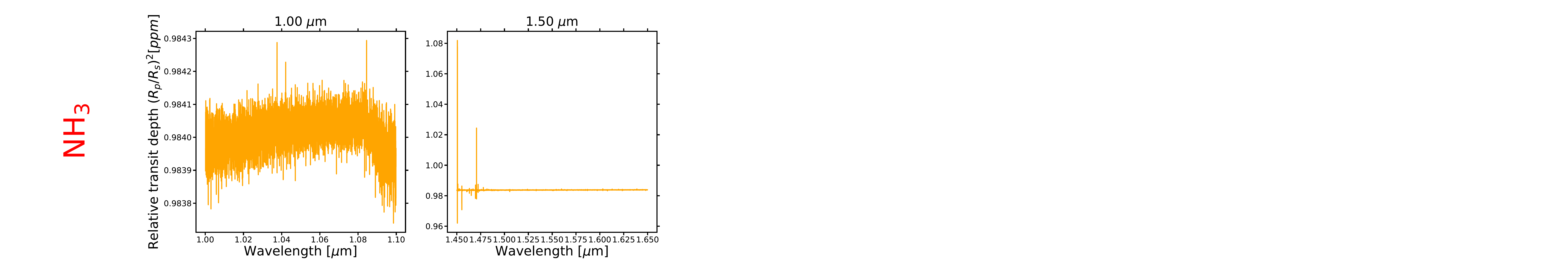}
        \end{minipage}
        \begin{minipage}[b]{\columnwidth}
            \includegraphics[width=\columnwidth]{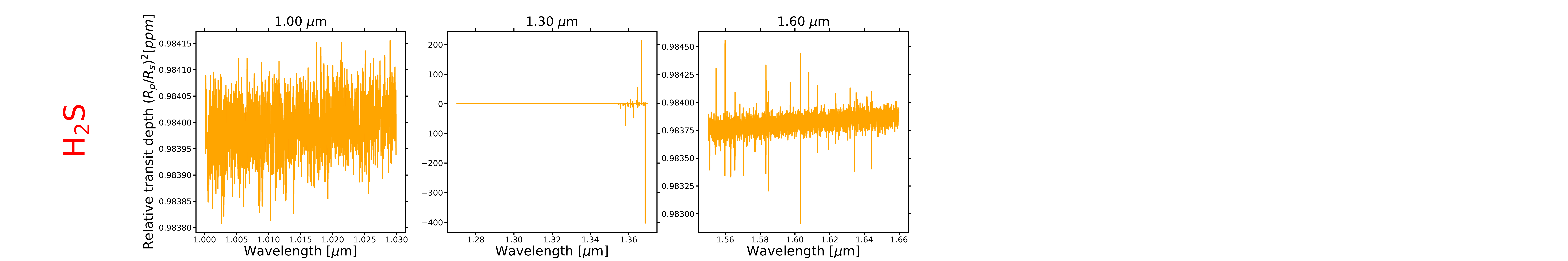}
        \end{minipage}
\caption{Same as Figure \ref{fig:band_spectrum_TNG_76b}, but for CAHA.}
\label{fig:band_spectrum_CAHA_76b}
\end{figure*}

\begin{figure*}
    \centering  
        \begin{minipage}[b]{\columnwidth}
            \includegraphics[width=\columnwidth]{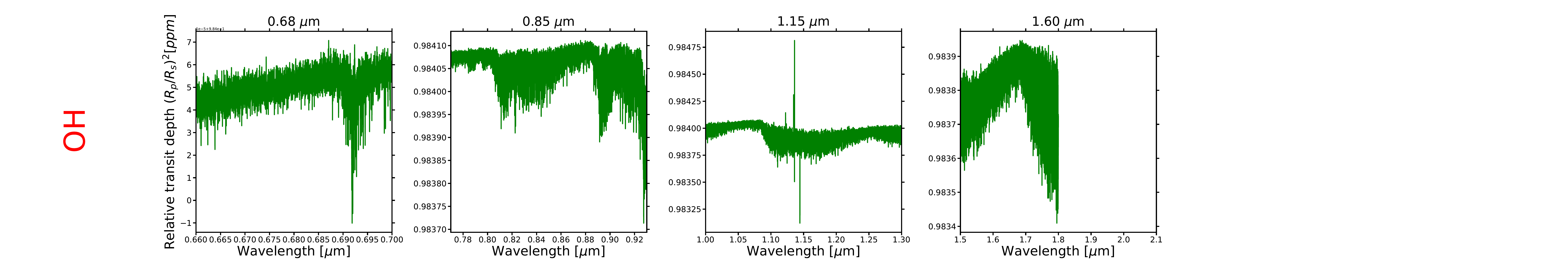}
        \end{minipage}
        %\columnbreak
         \begin{minipage}[b]{\columnwidth}
            \includegraphics[width=\columnwidth]{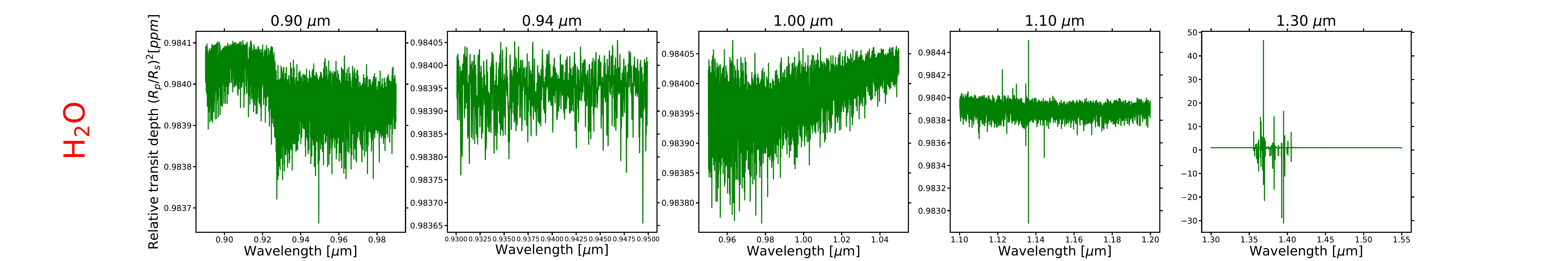}
        \end{minipage}
         \begin{minipage}[b]{\columnwidth}
            \includegraphics[width=\columnwidth]{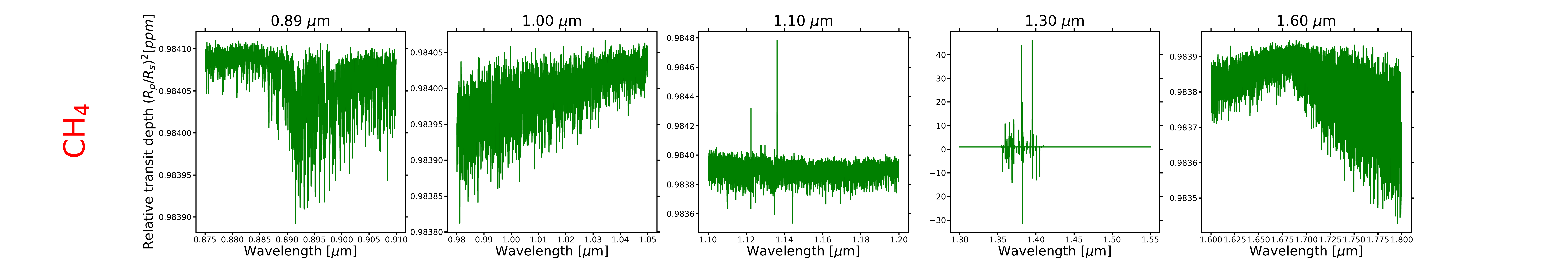}
        \end{minipage}
        \begin{minipage}[b]{\columnwidth}
            \includegraphics[width=\columnwidth]{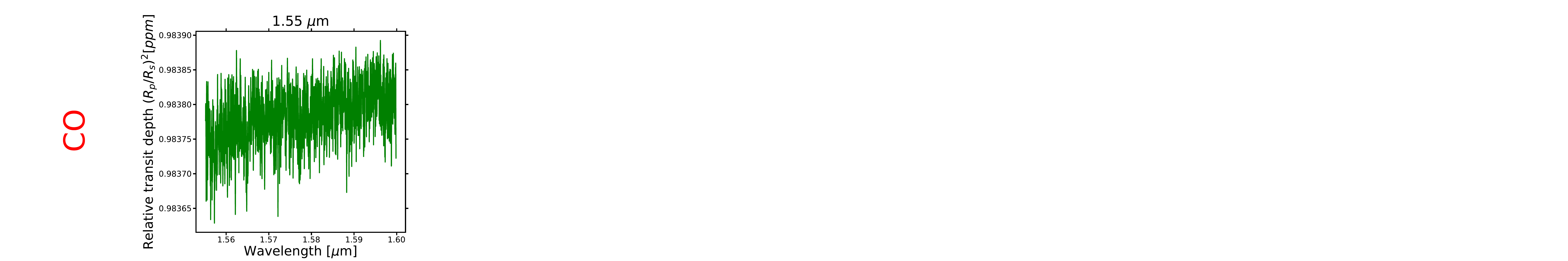}
        \end{minipage}
        \begin{minipage}[b]{\columnwidth}
            \includegraphics[width=\columnwidth]{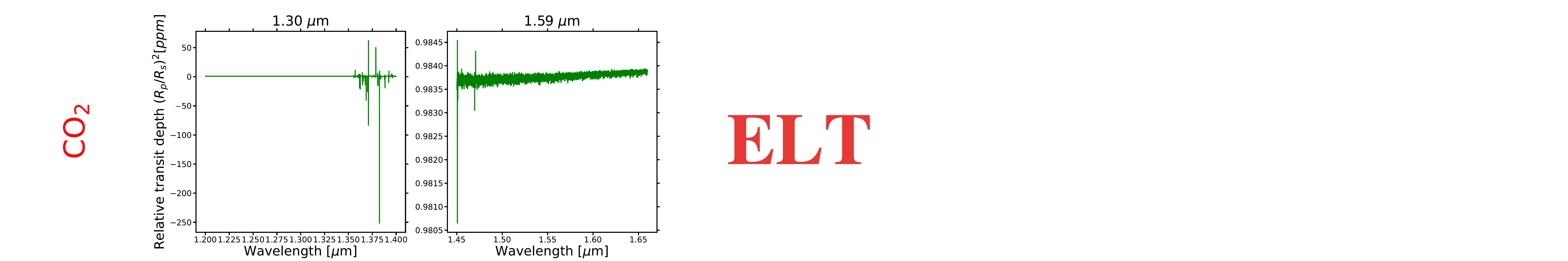}
        \end{minipage}
        \begin{minipage}[b]{\columnwidth}
            \includegraphics[width=\columnwidth]{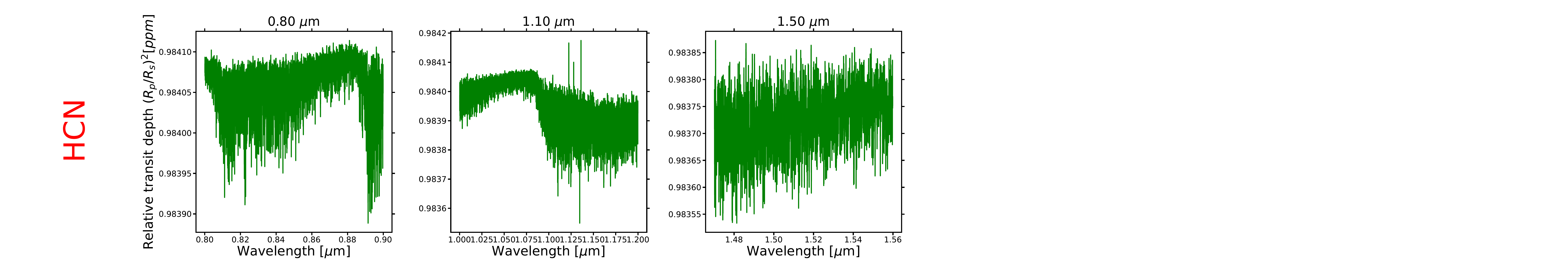}
        \end{minipage}
        \begin{minipage}[b]{\columnwidth}
            \includegraphics[width=\columnwidth]{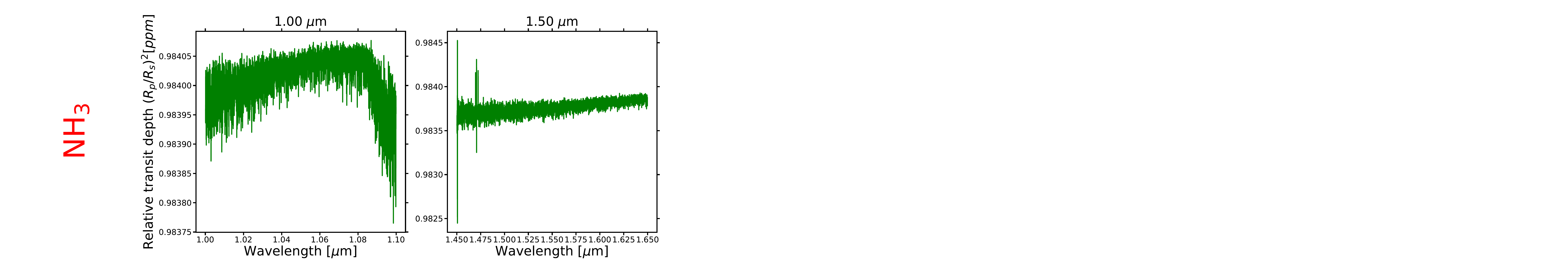}
        \end{minipage}
        \begin{minipage}[b]{\columnwidth}
            \includegraphics[width=\columnwidth]{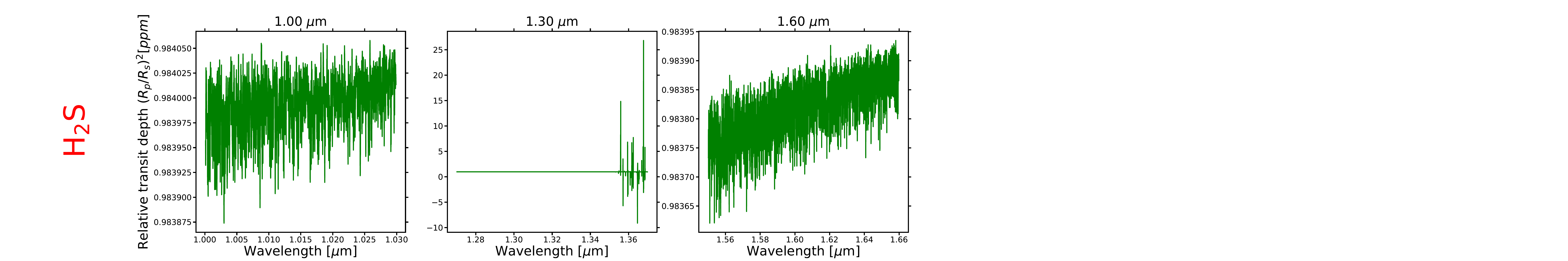}
        \end{minipage}
\caption{Same as Figure \ref{fig:band_spectrum_TNG_76b}, but for ELT.}
\label{fig:band_spectrum_ELT_76b}
\end{figure*}

\clearpage
%%%%%%%%%%%%%%%%%%%%%%%%%%%%%%%%%%%%%%%%%%%%%%%%%%%%%%%%%%%%%%%%%%%%%%%%%%%%%%%%%%%%%%%%%%%%%%%%%%%%%%%%%%%%%
%%%%%%%%%%%%%%%%%%%%%%%%%%%%%%%%%%%%%%%%%%%%%%%%%%%%%%%%%%%%%%%%%%%%%%%%%%%%%%%%%%%%%%%%%%%%%%%%%%%%%%%%%%%%%
\section{Molecular bands of \textbf{WASP-77 A b}}
\label{appendix_3}
\begin{figure}[H]
    \centering  
        \begin{minipage}[b]{\textwidth}
            \includegraphics[width=\columnwidth]{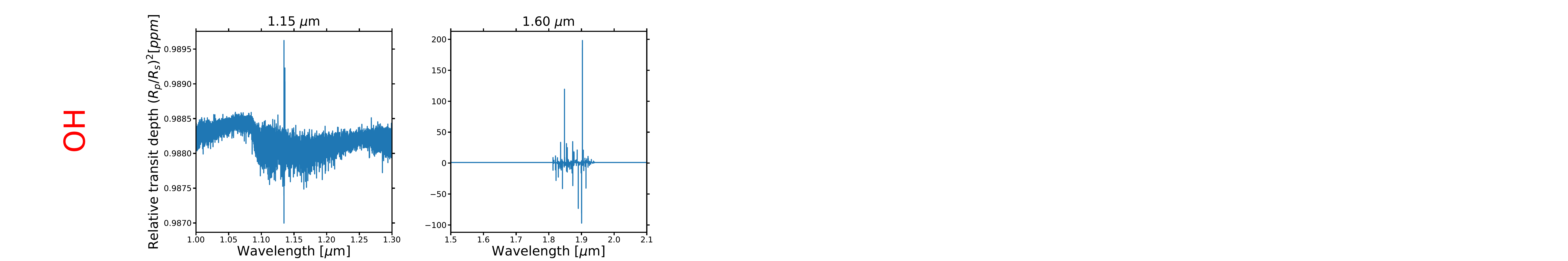}
        \end{minipage}
        %\columnbreak
         \begin{minipage}[b]{\textwidth}
            \includegraphics[width=\columnwidth]{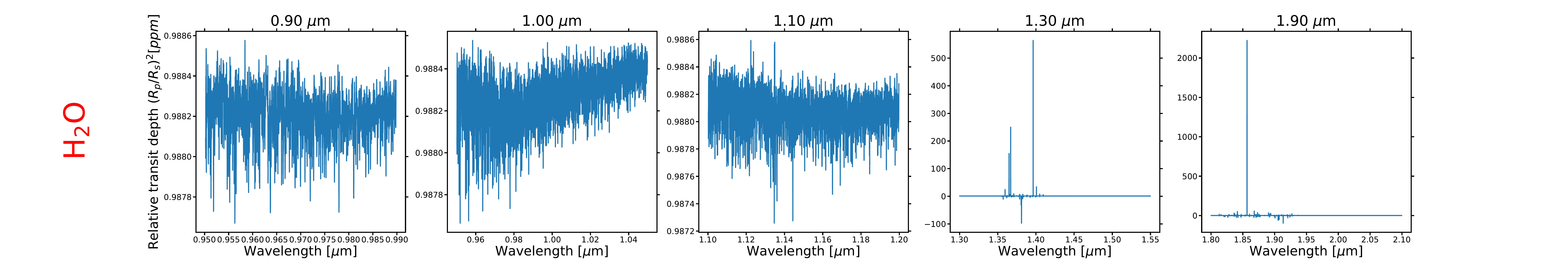}
        \end{minipage}
         \begin{minipage}[b]{\textwidth}
            \includegraphics[width=\columnwidth]{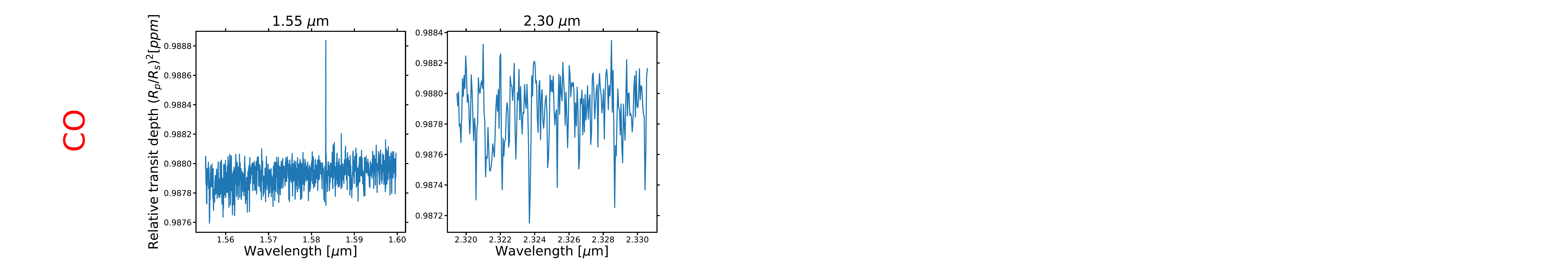}
        \end{minipage}
        \begin{minipage}[b]{\textwidth}
            \includegraphics[width=\columnwidth]{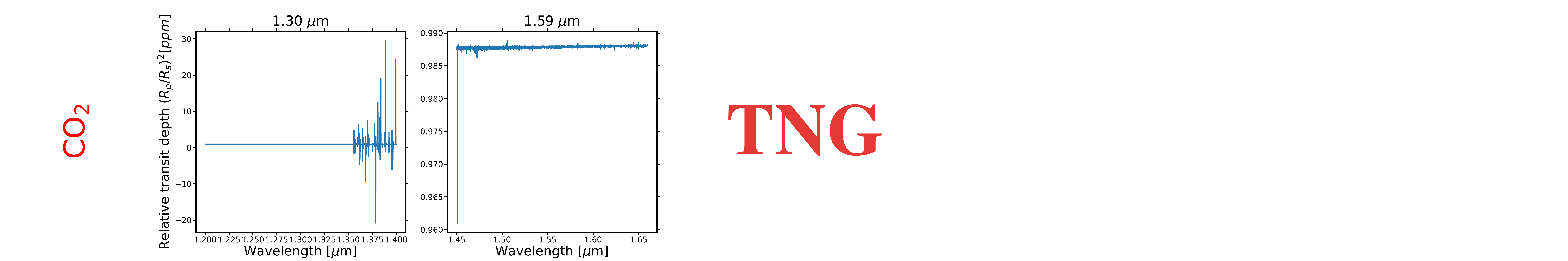}
        \end{minipage}
        \begin{minipage}[b]{\textwidth}
            \includegraphics[width=\columnwidth]{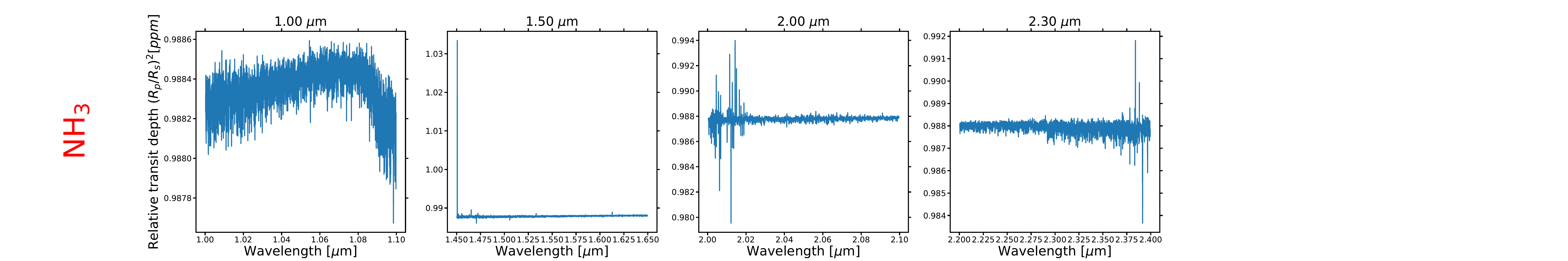}
        \end{minipage}
        \begin{minipage}[b]{\textwidth}
            \includegraphics[width=\columnwidth]{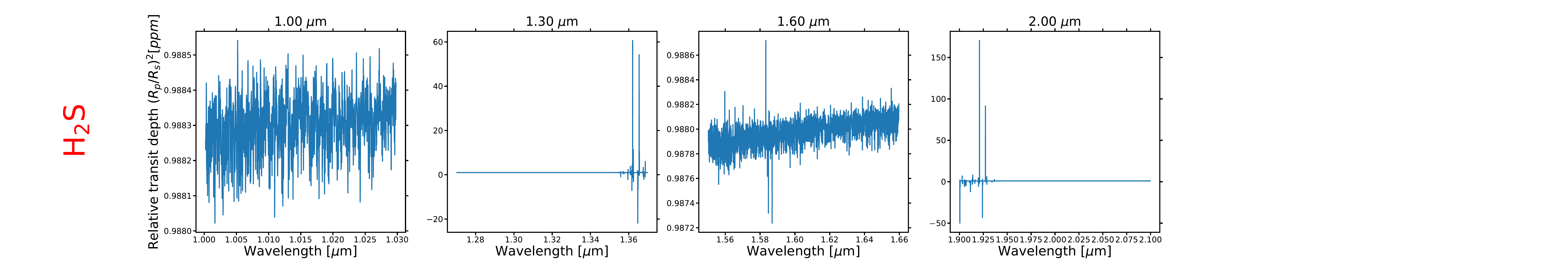}
        \end{minipage}
\caption{Continuum subtracted \textbf{WASP-77 A b} molecular bands using high pass filters, calculated for 3 transits for TNG.}
\label{fig:band_spectrum_TNG_77ab}
\end{figure}

\clearpage
\begin{figure*}
    \centering  
        \begin{minipage}[b]{\columnwidth}
            \includegraphics[width=\columnwidth]{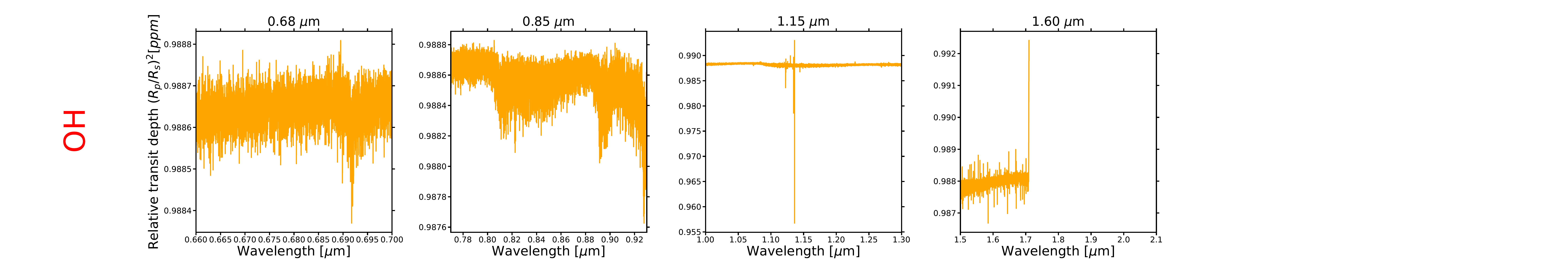}
        \end{minipage}
        %\columnbreak
         \begin{minipage}[b]{\columnwidth}
            \includegraphics[width=\columnwidth]{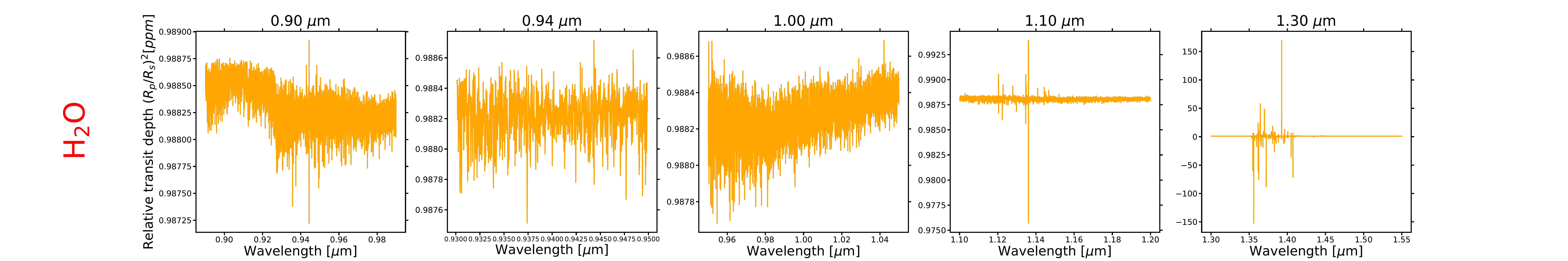}
        \end{minipage}
         \begin{minipage}[b]{\columnwidth}
            \includegraphics[width=\columnwidth]{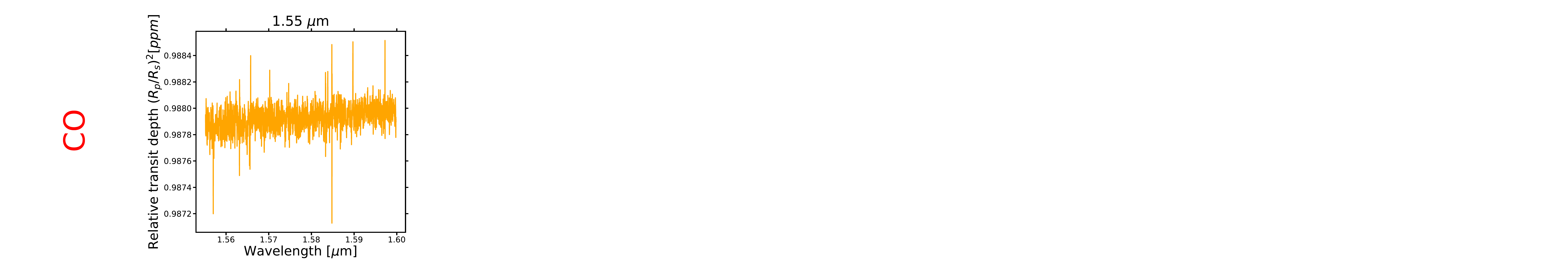}
        \end{minipage}
        \begin{minipage}[b]{\columnwidth}
            \includegraphics[width=\columnwidth]{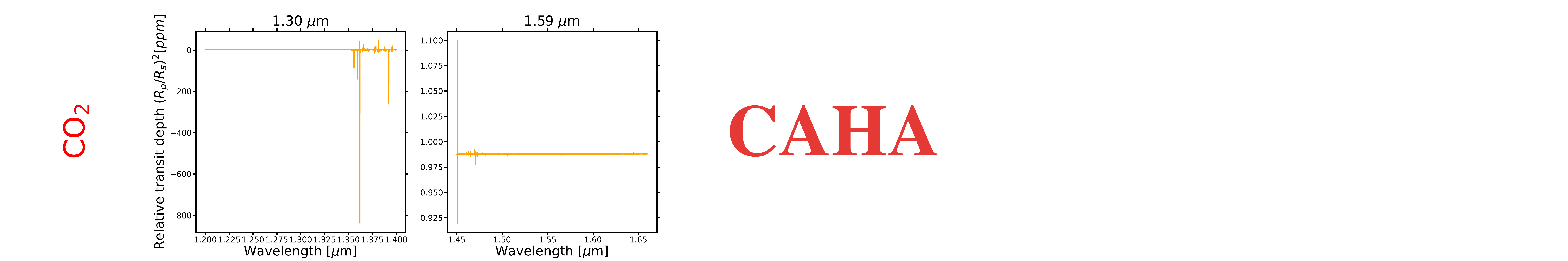}
        \end{minipage}
        \begin{minipage}[b]{\columnwidth}
            \includegraphics[width=\columnwidth]{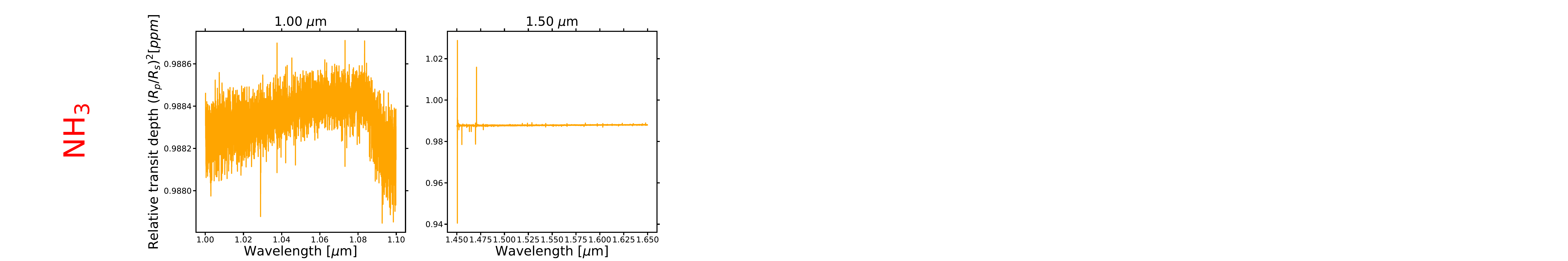}
        \end{minipage}
        \begin{minipage}[b]{\columnwidth}
            \includegraphics[width=\columnwidth]{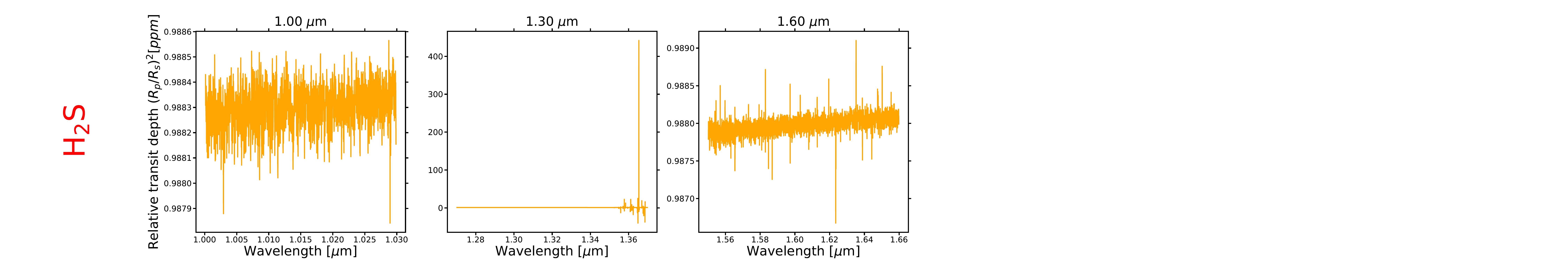}
        \end{minipage}
\caption{Same as Figure \ref{fig:band_spectrum_TNG_77ab}, but for CAHA.}
\label{fig:band_spectrum_CAHA_77ab}
\end{figure*}

\clearpage
\begin{figure*}
    \centering  
        \begin{minipage}[b]{\columnwidth}
            \includegraphics[width=\columnwidth]{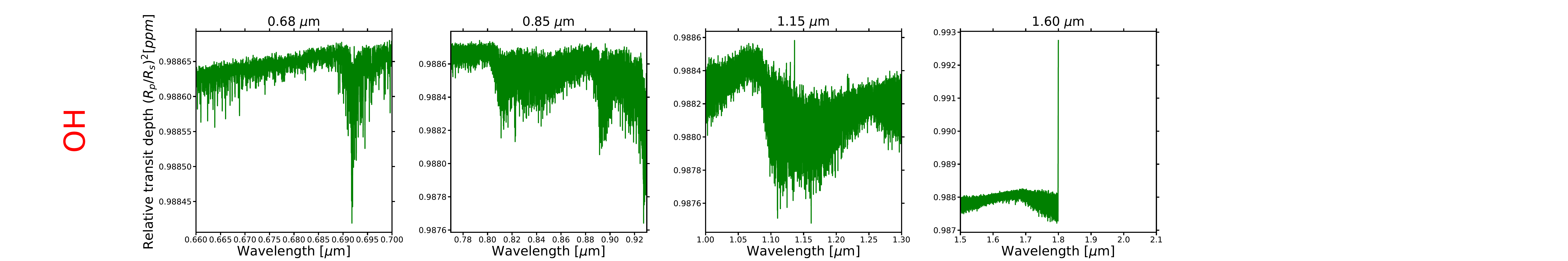}
        \end{minipage}
        %\columnbreak
         \begin{minipage}[b]{\columnwidth}
            \includegraphics[width=\columnwidth]{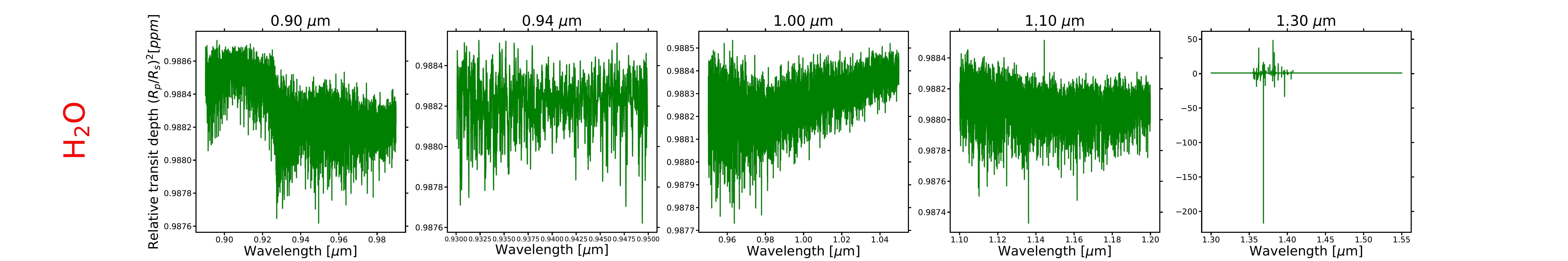}
        \end{minipage}
         \begin{minipage}[b]{\columnwidth}
            \includegraphics[width=\columnwidth]{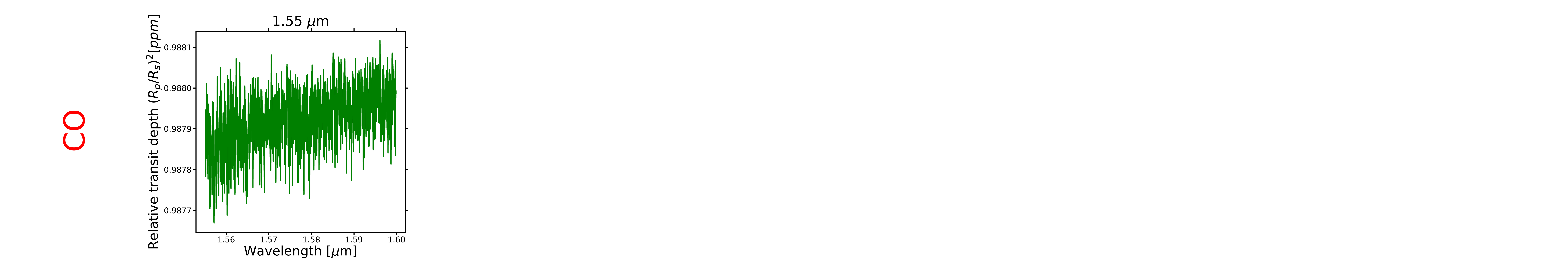}
        \end{minipage}
        \begin{minipage}[b]{\columnwidth}
            \includegraphics[width=\columnwidth]{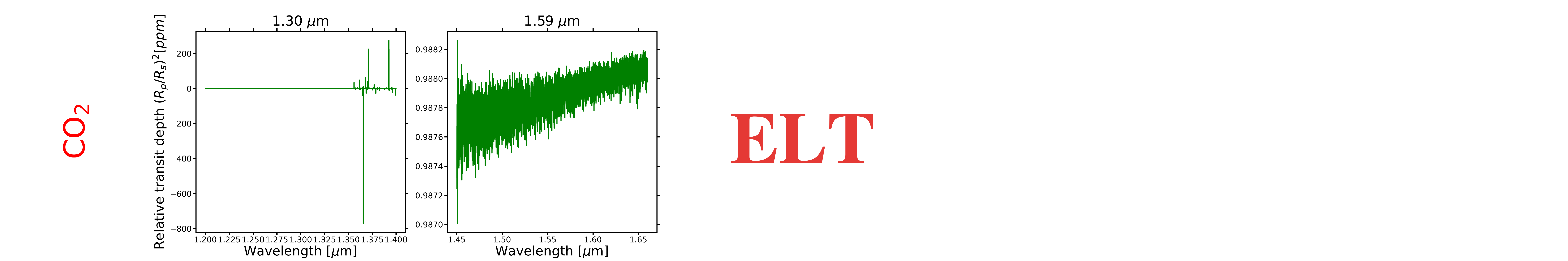}
        \end{minipage}
        \begin{minipage}[b]{\columnwidth}
            \includegraphics[width=\columnwidth]{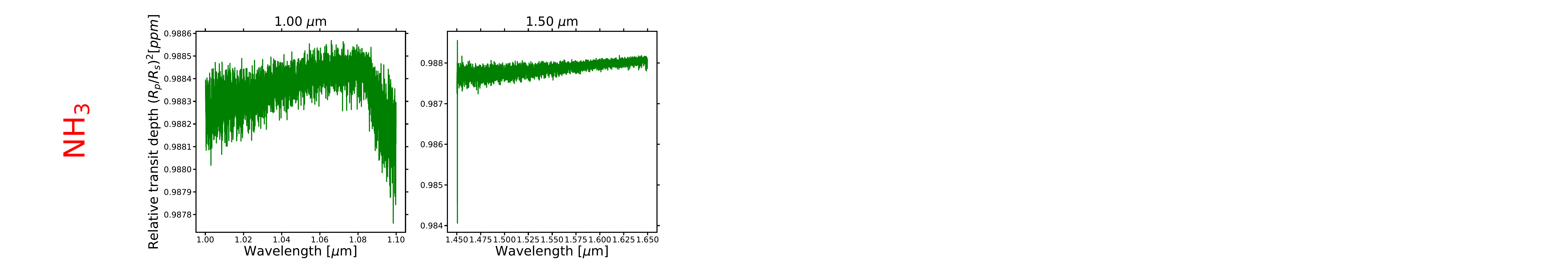}
        \end{minipage}
        \begin{minipage}[b]{\columnwidth}
            \includegraphics[width=\columnwidth]{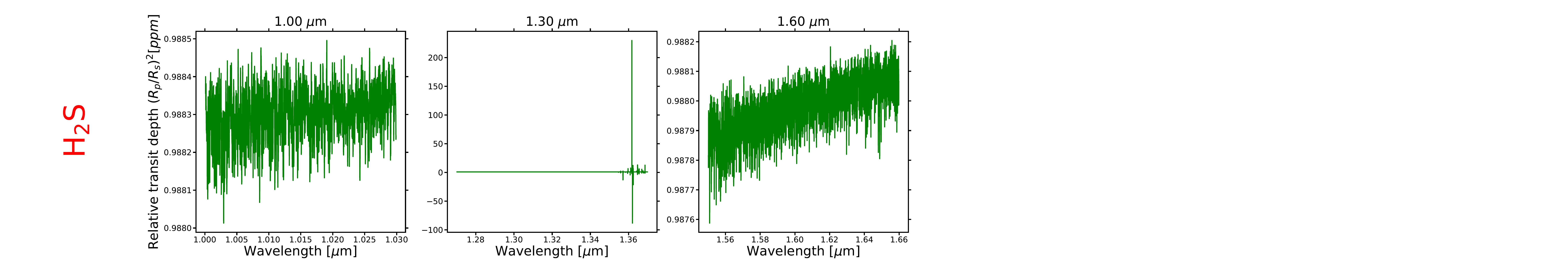}
        \end{minipage}
\caption{Same as Figure \ref{fig:band_spectrum_TNG_77ab}, but for ELT.}
\label{fig:band_spectrum_ELT_77ab}
\end{figure*}




\end{document}